\documentclass[twocolumn,tighten]{aastex61}

\usepackage{color}
\usepackage{xcolor}
\usepackage{natbib}
\usepackage{graphicx}	
\usepackage{amssymb}
\usepackage{amsmath}
\usepackage{indentfirst}
\usepackage{multirow}
\usepackage{rotating}
\usepackage{hhline}
\usepackage{url}
\usepackage{longtable}
\usepackage{resizegather}

\newcommand{\blue}{\color{black}}
\newcommand{\red}{\color{black}}

\newcommand{\OL}{\Omega_{\Lambda}}
\newcommand{\OM}{\Omega_{\rm M}}
\newcommand{\mB}{m_{\rm B}}
\newcommand{\tlam}{t,\lambda}

\newcommand{\FoMtot}{{\rm FoM}_{\rm tot}}
\newcommand{\FoMcurr}{{\rm FoM}_{\rm tot,curr}}
\newcommand{\FoMopt}{{\rm FoM}_{\rm tot,opt}}
\newcommand{\FoMstat}{{\rm FoM}_{\rm stat}}

\newcommand{\NSTRATEGY}{11} 
\newcommand{\SIMw}{-1}
\newcommand{\SIMwa}{0}
\newcommand{\SIMOM}{0.3}
\newcommand{\SIMOL}{0.7}

\newcommand{\SIMalpha}{0.14}
\newcommand{\SIMbeta}{3.1}
  
\newcommand{\Ctot}{{\cal C}_{\rm tot}}
\newcommand{\SCC}{S_{\rm CC}}

\newcommand{\WFIRST}{{\it WFIRST}} 
\newcommand{\SNANA}{{\sc snana}}          
\newcommand{\BBC}{{\tt BBC}}              
\newcommand{\CosmoMC}{{\tt CosmoMC}}      
\newcommand{\CosmoMCfast}{{\tt CosmoMC*}}

\received{2017 February 7}
\revised{2018 March 14}
\accepted{2018 March 23}

\begin{document}

\title[Simulations of the {\it WFIRST} SN Survey]{Simulations of the {\it WFIRST} Supernova Survey and Forecasts of Cosmological Constraints}

\correspondingauthor{R.~Hounsell}
\email{rhounsel@ucsc.edu}

\author{R.~Hounsell}
\affiliation{Department of Astronomy and Astrophysics, University of California Santa Cruz, 1156 High St., Santa Cruz, CA 95064, USA}
\affiliation{Department of Astronomy, University of Illinois Urbana Champaign, 1002 W Green St., Urbana, IL 61801, USA}

\author{D.~Scolnic}
\affiliation{Kavli Institute for Cosmological Physics at the University of Chicago, 5620 S Ellis Ave., Chicago, IL 60637, USA}

\author{R.~J.~Foley}
\affiliation{Department of Astronomy and Astrophysics, University of California Santa Cruz, 1156 High St., Santa Cruz, CA 95064, USA}

\author{R.~Kessler}
\affiliation{Kavli Institute for Cosmological Physics at the University of Chicago, 5620 S Ellis Ave., Chicago, IL 60637, USA}

\author{V.~Miranda}
\affiliation{University of Pennsylvania, Department of Physics \& Astronomy, 209 South 33rd St., Philadelphia, PA 19104-6396, USA}

\author{A.~Avelino}
\affiliation{Harvard-Smithsonian Center for Astrophysics, 60 Garden Street, Cambridge, MA 02138, USA}

\author{R.~C.~ Bohlin}
\affiliation{Space Telescope Science Institute, 3700 San Martin Dr., Baltimore, MD 21218, USA}

\author{A.~V.~Filippenko}
\affiliation{Department of Astronomy, University of California, Berkeley, CA 94720-3411, USA}
\affiliation{Miller Senior Fellow, Miller Institute for Basic Research in Science, University of California, Berkeley, CA 94720, USA}

\author{J.~Frieman}
\affiliation{Kavli Institute for Cosmological Physics at the University of Chicago, 5620 S Ellis Ave., Chicago, IL 60637, USA}
\affiliation{Fermi National Accelerator Laboratory, P.\ O.\ Box 500, Batavia, IL 60510, USA }

\author{S.~W.~Jha}
\affiliation{Department of Physics and Astronomy, Rutgers, the State University of New Jersey, 136 Frelinghuysen Rd., Piscataway, NJ 08854, USA}

\author{P.~L.~Kelly}
\affiliation{School of Physics and Astronomy, University of Minnesota, 116 Church Street SE, Minneapolis, MN 55455, USA}

\author{R.~P.~Kirshner}
\affiliation{Harvard-Smithsonian Center for Astrophysics, 60 Garden Street, Cambridge, MA 02138, USA}
\affiliation{Gordon and Betty Moore Foundation, 1661 Page Mill Road, Palo Alto, CA 94304, USA}

\author{K. Mandel}
\affiliation{Harvard-Smithsonian Center for Astrophysics, 60 Garden Street, Cambridge, MA 02138, USA}
\affiliation{Institute of Astronomy and Kavli Institute for Cosmology, Madingley Road, Cambridge, CB3 0HA, UK}
\affiliation{Statistical Laboratory, DPMMS, University of Cambridge, Wilberforce Road, Cambridge, CB3 0WB, UK}

\author{A.~Rest}
\affiliation{Space Telescope Science Institute, 3700 San Martin Dr., Baltimore, MD 21218, USA}
\affiliation{Department of Physics and Astronomy, The Johns Hopkins University, 3400 N.\ Charles St., Baltimore, MD 21218, USA}

\author{A.~G.~Riess}
\affiliation{Space Telescope Science Institute, 3700 San Martin Dr., Baltimore, MD 21218, USA}
\affiliation{Department of Physics and Astronomy, The Johns Hopkins University, 3400 N.\ Charles St., Baltimore, MD 21218, USA}

\author{S.~A.~Rodney}
\affiliation{Department of Physics and Astronomy, University of South Carolina, 712 Main St., Columbia, SC 29208, USA}

\author{L.~Strolger}
\affiliation{Space Telescope Science Institute, 3700 San Martin Dr., Baltimore, MD 21218, USA}

\begin{abstract}

The {\it Wide Field InfraRed Survey Telescope} ({\it WFIRST}) was the highest ranked large space-based mission of the 2010 {\it New Worlds, New Horizons} decadal survey.  It is now a NASA mission in formulation with a planned launch in the mid-2020s.  A primary mission objective is to precisely constrain the nature of dark energy through multiple probes, including Type Ia supernovae (SNe~Ia).  Here, we present the first realistic simulations of the {\it WFIRST} SN survey based on current hardware specifications and using open-source tools. We simulate SN light curves and spectra as viewed by the {\it WFIRST} wide-field channel (WFC) imager and integral-field channel (IFC) spectrometer, respectively.  We examine 11 survey strategies with different time allocations between the WFC and IFC, two of which are based upon the strategy described by the {\it WFIRST} Science Definition Team, which measures SN distances exclusively from IFC data. We propagate statistical and, crucially, systematic uncertainties to predict the Dark Energy Task Force figure of merit (FoM) for each strategy. 
{\red Of the strategies investigated, we find the most successful to be WFC-focused. However, further work in constraining systematics is required to fully optimize the use of the IFC}. 
Even without improvements to other cosmological probes, the {\it WFIRST} SN survey has the potential to increase the FoM by more than an order of magnitude from the current values. Although the survey strategies presented here have not been fully optimized, these initial investigations are an important step in the development of the final hardware design and implementation of the {\it WFIRST}  mission.

\end{abstract}



\section{Introduction} \label{intro}

The {\it Wide-Field InfraRed Space Telescope} ({\it WFIRST}) is a NASA mission that will constrain the nature of dark energy through multiple probes. It was the top large space-based mission from {\it New Worlds, New Horizons}, the most recent U.S.\ astronomy and astrophysics decadal survey \citep{NAP12951}. As its name suggests, {\it WFIRST} is optimized for near-infrared (NIR) observations and it possesses a large field of view (FoV). The mission is in formulation at NASA, and several concepts have been suggested so far \citep{Spergel15}. The current design utilizes a telescope that was donated in 2012 by the National Reconnaissance Office. The aperture of the telescope is the same as that of the {\it Hubble Space Telescope} ({\it HST}), both having 2.37-m primary mirrors. Two main instruments are proposed for {\it WFIRST}: a coronagraph, which will be used for exoplanet and planetary disk studies, and a wide-field instrument, which will be used to probe dark-energy models. The wide-field instrument is itself composed of a wide-field channel (WFC) imager and integral-field channel (IFC) spectrometer.

Two major {\it WFIRST} goals are to measure the cosmological growth of the Universe and to probe its geometry on large scales. To achieve these milestones, {\it WFIRST} will conduct multiple observational programs, one of which is a supernova (SN) survey. Type Ia supernovae (SNe~Ia) have played a critical role in the discovery of  acceleration {\red in the expansion} of the Universe \citep{Riess98:Lambda, Perlmutter99}. Recent analyses using multiple cosmological probes \citep[e.g.,][]{Betoule14, Planck15, Alam16} 
are all consistent with a Universe that is geometrically flat, and that is filled with {\red cold dark matter and} dark energy that behaves like a cosmological constant  
\citep[the $\Lambda$CDM model; e.g.,][]{Peebles84, Efstathiou90, Frieman08:de}. There remain, however, theoretical arguments for alternatives to the cosmological constant \citep[e.g.,][]{Weinberg89, Frieman08:de}, which can serve as additional motivation for a new generation of experiments.   

The dark energy equation of state can be used to distinguish between many alternative explanations for the accelerated expansion of the Universe \citep[e.g., see][for a review of dark energy and modified gravity]{Joyce16}, and it is parameterized as
\begin{equation}
  P = w \rho c^{2},
\label{pde}
\end{equation}
where $P$ and $\rho$ are the dark-energy pressure and energy density, respectively, and $w$ is its equation-of-state parameter. In some models, the dark-energy equation of state evolves with time, and one common parameterization \citep[proposed by][]{Chevallier01, Linder03}, that we adopt in this work, is
\begin{equation}
  w = w_{0} + (1 - a) w_{a},
\label{w0}
\end{equation}
where $a = (1 + z)^{-1}$ is the scale factor of the Universe, $w_{0}$ is the current value of the equation-of-state parameter, and $w_{a}$ parameterizes its evolution. For a cosmological constant, $w_{0} \equiv -1$ and $w_{a} \equiv 0$.

Given the importance of measuring $w$, the Dark Energy Task Force \citep[DETF;][]{Albrecht06} suggested the use of a figure of merit (FoM) defined as the inverse of the area enclosed within the 95\% confidence contour in the $w_{0}$--$w_{a}$ plane, to compare the capabilities of different surveys in constraining the dark-energy equation of state. Current constraints on $(w_{0}; w_{a})$ are
\begin{equation}
   (w_{0}; w_{a}) = (-0.91 \pm 0.10; -0.39 \pm 0.34),
\label{w02}
\end{equation}
which correspond to a FoM of 32.6 in \citet{Alam16} \citep[see also][where FoM = 31.3]{Betoule14}. This FoM value includes the use of SNe; without SNe, \citet{Alam16} obtain ${\rm FoM} = 22.9$.

{\blue An alternative parameterization of the same linearly evolving $w_{a}$
model is} expressed as $w_{p}$, and its relation to ($w_{0};w_{a}$) and to the FoM described above are defined as 
\begin{equation}
   w_{p} = w_{0 }+ (1 - a_{p}) w_{a} \label{wpr}, \\
   {\rm FoM} \propto [\sigma(w_{a})\sigma(w_{p})]^{-1}, \label{wp}
\end{equation}
where a$_{p}$ is the pivot value of $a$ and represents the point at which the uncertainty in $w_{a}$, for a given data model, is minimized \citep{Albrecht06}. Details of $w_{p}$ and its application within SN surveys can be found within \citet{Astier14}. For the purposes of our paper, however, we have chosen to investigate the $w_{0}$-$w_{a}$ plane only.  

Understanding the nature of the largest component of the Universe is an important goal, one in which the community has invested significant resources. The DETF identified different ``stages" of dark-energy experiments starting with initial studies (Stage 1) and progressing toward Stage 4 surveys in the mid 2020s. Stage 3 experiments are currently underway \citep[e.g., the Dark Energy Survey][]{DES05}\footnote{See \url{http://www.darkenergysurvey.org}} and are expected to increase the FoM by a factor of 3--5 over Stage 2 experiments. 
{\red WFIRST is a Stage 4 experiment, and it is designed
to reach a factor of ten gain over Stage 2 experiments (i.e.,
$\textrm {\rm FoM} \geq 320$) via a combination of larger statistical samples and a reduction of systematic uncertainties.}

In order for the combined probes from Stage 3 and Stage 4 experiments to reach their projected constraints, SNe~Ia are critical. Several surveys have been working to gather data on SNe~Ia over a broad range of redshifts. Low-redshift ($0.01 < z < 0.1$) SN~Ia data have been obtained by groups/surveys such as the Center for Astrophysics 1--4 \citep[CfA;][]{Riess99:lc,Jha06:lc,Hicken09:lc,Hicken09:de,Hicken12}, the Carnegie Supernova Project \citep[CSP;][]{Contreras10,Folatelli10,Stritzinger11}, the Lick Observatory Supernova Search \citep[LOSS;][]{Ganeshalingam10,Ganeshalingam13}, and the Foundation SN survey \citep{Foley17}. SNe~Ia at higher redshifts  ($1.0 < z < 1.1$) have been examined by surveys including ESSENCE \citep{Miknaitis07, Wood-Vasey07, Narayan16:essence}, the SuperNova Legacy Survey \citep[SNLS;][]{Conley11,Sullivan11:cosmo}, the Sloan Digital Sky Survey \citep[SDSS;][]{Frieman08, Kessler09:cosmo, Sako16}, and Pan-STARRS1 \citep[PS1;][]{Rest14, Scolnic14:ps1}. To date, some of the highest redshift ($z>1.0$) SNe~Ia have been observed by the Supernova Cosmology Project \citep[SCP;][]{Suzuki12}, GOODS \citep{Riess07}, the Cosmic Assembly Near-infrared Deep Extragalactic Legacy Survey \citep[CANDELS;][]{Rodney14}, and the Dark Energy Survey's SN program \citep[DES-SN;][]{Bernstein12}. These surveys form our current state-of-the-art cosmology sample, consisting of over 1000 spectroscopically confirmed SNe~Ia, and extending the Hubble diagram out to $z\approx2$.

{\red Each SN~Ia light curve must be corrected for both color and shape {\red (``stretch'')} in order to standardize the SN brightness and to reduce the Hubble residual dispersion. In addition, redshift-dependent bias corrections are needed, particularly at higher redshifts where fainter SNe are excluded from the sample. Finally, systematic uncertainties must be evaluated and propagated to the inference of cosmological parameters.}

Using a simple model for statistical and systematic uncertainties, the {\it WFIRST} Science Definition Team (SDT) outlined a baseline 6-year mission, including a 2-year SN survey, corresponding to 6 months of ``on-sky" time \citep{Spergel15}. {\red A key SDT assumption is that systematic uncertainties can be characterized in discrete, {\red independent} redshift bins ($\Delta z = 0.1$). {\red For many systematics, however this is an oversimplification, and correlations are found across much broader redshift ranges (e.g., calibration and SN color).}
The focus of our paper is to expand the discussion of survey strategies, 
progress toward a more optimized {\it WFIRST} SN strategy, and include systematic uncertainties with state-of-the-art analysis tools. Using a covariance matrix approach to systematics, we account for correlations among all redshifts.}

A greater understanding of systematic uncertainties and {\red their impact on cosmological constraints} is obtained by accurately simulating the survey with sophisticated analysis software {\red called} the SuperNova ANAlysis \citep[SNANA;][]{Kessler09:SNANA} package. 
\SNANA\ is designed to generate highly realistic simulations of SN surveys, and to model the impact of systematic uncertainties.  {\red The best SN~Ia cosmology constraints \citep{Scolnic17, Rodney14, Rest14, Betoule14}  are from analyses where \SNANA\ has been used} to perform light-curve fitting and predict bias corrections for a variety of surveys including those at low redshift, SDSS, PS1, SNLS, and {\it HST}. It is routinely updated with the most current techniques for simulations and analysis. Using \SNANA\ in addition to several other open-source tools, we have designed and evaluated various {\it WFIRST} SN survey strategies, {\red created detailed simulations, and conducted a thorough investigation of uncertainties}. Our simulations are the first of their kind for the {\it WFIRST} mission, and they allow us to predict and compare the potential scientific impact of each strategy. Furthermore, our work acts as a reference for future simulations and provides a guide for the ongoing planning of the {\it WFIRST} mission.

This paper is structured as follows. We describe {\it WFIRST} and its instruments in Section~\ref{obs}. Section~\ref{currentsn} presents an outline of the SDT SN survey strategy, while Section~\ref{simulation} provides a comprehensive description of how we applied all tools to create the various SN simulations. Additional survey strategies as well as analyses of those strategies examined are presented in Section~\ref{results}. We explore different assumptions for various systematic uncertainties and outline their impact on the FoM measured by {\it WFIRST} simulated SN surveys in Section~\ref{sysun}. Section~\ref{compss} compares the simulated survey strategies described in this work, with Section~\ref{diss} providing a discussion of future considerations for the optimization of the {\it WFIRST} SN survey. Section~\ref{con} presents our conclusions.

\section{{\it WFIRST} Hardware}
\label{obs}
Planned for launched in the mid 2020s, {\it WFIRST} is expected to be placed into an L2 orbit (1.5 million km away from Earth at the second Lagrange point), where it will reside for the duration of its {\red mission}. Analogous to {\it HST}, {\it WFIRST} consists of a primary mirror that is 2.37~m in diameter. Light from the primary is reflected to the on-axis secondary mirror, which then feeds into the paths of its various instruments. The design of the telescope is not yet finalized; however, current plans call for both a wide-field instrument (WFI) and a coronagraph\footnote{For more information on the coronagraph, see \url{http://wfirst.gsfc.nasa.gov/observatory.html}}. For the purposes of this paper we focus only on the WFI. When preparing our simulations we used the best-available WFI hardware specifications; these were taken from the 2017 July 30 (Cycle 7) spacecraft and instrument parameter release\footnote{\url{https://wfirst.ipac.caltech.edu/sims/Param_db.html}}, and an operational temperature of 260~K is assumed.  

\subsection{The Wide Field Instrument}
The WFI has two optical channels: the first is a Wide Field Channel (WFC), the second an Integral-Field Channel (IFC). The WFC possesses an imager and has the ability to perform slitless grism spectroscopy, while the IFC has two small-field integral-field units (IFUs). Combined, these instruments will be used to perform the dark-energy survey, as well as the microlensing and high-latitude surveys. \\

{\bf The Wide Field Channel:}
In its most simplified form, the optical layout of the WFC consists of three mirrors, two fold mirrors, and a {\red nine}-slot filter wheel. Currently, {\red seven} of these slots are dedicated to imaging filters, one is for a grism that will provide low-resolution spectra of the full WFC FoV, {\red and the last is a blank position dedicated to dark and flat-field calibration.}

Eighteen 4k $\times$ 4k HgCdTe detectors (H4RG-10) will be used by the WFC, and will be arranged into a $6 \times 3$ array to generate an effective FoV\footnote{See~\url{https://wfirst.ipac.caltech.edu/sims/Param_db.html?csvfile=WFirstParameters_v5.0.csv} for a list of more detailed WFI parameters.} of 0.281~deg$^{2}$.

\begin{figure*}
\centering
\includegraphics[keepaspectratio=true,  scale=0.75]{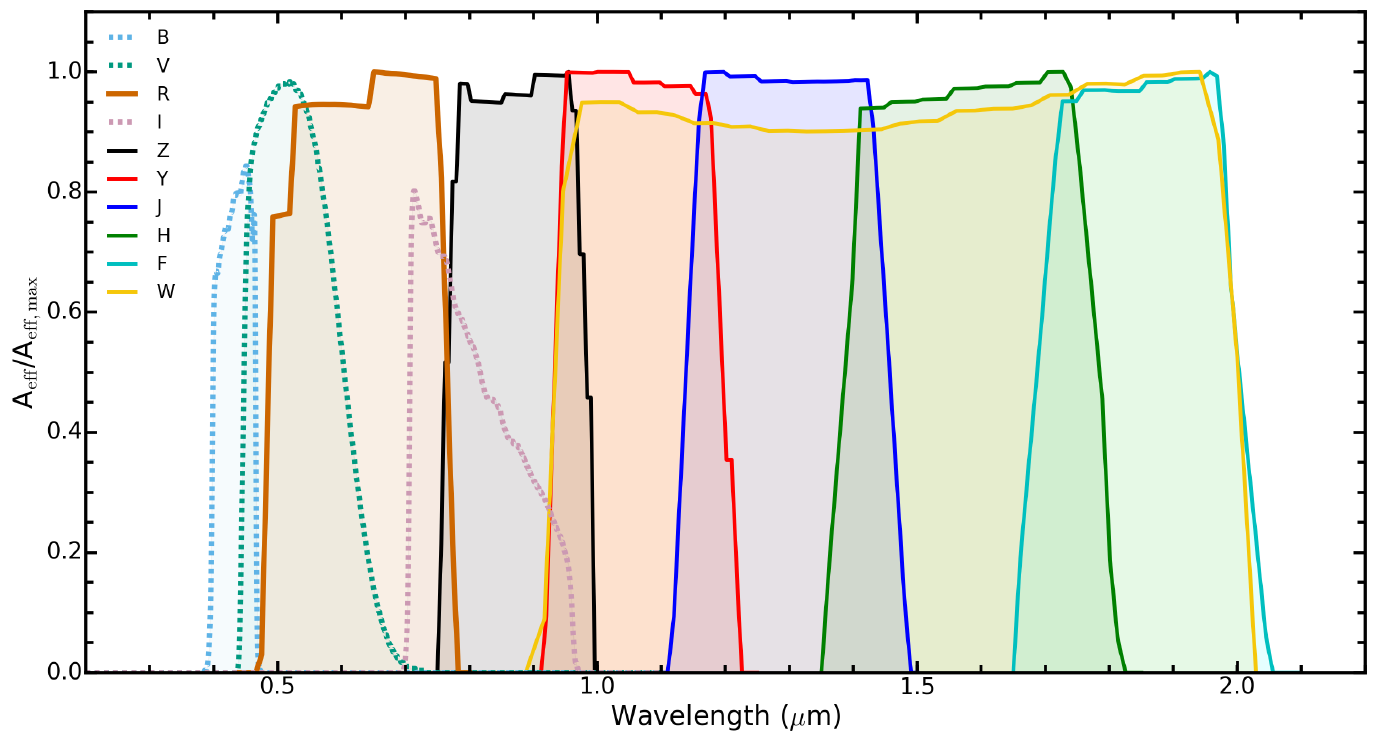}
\caption{{\it WFIRST} WFC imaging filter bandpass effective areas, $A_{\rm eff}$, divided by the maximum effective area (solid lines) as described by the {\it WFIRST} Cycle 7 instrument parameter release. Also shown are the {\it HST} WFC3 filters used for this work (dotted lines). The WFC3 throughputs presented here have been scaled for comparison.}
\label{bands}
\end{figure*}

The {\red seven} imaging filters of the WFC are named $F062$, $Z087$, $Y106$, $J129$, $H158$, $F184$, and $W149$ (a very wide filter); hereafter, these filters will be referred to in the text as $R$, $Z$, $Y$, $J$, $H$, $F$, and $W$. The central wavelengths of these filters are 0.62, 0.87, 1.09, 1.30, 1.60, 1.88, and 1.40~$\mu$m (respectively), and combined they cover the 0.44--2.0~$\mu$m range, as illustrated in Figure~\ref{bands}\footnote{More filter information is provided within \url{https://wfirst.gsfc.nasa.gov/science/sdt_public/wps/references/instrument/WFIRST-WFI-Transmission_160720.xlsm} -- pages 5 through 10.}. 

The spatial resolution of the imaging component of the WFC is $\sim0.11$\arcsec~pixel$^{-1}$ with an inter-pixel capacitance {\red (IPC) of 0.02 in each of the four neighboring pixels. 
IPC is a form of crosstalk in NIR detectors, in which some of the charge from one pixel will transfer to a neighboring pixel during readout. The effect of IPC is to redistribute
charge, {\red which can alter the full width at half-maximum intensity (FWHM) and change the impact of cosmic rays and hot pixels}. IPC must therefore be taken into account when calculating the point-spread function (PSF)  FWHM of a source for each {\it WFIRST} filter}. 

The gain for the WFC is assumed to be unity. A more detailed description of the WFC filters, including their zero-points and 
FWHM\footnote{As the WFIRST point-spread function (PSF) is non-Gaussian, the PSF FWHM values presented and used for this analysis are derived from the noise-equivalent areas.}, 
can be found in Table~\ref{zpts}.\\

\begin{deluxetable}{ccccc}
\tablecaption{The WFC imaging filters. \label{zpts}}
\tablecolumns{5}
\tablewidth{0pt}
\tablehead{
\colhead{Filter} &  \colhead{Central}  & \colhead{Filter} & \colhead{AB} & \colhead{PSF}\\
\colhead{} & \colhead{Wavelength} &  \colhead{FWHM} & \colhead{Zero-point\tablenotemark{a}} & \colhead{FWHM}\\ 
\colhead{} &  \colhead{($\mu$m)} & \colhead{($\mu$m)}  & \colhead{} & \colhead{(pixel)}
}
\startdata
$F062$ & 0.62 & 0.28 & 26.99 & 1.68\\
$Z087$ &  0.87 & 0.22 & 26.39 & 1.69\\
$Y106$ & 1.09 & 0.27 & 26.41 & 1.86\\
$J129$  & 1.30 & 0.32 & 26.35 & 2.12\\
$H158$ & 1.60 & 0.40 & 26.41 & 2.44\\
$F184$  & 1.88 & 0.31 & 25.96 & 2.71\\
$W149$ & 1.40 & 1.1 & 27.50 & 2.19\\
\enddata
\tablenotetext{a}{Here the zero-point is calculated using each filter's effective area; it is equivalent to the magnitude that results in one count per second for an {\red infinite detection aperture}.}
\end{deluxetable}

The WFC grism is designed such that it provides spectroscopic coverage within the 1.00--1.89~$\mu$m range. It possesses a dispersion of 1.04--1.14~nm~pixel$^{-1}$, with a spectral resolving power of $\lambda/\Delta\lambda \approx  435$--865 (2 pixels). However, we do not focus on the use of the grism in this paper.\\

\begin{figure}
\centering
\includegraphics[keepaspectratio=true,  width=0.5\textwidth]{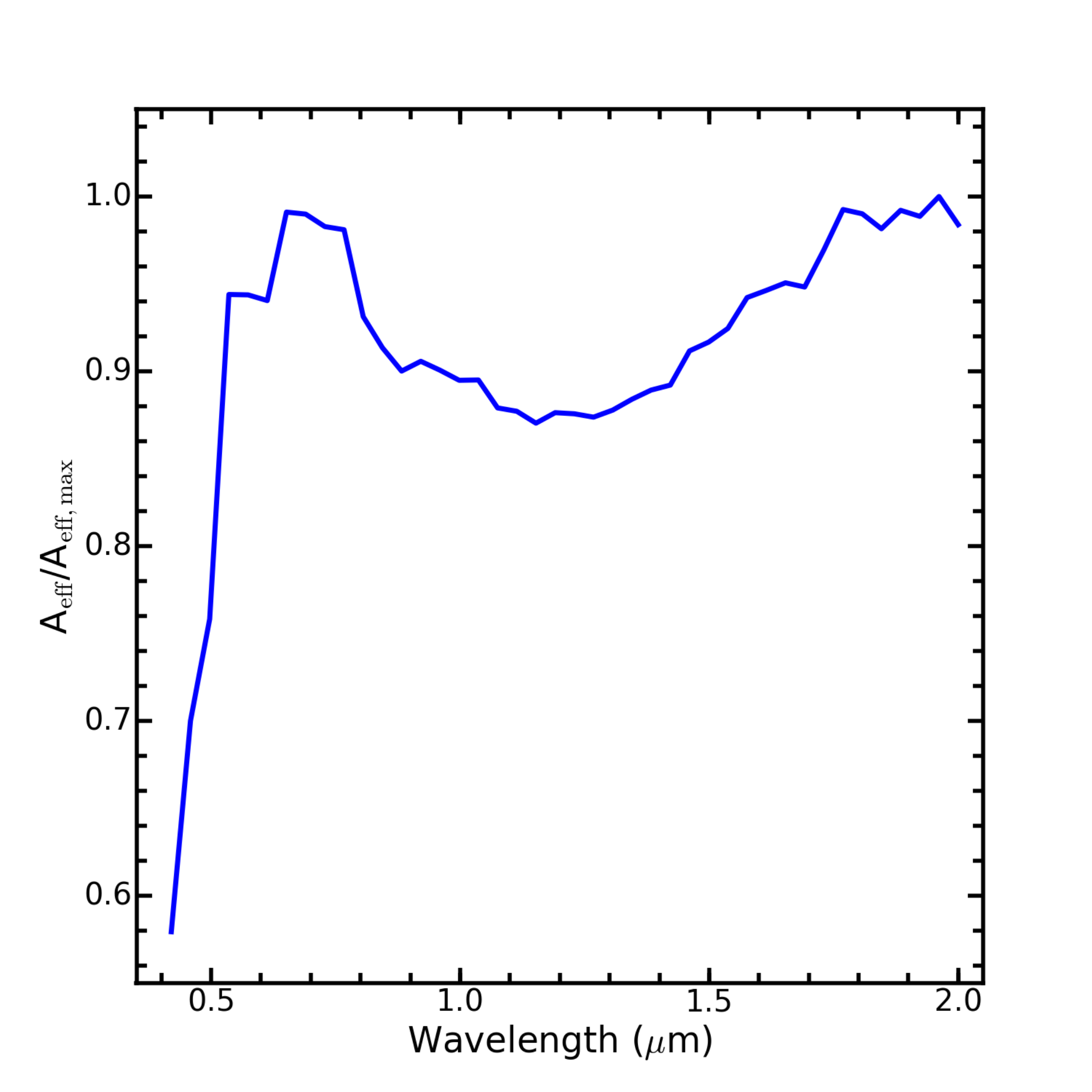}
\caption{{\red Relative throughput efficiency ($A_{\rm eff}/A_{\rm eff,max}$) vs. wavelength for the {\it WFIRST} IFC-S}.Wavelengths beyond those displayed have not had their throughputs calculated.}
\label{bandsifu}
\end{figure}

{\bf The Integral Field Channel:} 
The IFC contains two image slicers that feed a spectrograph. Each image slicer corresponds to a different FoV: the smaller FoV, higher spatial-resolution IFC-S, which is designed for SN observations, and the larger FoV, lower spatial resolution IFC-G, which is designed for galaxy observations (unrelated to the SN survey). The IFC-S has a 3.00\arcsec\ $\times$ 3.15\arcsec\ FoV that is composed of  0.15\arcsec\ wide slices, a 0.05\arcsec~pixel$^{-1}$ scale, and a wavelength range of 0.42--2.0~$\mu$m. The instrument has a  spectral resolution of $\lambda/\Delta\lambda \approx 70$--225 (per two-pixel resolution element) and like the WFC contains H4RG detectors. 
{\red The IFC-S consists of 352 spectral bins, and the properties of each bin (wavelength range, PSF, noise) are given in Table~\ref{ifuextend}} of Appendix~\ref{ap1}. The resolution of the IFC-S is based on the design described within \citet{Content13}, but with two recent modifications: the extension of the IFC-S blueward of 6000~\AA\ to 4200~\AA, and the use of H4RG detectors which affect the pixel scale. The PSF FWHM values presented in Table~\ref{ifuextend} were calculated using an Airy disc approximation for each bin. The wavelength coverage of the IFC-S is illustrated in Figure~\ref{bandsifu}\footnote{See page 13 of \url{https://wfirst.gsfc.nasa.gov/science/sdt_public/wps/references/instrument/WFIRST-WFI-Transmission_160720.xlsm} for more IFC-S information.}.\\

\section{An Outline of the SDT SN Survey Strategy}
\label{currentsn}

The {\it WFIRST} SDT final report \citep{Spergel15} presents a SN survey strategy in which the imaging component of the WFC is used for SN discovery and the IFC-S for classification and obtaining distances. An outline of this strategy is described {\red below}.

\begin{itemize}
\item{The SN survey {\red spans 2 years} in a selected SN field, with a 5-day cadence. There are therefore 146 visits to the SN field.}
\item{Each visit, or epoch of observation, is 30 hours long {\red including overhead, resulting in} a total survey time of 4380 hours (6 months).}
\item{Within each visit, 8 hours of imaging is used exclusively for SN discovery. These data are obtained every 5 observer-frame days.}
\item{The imaging is split into 3 subsurveys (hereafter referred to as tiers) of differing area/depth, and using different discovery filters (see Table~\ref{filtttt}).}
\item{The remaining 22 hours in each visit are for IFC-S observations, used to classify {\red SN candidates} and to synthesize broadband photometry.}
\item{IFC-S observations are designed to be taken at a cadence of roughly 5 rest-frame days, with the goal of obtaining spectrophotometry to measure distances.}
\item{There are 3 different IFC-S exposure {\red times}: typical short exposures, medium classification exposures, and long ``deep" exposures. These 3 exposure {\red times are the} first three IFC-S spectra taken for each SN {\red candidate}.}
\item{The {\red short- and medium-exposure} spectra are used for initial classification, and if these spectra meet certain criteria (outlined in more detail below) the IFC-S obtains a long exposure through which a final classification is obtained. If {\red classified} as a SN~Ia, further follow-up observations are initiated.}
\item{The follow-up observations consist of six {\red short-exposure} spectra plus one medium exposure of the host galaxy, taken after the SN has faded, to use as a template.}
\item{The exposure times for the long and {\red medium-exposure} spectra are on average approximately 3.9 and 1.9 times longer than the short exposure, respectively.}
\item{The total set of observations for any given SN~Ia {\red(i.e., excluding the host template)} is equivalent to an average of $\sim$12.8 short exposures. The exposure times are set by the redshift of the SN.}
\end{itemize}

The SN survey strategy proposed by the SDT report is designed {\red to achieve} a relatively flat redshift distribution {\red using a 3-tier survey, where each successive tier is deeper and covers less area than the previous tier.} 
The first tier consists of a shallow wide field for SNe with $z < 0.4$, over an area of 27.44~deg$^{2}$, using the $Y$+$J$ filters for discovery. The second is a medium tier for SNe with $0.4 \leq z < 0.8$, over a moderate 8.96~deg$^{2}$ area, using the $J$+$H$ filters. Finally, the last is a deep tier for SNe with $z \leq 1.7$, over a small 5.04~deg$^{2}$ area, again using the $J$+$H$ filters. Table~\ref{exptim} lists the exposure times for each of the three tiers and the number of spacecraft pointings required to make up their designated areas. The different filter combinations for each tier were chosen in order to probe similar rest-frame wavelengths. However, for the shallow tier the $Z$-band filter is the only band that covers a rest-frame wavelength range which is sufficiently modeled for  cosmological analysis. One might assume, therefore, that redder wavelengths (i.e., $>$7000~\AA\ in the rest frame) will be accurately trained either with data from {\it WFIRST} or from precursor data.\\

Of the 146 planned visits, the discovery search will be implemented in only 132. The remaining survey time will be used for host-galaxy follow-up observations, acquiring a template. The host-galaxy template spectrum is to be taken a year after the peak brightness of the SN, when the relative amount of light from the SN compared to the galaxy is negligible. Thus, in the first year only 27 of the total 30 hours in each 5-day visit will be used, with the remainder deferred to year 2. SNe discovered during the second year will have their galaxy reference spectrum taken in year 3, after the discovery component of the 2-year SN survey has concluded. \\

\begin{deluxetable*}{lccccc}
\tablecaption{Description of the three-tier WFC survey as outlined in the SDT report. \label{filtttt}}
\tablecolumns{6}
\tablewidth{0pt}
\tablehead{
\colhead{Survey} & \colhead{Redshift} & \colhead{Area} & \colhead{Discovery} & \colhead{Depth per} & 
     \colhead{Total Depth\tablenotemark{a,b}}\\
\colhead{Tier} & \colhead{Range} & \colhead{(deg$^{2}$)} & \colhead{Filters} & \colhead{Exposure\tablenotemark{a} (mag)} & 
      \colhead{(mag)} 
}
\startdata
Shallow & $0.1 \leq z <    0.4$  &  27.44 & $Y,J$  & {\blue 22.0}, {\blue 22.0} & {\blue 24.7}, {\blue 24.7}\\
Medium & $0.4 \leq z <   0.8$   &  8.96   & $J,H$ & {\blue 24.8}, {\blue 24.8} & {\blue 27.5}, {\blue 27.5}\\
Deep     & $0.8 \leq z \leq 1.7$ &  5.04   & $J,H$  & {\blue 26.2}, {\blue 26.2} & {\blue 28.9}, {\blue 28.9}\\
\enddata
\tablenotetext{a}{{\red Accounts for overhead from slew-and-settle time.}}
\tablenotetext{b}{{\red Total depth is for co-add over all 146 visits.}}
\end{deluxetable*}

\begin{deluxetable}{lccccc}
\tablecaption{WFC exposure times ($t_{\rm exp}$) and number of pointings ($N_{p}$) for each filter and each tier.\tablenotemark{a}} \label{exptim}

\tablecolumns{6}
\tablewidth{0pt}
\tablehead{
\colhead{Survey} & \colhead{$Y$ band} & \colhead{$J$ band} & \colhead{$H$ band} & \colhead{$N_{p}$} & \colhead{$t_{\rm tot}$}\\
\colhead{Tier} & \colhead{$t_{\rm exp}$ (s)} &  \colhead{$t_{\rm exp}$ (s)} & \colhead{$t_{\rm exp}$ (s)} &\colhead{} & \colhead{(hr)}
}
\startdata
Shallow  & 13   & 13     & 0     & 98 & 3.0\\
Medium  & 0     & 67    & 67    & 32 & 2.0\\
Deep      & 0     & 265  & 265  & 18  & 3.0\\  
\enddata                                                       
\tablenotetext{a}{The exposure times listed here do not include the 42~s slew. 
{\red However, slew-and-settle times are incorporated into our overall calculations, so that the total observatory time allocated to the SN survey is still 6 months.}}
\end{deluxetable}

The spectroscopic observations planned in the SDT report are designed to observe one SN at a time, using the IFC-S. The exposure times were tailored to achieve a signal-to-noise ratio (SNR) high enough to clearly identify key spectral features. The longest exposure times are therefore required for the highest redshift SNe, i.e., $z \approx 1.7$ events. 
For each SN classified as SN~Ia, a series of 10 spectra is obtained. The first three 
spectra vary in exposure time and are used not only for obtaining time-critical data on the SN, but also for selection and identification purposes.
It is expected that by the third spectrum, core-collapse (CC) SNe are eliminated from the sample (see Section~\ref{classification}). A list of exposure times for each SN (excluding the host-galaxy template) per redshift tier is given in Table~\ref{ifutime}. {\red The exposure times listed here are based on initial estimates provided
by contributing SDT report authors, but adjusted for the number of spectra per SN (10) as specified within the SDT report}.  

\begin{deluxetable}{ccccc} 
\tablecaption{Exposure times per 0.1 redshift bin for the {\it WFIRST} IFC-S component. \label{ifutime}}
\tablecolumns{5}
\tablewidth{0pt}
\tablehead{
\colhead{Mean} & \colhead{Short} & \colhead{Medium} & \colhead{Long} & 
     \colhead{{\red Total}\tablenotemark{a}}  \\
\colhead{Redshift} & \colhead{$t_{\rm exp}$ (s)} & \colhead{$t_{\rm exp}$ (s)} & \colhead{$t_{\rm exp}$ (s)} &
      \colhead{$t_{\rm tot}$ (s)} 
}  
\startdata
0.15	& 30.4    & 47.0	     & 76.0   & 335.8\\ 
0.25	& 52.1    & 83.8	     & 143.7 & 592.2\\ 
0.35	& 80.5    & 134.1   & 241.6  & 939.2\\ 
0.45	& 118.5  & 205.2   & 	387.5  & 1422.2\\ 
0.55	& 162.7  & 291.5   &	571.8  & 2002.2\\ 
0.65	& 184.5  & 337.2   &	675.7  & 2304.4 \\ 
0.75	& 208.6  & 386.3   &	785.1  & 2631.6\\
0.85	& 229.5  & 428.7   &	879.0  & 2914.2\\
0.95	& 267.6  & 508.5   &	1060.8 & 3442.5\\
1.05	& 319.9  & 621.8   &	1325.7 & 4186.8\\
1.15	& 368.9  & 729.0   &	1578.0 & 4889.3\\
1.25	& 427.9  & 862.2   &	1899.8 & 5757.3\\
1.35	& 493.3  & 1012.3 &	2268.4 & 6733.8\\
1.45	& 550.4  & 1146.9 &	2604.2 & 7603.9\\
1.55	& 603.5  & 1274.1 &	2926.5 & 8425.1\\
1.65	& 629.9  & 1336.1 &	3081.0 & 8826.4\\
\enddata    
\tablenotetext{a}{Total time observing one SN within a given 0.1 redshift bin (not including the template host-galaxy spectrum).}
\end{deluxetable}

{\red Both the SDT report and \citet[][an earlier SDT publication]{Spergel13b} assumed a combined instrumental slew-and-settle time of 42 s. Although these overheads were mentioned by \citet{Spergel13b}, they were not incorporated within the SDT's SN strategy, and as such the exposure times and search depths presented for each imaging tier were overestimated. 
In our simulations the slew-and-settle overheads are incorporated within each strategy, and we present\footnote{Derived from ETC calculations, see \url{https://wfirst.ipac.caltech.edu/sims/ETC.html}} updated depths in Table~\ref{filtttt}. 
Note that this 42 s overhead is a severe underestimate of the actual value {\red (the most recent estimates of slew-and-settle time from mission HQ are almost double that quoted here)}, and therefore these overheads remain an uncertain aspect of mission performance.}

The total time ($t_{\rm tot}$) listed per imaging tier in Table~\ref{exptim}, including overheads, is therefore calculated as
\begin{equation}
  t_{\rm tot}~({\rm s}) = (t_{\rm exp} + t_{\rm oh}) \times N_{\rm f} \times N_{\rm p},
\end{equation} 
where $t_{\rm exp}$ is the exposure time on the sky in seconds, $t_{\rm oh}$ is the 42 s overhead, $N_{\rm f}$ is the number of filters used (which for discovery is always 2), and $N_{\rm p}$ is the number of pointings.

\subsection{SDT Detection and Classification}
\label{classification}

The detection and selection of SNe~Ia for follow-up observations as outlined in the SDT report is a complex process, 
influenced by the cost of single-object follow-up observations with the IFC-S. The process starts with all possible SN candidates, including both SNe~Ia and CC SNe, and then progressively removes SNe which do not satisfy certain conditions. The first part of this selection procedure involves a SNR requirement.
It is not clear within the SDT report if their SNR is based on image-subtracted data, {\red but in this paper the
SNR includes noise from both the search and template images.}
Note also that pre-existing spectroscopic redshifts for all host galaxies are assumed by the SDT report, thus enabling the classification procedure outlined. 
At each stage of the selection process SN candidates are removed, and cannot re-enter. Therefore, each step in the selection process is considered a selection cut, which we list below.

\begin{itemize}
\item{\it Cut 0:} {\red An object is ``detected''} if ${\rm SNR} \geq 4$ in both imaging discovery bands ($Y$+$J$ or $J$+$H$), within a single epoch. {\red Of these objects, those that have a discovery-epoch color inconsistent with a SN~Ia at their host-galaxy redshift are removed.  
All remaining objects are scheduled for a short-exposure IFC-S spectrum during the next visit to the SN field.}
\item{\it Cut 1:} {\red An object is removed if its flux does not increase between the first and second epochs (in both filters), or if the color at the second epoch is inconsistent with a SN Ia at its host-galaxy redshift.} All remaining objects are scheduled for a medium-exposure IFC-S spectrum.
\item{\it Cut 2:} {\red An object is removed if its flux does not increase between the second and third epochs (in both filters), or if the color at the third epoch is inconsistent with a SN Ia at its host-galaxy redshift.} 
\item{\it Cut 3:} If the medium-exposure spectrum does not resemble that of a SN Ia (see Section~\ref{simulation} for how we determine this), the object is removed. All remaining objects are scheduled for a long-exposure IFC-S spectrum.
\item{\it Cut 4:} An object that is not {\red identified (again, see Section~\ref{simulation} for how we determine this)} as a SN~Ia with the long-exposure IFC-S spectrum is removed. Remaining objects are scheduled for additional IFC-S observations and are included in the final cosmology sample.
\end{itemize}

\subsection{SDT Statistical and Systematic Uncertainties}
\label{currentstatsys}

\begin{figure*}
\centering
\includegraphics[keepaspectratio=true, width=0.9\textwidth]{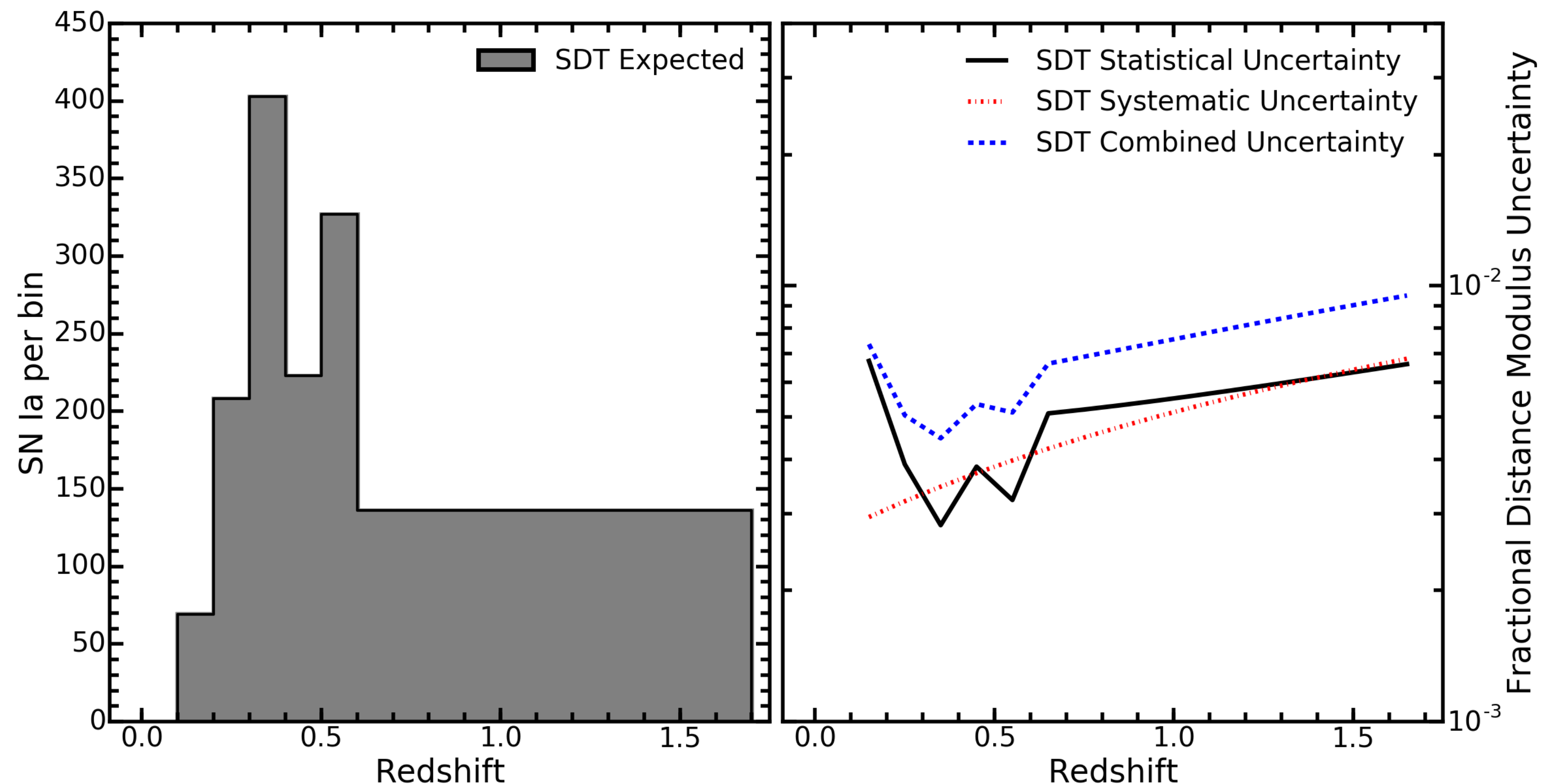}
\caption{{\it Left}: Redshift distribution of the {\it WFIRST} SN survey presented (and assumed) by the SDT report. {\it Right}: Fractional statistical (black curve), systematic (red dot-dashed curve), and total (blue dashed line) distance uncertainty per $\Delta z = 0.1$ bin as assumed in the SDT report.}
\label{hist22}
\end{figure*}

The SDT survey strategy is designed such that statistical uncertainties {\red match an assumed systematic} uncertainty budget. This means that the assumptions about systematic uncertainties {\red motivate the whole SN survey strategy, as these assumptions dictate the desired sample size, which in turn sets the required discovery rate and redshift distribution.} The final distribution of SNe~Ia per 0.1 redshift bin, as expected by the SDT report, is shown in Figure~\ref{hist22} (left panel). 

The systematic uncertainties presented in the SDT report for the {\it WFIRST} SN survey follow the description of distance modulus uncertainties used for the SNAP design outlined by \citet{Kim04} \citep[see also][]{Perlmutter03, Frieman03}. In the SDT report, the magnitude of the uncertainties was reduced roughly by a factor of two compared to the SNAP design. The formulation assumes that the systematic uncertainties are uncorrelated on scales larger than $\Delta z = 0.1$ and can be treated equivalently to statistical uncertainties. The total systematic uncertainty is assumed to increase with redshift, following
\begin{equation}
  \sigma_{\rm sys} = \frac{0.01 (1 + z)}{1.8}~\rm{(mag)}.
\label{sys}
\end{equation}
However, there are known systematics which contradict this assumption. Specifically, uncertainties related to calibration and SN color are correlated across a wide redshift range. The SDT systematic model (Equation~\ref{sys}) is overly simplistic and not used for our analysis. The SDT functional form for the systematic uncertainty model also drives the broad, flat redshift distribution seen in Figure~\ref{hist22} (right panel).

The SDT report assumes that the distance precision per SN is $\sigma_{\rm meas} = 0.08$~mag, and this includes both statistical measurement uncertainties and statistical model uncertainties. This uncertainty is a constant since the SDT strategy adjusts the exposure time for each SN observation based on redshift so that all SNe have approximately the same distance uncertainty. The intrinsic scatter in corrected SN~Ia distances is set to be $\sigma_{\rm int} = 0.09$~mag. This value is more optimistic than what is currently measured for optical data where $\sigma_{\rm int} \simeq 0.13$~mag \citep[see Section 7.1 of][]{Kessler16}. The lensing uncertainty is modeled as $\sigma_{\rm lens} = 0.07 \times z$~mag, which is an average of the values derived by \citet{Holz05}, \citet{Gunnarsson06}, and \citet{Jonsson10}. The total statistical uncertainty for a given redshift bin is therefore given in the SDT report as
\begin{equation}
  \sigma_{\rm stat}= \frac{(\sigma^{2}_{\rm meas}+ \sigma^{2}_{\rm int} + \sigma^{2}_{\rm lens})^{1/2}}{N^{1/2}_{\rm SN}}~\rm{(mag)},
\label{stat}
\end{equation}
where $N_{\rm SN}$ is the number of SNe~Ia in a given redshift bin. 

The statistical, systematic, and combined uncertainty budgets of the SDT report SN survey are illustrated in Figure~\ref{hist22} (right panel). To be clear, the SDT analysis is not based on SN simulations or light-curve analysis, but instead is based on assumptions about statistical and systematic uncertainties that would arise from such an analysis.

\section{Simulation and Analysis Tools}
\label{simulation}

Within this paper, we test the various assumptions made in the SDT report for the SN survey, evaluate its statistical and systematic uncertainty budget, and develop a framework to explore other strategies and optimize parameters for the future {\it WFIRST} mission. To accomplish this, we simulate and analyze a realistic survey and include the most significant uncertainties. Here we describe software tools that we have used to implement the simulation, apply selection criteria, and determine cosmological constraints used to compute the FoM.

To examine a variety of possible {\it WFIRST} survey strategies, we used the \SNANA\ simulation package \citep{Kessler09:SNANA}\footnote{\url{http://snana.uchicago.edu}}. 
\SNANA\ has been extensively used for the simulation of SN surveys and analysis of SN samples \citep[see, e.g.,][]{Betoule14, Scolnic14:ps1}. The goal of the {\it WFIRST} 
simulation is to provide the same fidelity as an ideal image-level simulation by using image properties (zero-points, sky noise, PSFs) rather than images themselves. As this is a ``catalog-level" simulation rather than a pixel-level simulation, we assume that Poisson noise correctly describes the uncertainties from the image subtraction. 

To characterize a {\it WFIRST} SN strategy, we provide \SNANA\  with information about the observatory  (e.g., filter/spectrograph properties and  noise sources), the survey (e.g., cadence, exposure time, selection requirements), and the physical Universe (e.g., SN spectral models, SN rates, cosmological parameters, lensing assumptions). 
Each of these components is described below, {\red along with other analysis tools needed to determine the FoM.} Our analysis has resulted in several publicly available upgrades to {\SNANA}.\\

\noindent{\bf Imaging filters and spectroscopic bins:} Tables~\ref{zpts} and \ref{ifuextend} (in Appendix~\ref{ap1}) describe the WFC imaging filters and IFC-S wavelength bins used within our simulations. \SNANA\ was originally designed only to simulate broadband SN light curves. In order to simulate the IFC-S, we added a new \SNANA\ module for simulating spectra and ``synthetic" broadband filters. 

While it may be possible to directly infer distances from SN spectral time series, examination of that approach is beyond the scope of this paper. Instead, we implement the SDT report's IFC-S strategy in \SNANA\  by integrating each simulated spectrum into a set of 52 synthetic filters. These synthetic filters were determined by binning together the 352 spectral elements of the IFC-S by a factor of $\sim$7, and taking the upper and lower wavelength limits. The \SNANA\ software allows up to 62 broadband filters, ten of which are used for broadband imaging filters, leaving 52 for the IFC-S synthetic filters (note that there is no limit on spectral binning within {\SNANA}). 
Once binned, \SNANA\ treats the resulting ``synthetic photometry" in a similar manner to any broadband photometry for estimating distances. As the SDT analysis only uses spectral data from the rest-frame optical (3000--8000~\AA), we have limited our IFC-S simulations/data accordingly. This choice likely limits the full capability of the IFC-S, however the various published analyses of IFU data have only probed SNe~Ia in the rest-frame optical \citep[e.g.,][]{Saunders15, Fakhouri15}. Note, however, that the SDT discovery imaging still makes use of the NIR filters to enable follow-up spectroscopy with the IFC-S.\\

\noindent{\bf Cadence and exposure time:} The cadence of both the WFC and IFC-S components of the SN survey are described in Section~\ref{currentsn}. The exposure time per imaging tier of the survey is given in Table~\ref{exptim}, with IFC-S redshift-dependent times presented in Table~\ref{ifutime}. The exposure time of the IFC-S within a given 0.1 redshift bin is identical between imaging tiers. Our simulations do not make adjustments to account for the mean SN brightness shifting slightly within a redshift bin (i.e., changes in brightness at $z = 0.45$ to $z = 0.46$, etc.,) as it is unlikely that any actual SN survey executed would have specific exposure times for individual objects of interest.\\

\noindent{\bf Sources of noise:} For all simulated SN observations, we include four sources of noise: zodiacal light, thermal background, dark current, and read noise. The contributions from each of these sources are presented in Tables~\ref{zpts2}, \ref{RNval}, and \ref{ifuextend}, within Appendix~\ref{ap1}. Host-galaxy Poisson noise is also included in both the SN-search and template observations, where possible.

The zodiacal light is calculated using a broken power law as described by \citet{Aldering01}. Thermal noise contributions are calculated using code developed by D.\ Rubin (private comm.) under the assumption of a 260~K operating temperature, and are comparable to values produced when using the {\it WFIRST} ETC\footnote{See \url{https://wfirst.ipac.caltech.edu/sims/ETC.html}}. The zodiacal and thermal noise for the IFC-S, as a function of wavelength, are presented in Table~\ref{ifuextend} in Appendix~\ref{ap1}. The higher resolution of the IFC-S leads to smaller zodiacal and thermal noise contributions when compared to the WFC.

From \cite{Hirata14}, we assume a dark current for the WFC to be
$0.015~e^{-}~{\rm s}^{-1}{\rm pixel}^{-1}$,
and for the IFC-S it is
$0.003~e^{-}~{\rm s}^{-1}{\rm pixel}^{-1}$ (a conservative estimate based on current measurements of 0.001~$e^{-}$~s$^{-1}$pixel$^{-1}$). The read noise is a function of exposure and readout time and is calculated using a modified version of the expression described by \citet{Rauscher07}. For any given WFC exposure time, the read noise, $\sigma_{\rm read}$, is
\begin{equation}
  \sigma_{\rm read}~{\rm (e/s)} = \sqrt{25 + 4800 \times \frac{(t_{\rm exp}/t_{\rm read} - 1)}{(t_{\rm exp}/t_{\rm read})} \times \frac{1}{(t_{\rm exp}/t_{\rm read}+1)}},
\label{RN}
\end{equation}
where $t_{\rm exp}$ is the exposure time of the observation in seconds and $t_{\rm read}$ is the read time in seconds, which is taken as 2.825 s.

\begin{deluxetable}{ccc}
\tablecaption{WFC imaging filters: Sources of noise. \label{zpts2}}
\tablecolumns{3}
\tablewidth{0pt}
\tablehead{
\colhead{Filter}  & \colhead{Zodiacal Noise} & \colhead{Thermal Noise}\\
\colhead{} & \colhead{($e^{-}$~s$^{-1}$~pixel$^{-1}$)} & \colhead{($e^{-}$~s$^{-1}$~pixel$^{-1}$)} 
}
\startdata
$F062$ & 0.44 & 0\\
$Z087$ & 0.34 & 0\\
$Y106$ & 0.38 & 0\\
$J129$ & 0.36 & 0\\
$H158$ & 0.35 & 0.005\\
$F184$ & 0.20 & 0.125\\
$W149$ & 0.97 & 0.099\\
\enddata
\end{deluxetable}

\begin{deluxetable}{cc}
\tablecaption{Read noise, WFC imaging survey. \label{RNval}}
\tablecolumns{2}
\tablewidth{0pt}
\tablehead{
\colhead{Survey Tier} & \colhead{Read Noise} \\
\colhead{} & \colhead{($e^{-}$ pixel$^{-1}$)}
}
\startdata
Shallow  & 26.38\\
Medium  & 14.53\\
Deep      & 8.67\\
\enddata
\tablenote{b}{Calculated via Equation~\ref{RN}.} 
\end{deluxetable}

For each SN, the underlying sky and host-galaxy flux is constant in time, meaning that the associated ``template'' noise for a SN is coherent across exposures.  The inclusion of this noise source is particularly important to the analysis of IFC-S observations. In the SDT report each template is planned to be a single medium exposure.  As this exposure is not particularly long (and shorter than the long exposures), it adds significant noise to the template-subtracted SN spectrophotometry.  On the other hand, this source is negligible for the WFC photometry, as imaging templates can be generated from several images, significantly reducing the template noise.

For each WFC simulated SN, we draw an underlying host-galaxy flux from a distribution determined from the high-$z$ {\it HST} SN survey portion of the CANDELS \citep{Grogin11, Koekemoer11} program.  From the CANDELS SN sample, we determine the host-galaxy surface brightness at the SN position for a 0.2\arcsec\ radius aperture in the $F606W$, $F775W$, $F850L$, $F105W$, $F125W$, $F140W$, and $F160W$ {\it HST} filters.  We then fit spectral models to the host-galaxy measurements.  From this sample, we determine the expected flux through each {\it WFIRST} filter as a function of redshift.  \SNANA\  has the ability to {\red include host-galaxy noise in simulations} for a variety of galaxy profiles and brightnesses. Since we have measured the {\red host-galaxy flux} at the SN position, we force the SN position to be at the center of an appropriate-brightness galaxy with a Sersic profile of index 0.5; {\red see Appendix~\ref{ap1a} for more details}.

{\red \SNANA\ includes the impact of host-galaxy Poisson noise when calculating the SN photometry, but it cannot currently simulate the same effect on IFC-S spectrophotometry. 
Investigations of this noise}
in WFC simulations shows that this is a negligible ($<$ 5\%) source of uncertainty for these observations. \\

\noindent{\bf Volumetric SN Rates:} To accurately determine the number of SNe~Ia (and CC SNe) that can be discovered by {\it WFIRST}, we parameterize the rate as a function of redshift, and fit to rate measurements that extend to $z = 2.5$ from \citet{Rodney14}, and \citet[][and references therein]{Graur14}. For SN~Ia the volumetric rates used are
\begin{equation}
{\rm R}_{\rm Ia} (z) = 
  \begin{cases}
  2.5 \times (1 + z)^{1.5} ~(10^{-5} {\rm ~yr}^{-1} {\rm ~Mpc}^{-3}), & {\rm for}~ z < 1.~ \\
  9.7 \times (1 + z)^{-0.5} ~(10^{-5} {\rm ~yr}^{-1} {\rm ~Mpc}^{-3}), & {\rm for}~ 1 < z < 3. 
  \end{cases}
\label{SNP1}
\end{equation}
{\red For CC SNe, we use the volumetric rate from \citet{Strolger15} (green line, Figure.~6, Equation~9).}
As the expected detection rate for $z > 3$ SNe is low, we do not attempt to simulate SNe at those redshifts.\\

\noindent{\bf Spectral model for SN~Ia:} We base all of our SN~Ia simulations on the SALT2 spectral model \citep{Guy07,Guy10}. Accurate spectrophotometry can be produced from this model, covering a range of phases and light-curve shapes. 
{\red The SALT2 model flux ($F$) in the rest frame is parameterized by \citet{Guy07} as
\begin{eqnarray}
    F(\tlam) & = & X_0 \left[ M_0(\tlam)  + {\red x_1}M_1(\tlam) \right] \times e^{c\cdot CL(\lambda)}, 
        \label{eq:salt2flux}  \\
     X_0 & = & 10^{0.4(\alpha x_1 - \beta c {\red - M})}.
     	\label{equ:x0}
\end{eqnarray}

The two SN-dependent parameters are the light-curve shape parameter, $x_1$, which characterizes the brightness as a function of phase, and the color, $c$, which describes the slope of an empirically determined color law that changes the spectral shape but does not depend on phase \citep[typical values of $x_1$ and $c$ are between $\pm$3 and $\pm$0.3, respectively, as seen in Table.6 of][]{Betoule14}. {\red The parameter M, in Equation~\ref{equ:x0}, represents the magnitude for which $x_1 = c = 0$.} 
The overall scale, $X_0$, depends on the global standardization parameters taken from the
JLA analysis \citep{Betoule14}:  $\alpha = \SIMalpha$ and $\beta=\SIMbeta$.
The fixed parameters from light-curve training are the spectral surfaces ($M_0$ and $M_1$),
which describe the SED versus phase, and the color law ($CL$), which describes the
wavelength dependence. For each epoch, the flux in Equation~\ref{eq:salt2flux} is redshifted,
integrated over the {\WFIRST} {\red passband throughput}, and dimmed according to the distance modulus
based on the cosmology parameters given below.

For our simulations and analysis}, the SN~Ia spectral model is an extension of the SALT2 model in \citet{Betoule14}. While {\it  WFIRST} will observe SNe in the rest-frame NIR, the fiducial SALT2 model is limited to optical wavelengths below {\red 9200~\AA.} 
To extend the model into the NIR {\red up to 25,000~\AA}, we follow the procedure used to simulate SNe for the CANDELS and the Cluster Lensing And Supernova survey with Hubble (CLASH) 
\citep{Rodney14,Graur14,Strolger15}, which is implemented with the {\tt SNSEDextend} software package \citep{Pierel18a}.

This SALT2 extension uses a compilation of 118 well-sampled, low-$z$ SNe~Ia with both optical and NIR light curves (Avelino et al.,\ in prep.; Friedman et al.,\ in prep.). NIR light-curve data are obtained from nearby SN surveys, principally from 
CfA IR1-2 \citep{Wood-Vasey08,Friedman15}, and 
CSP \citep{Contreras10,Stritzinger11}, 
as well as other sources 
\citep[see Table 3 of][and references therein]{Friedman15}. 
Corresponding optical photometry comes largely from 
CfA1-- 4 \citep{Riess99:lc, Jha06:lc, Hicken09:lc, Hicken12}, 
CSP \citep{Contreras10, Stritzinger11}, and 
LOSS \citep{Ganeshalingam10}. Each SN light curve in this sample is used to generate a spectrophotometric model by warping the SN~Ia spectral template from \citet{Hsiao07} to match the observed photometric colors at each epoch.

From the resulting set of 118 warped spectral time series models, a median spectral-energy distribution (SED) is derived for each phase, and smoothly joined with the $0^{\rm th}$-order component of the SALT2 model \citep[the $M_{0}$ component in][]{Guy07}. The higher order SALT2 model components, including variance and covariance terms, are extrapolated using flat-line extensions\footnote{For more details, see \url{http://github.com/srodney/wfirst}}. This model has not yet been calibrated to produce accurate distance estimates from real data. However, this SALT2 extrapolation is sufficient for producing realistic simulations for the purposes of investigating the {\it WFIRST} SN survey optimization.

Finally, the SALT2 color law was extended to infrared wavelengths using a modification of the polynomial function from \citet{Guy10}. The polynomial coefficients were set so that the effective color law approximately matches the extinction curve of \citet{Cardelli89}, with $R_{V} = 3.1$.

{\red To model intrinsic scatter, we use the spectral variation model in \citet{Kessler13} that is derived from the uncertainty model of \citet{Guy10}. This model results in 0.13~mag of scatter to the Hubble residuals: $\sim 70\%$ of this scatter is from achromatic luminosity variation at all wavelengths and phases, and $\sim 30\%$ of the scatter is from color variation.} 

While the SDT report assumes the intrinsic scatter is entirely achromatic, the scatter model used here does not. The population parameters for the color and stretch distributions of our simulations are those derived by \citet{Scolnic16} for the high-$z$ SN sample.\\

\noindent{\bf Spectral model for CC SNe:}
{\red The CC SED models used in our simulations are described by \citet{Kessler17} and \citet{Kessler10}, and were generated from a combination of SDSS \citep{Sako14} and CSP \citep{Hamuy06} light-curve data using $u,g,r,i,z$ filters. These optical CC templates have been extended into the NIR by warping a CC SED model\footnote{See \url{https://c3.lbl.gov/nugent/nugent_templates.html}} to match the $V-H$ and $V-K$ colors in \citet{Bianco14}. The synthetic colors are $V-H = V-K = 1$ for Types Ib and II; and $V-H=V-K = 1.5$ for Type Ibc \citep[see][for a review of SN spectral classification]{Filippenko97}.}\\

\noindent{\bf Selection  requirements (cuts):} Within the SDT report, a SN (both Ia and CC) is detected if it has an observation with 
${\rm SNR} \geq 4$ in both of the discovery filters ($Y$+$J$ or $J$+$H$), within the same epoch. 
To reduce CPU time without loss of accuracy, we simulate a trigger that requires ${\rm SNR} \geq 3$ in both discovery bands on the same epoch {\red (a trigger of ${\rm SNR} \geq 3$ rather than 4 was chosen in order to prevent the loss of SNe close to the detection limit).}

{\red After the simulation has generated light curves satisfying the 2-band trigger, we apply the photometric selection criteria defined in Section~\ref{classification}}. Spectra of the objects that successfully pass these criteria are analyzed via a modified, NIR-enabled version of the Supernova Identification \citep[SNID;][]{Blondin07} package. SNID compares each input SN spectrum to a library of template spectra and determines how closely template spectra match the input. 
In our SN classification analysis, we define a ``good SN~Ia'' if $\>80\%$ of the matches and the top match are a SN~Ia at the correct redshift, and if the SN is discovered roughly 7-12 days before peak. This SNID spectral analysis is used to implement the spectroscopic cuts described in Section~\ref{classification}.

For imaging-only strategies the selection criteria for the final sample occur only in the final analysis; no choices are made during the survey itself. First, we require that each SN have at least one epoch with ${\rm SNR} \geq 10$ and at least two epochs with ${\rm SNR} \geq 5$. 
Next, we require that the light-curve parameters of each SN fall within a 
``well trained'' range of color and stretch values such as those defined by 
\citet[][and references therein]{Betoule14}, i.e.,  $-3 < x_{1} < 3$ and $-0.3 < c < 0.3$. {\red Note that no cut on distance uncertainty is applied.}\\

\noindent{\bf Lensing:} {\red  We simulate distance dispersion from line-of-sight gravitational lensing. 
For SNe at $z>1.4$ we use the log-normal approximation in \citet{Marra13}. 
For $0.4 < z < 1.4$, lensing is computed from a 900 deg$^{2}$ patch of the MICECATv1\footnote{\url{https://cosmohub.pic.es}} simulation. 
For $z<0.4$, the distance dispersion at $z=0.4$ is scaled by $z/0.4$. 
The resulting root-mean square (RMS) dispersion is $0.04\times z$, which is on the low side of predictions \citep[e.g.,][]{Jonsson10}. 
We therefore scale the dispersion by $\sim$1.4 to achieve the average predicted dispersion of $0.055 \times z$.}\\

\noindent{\bf Galactic Extinction:} Since the SN fields have not been chosen, we assume that  each field will have a low value of $E(B-V) = 0.015$~mag.\\

\noindent{\bf Low-Redshift Sample:}
To provide an anchor for the SN~Ia Hubble diagram,
we include 800 simulated SNe~Ia with $z < 0.1$ 
from a source other than {\it WFIRST}, which we model as having the characteristics of the Foundation SN survey \citep{Foley17}. The Foundation survey uses the PS1 telescope and observes low-$z$ ($0.01 < z < 0.1$) SNe in $griz$ every 5 days with typical {\red distance modulus uncertainties} {\red due to measurement error $<0.1$~mag}. A similar external low-$z$ SN~Ia sample is a requirement specified in the SDT report. \\

\begin{figure*}
\centering
\includegraphics[width=1.7in]{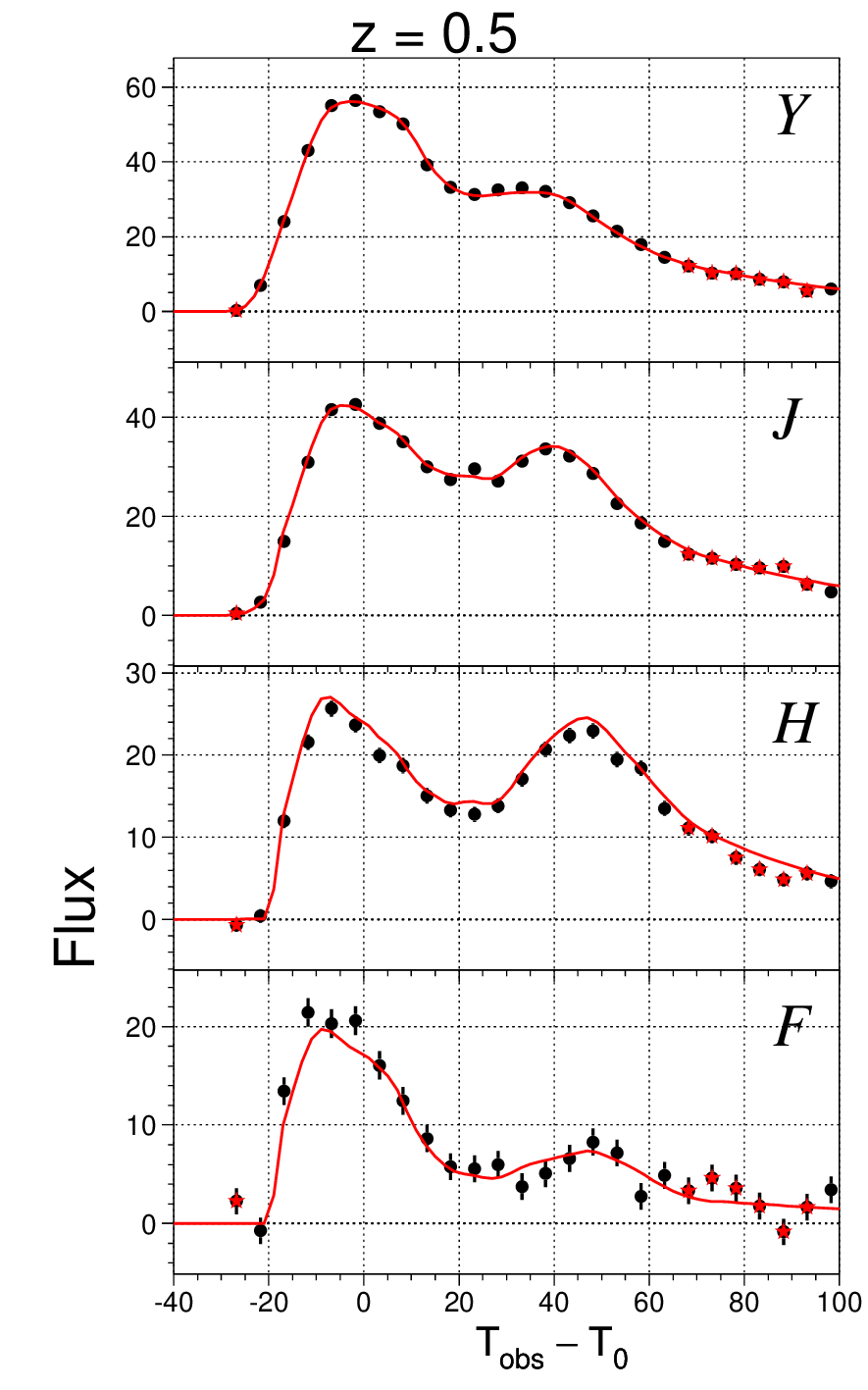}
\includegraphics[width=1.7in]{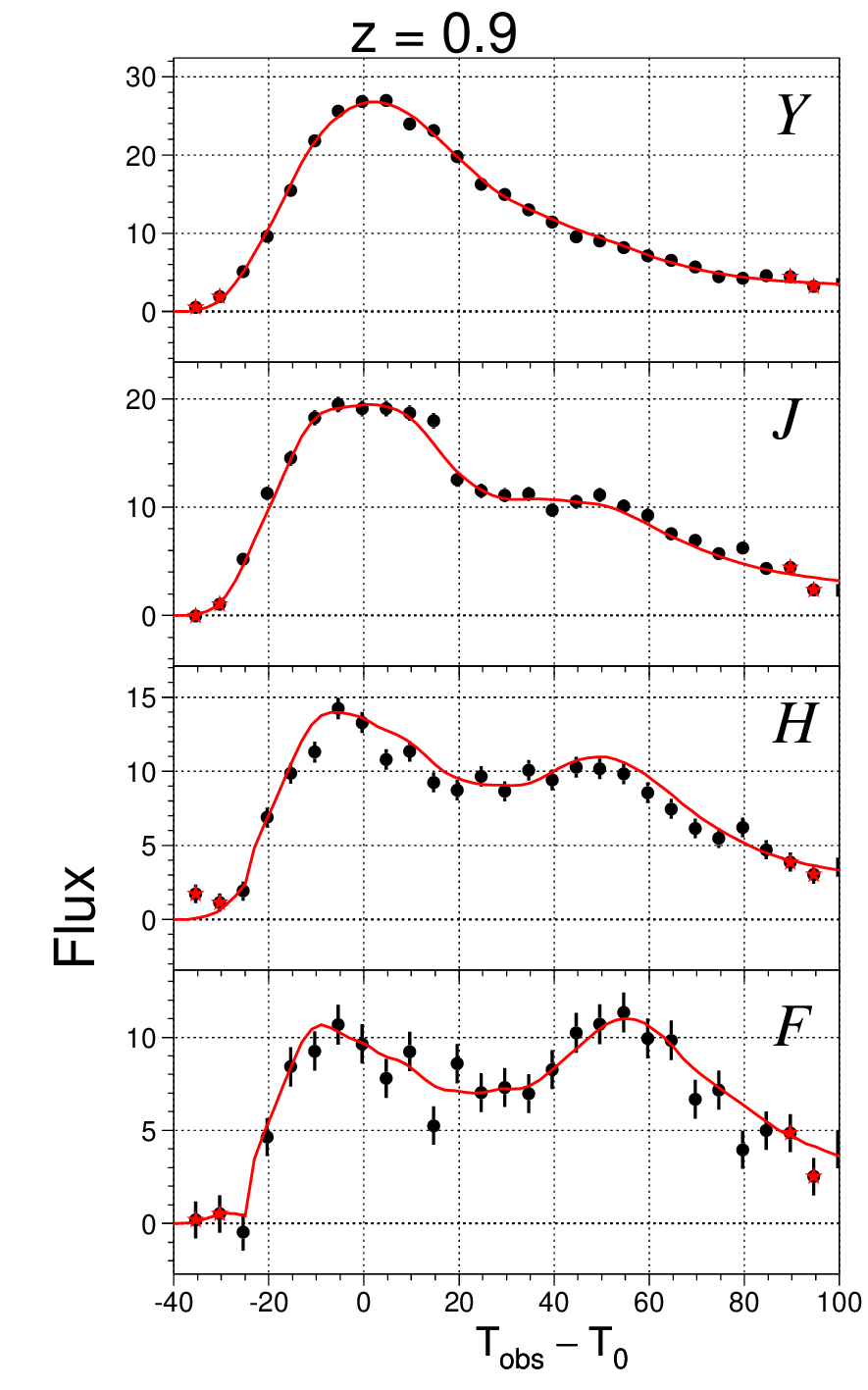}
\includegraphics[width=1.7in]{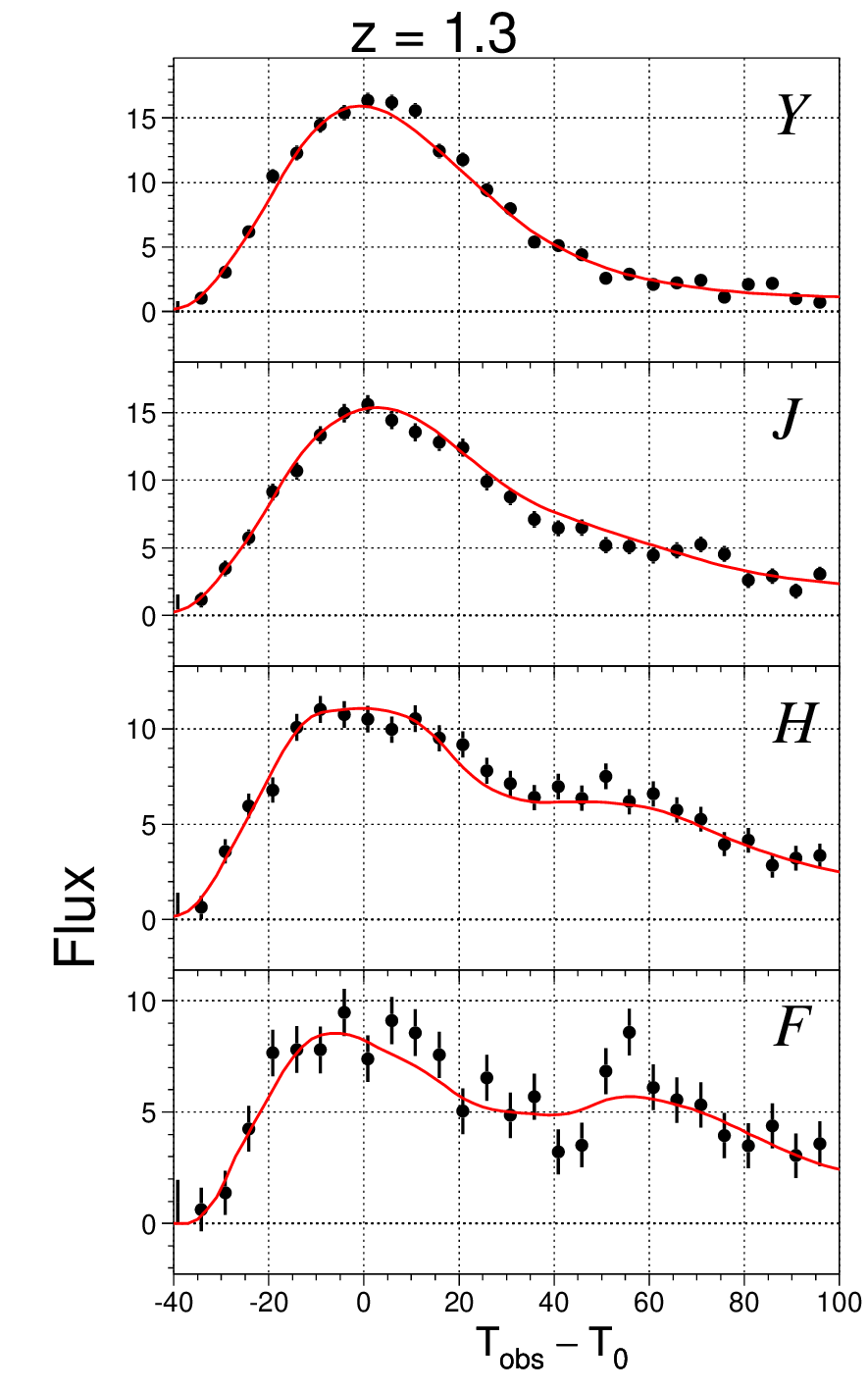}
\includegraphics[width=1.7in]{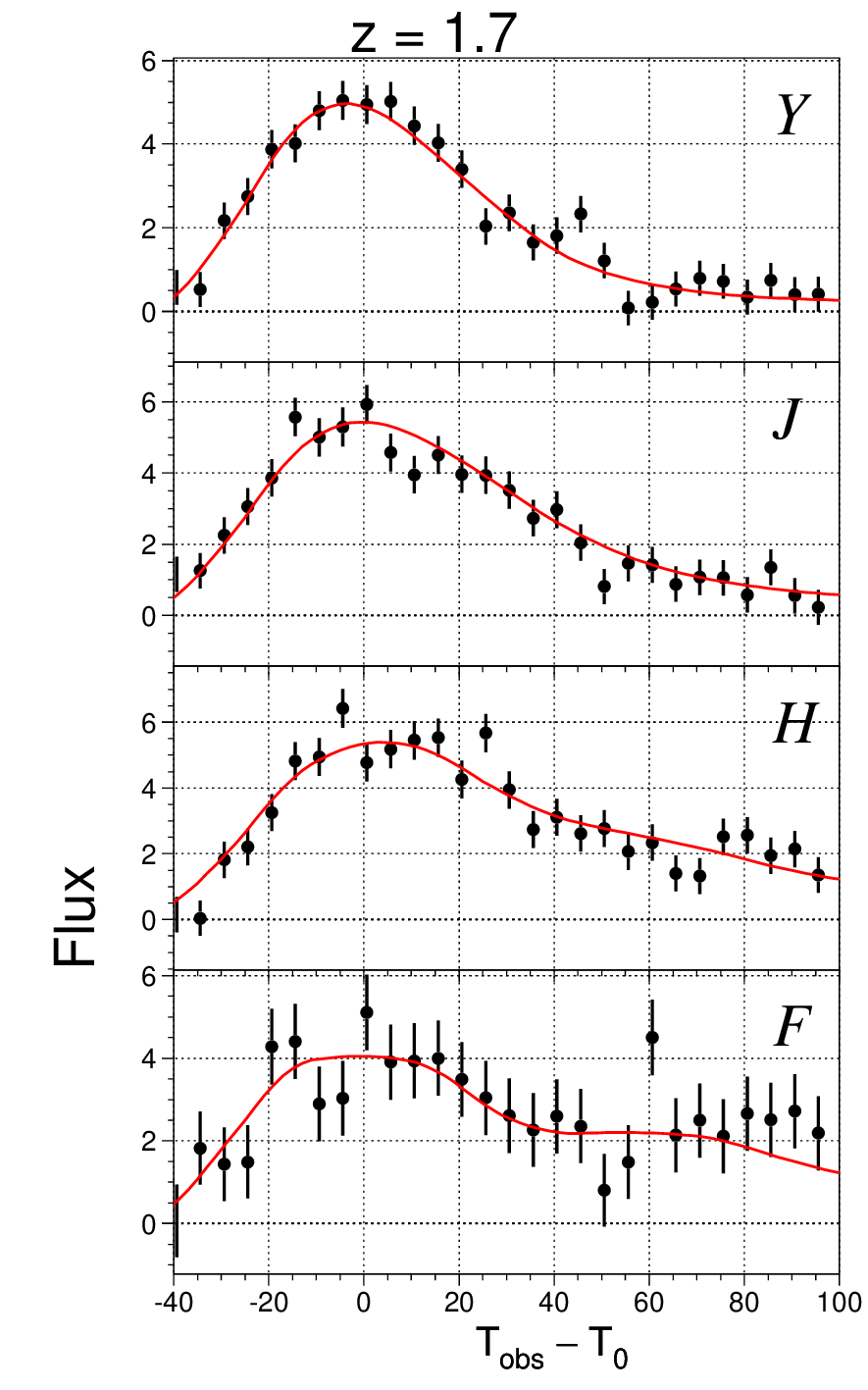}
\caption{Example {\it WFIRST} broadband ($YJHF$) simulated light curves (black circles) and best-fit light curves (smooth curve) for SNe at redshifts 0.5, 0.9, 1.3, and 1.7. Magnitudes are $27.5 - 2.5 \times {\rm log}_{10}(\rm Flux)$; e.g., ${\rm m} = 25.0$ mag for ${\rm Flux} = 10$. These data are generated using the medium imaging exposure time of 67 s. Red stars seen in some of the plots indicate a point that is excluded from the SNANA fit.}
\label{RICK}
\end{figure*}

\noindent{\red{\bf Cosmology Parameters:}
Distance moduli are generated with the following $w$CDM model parameters:
$\OM = \SIMOM$, $\OL = \SIMOL$, $w_0 = \SIMw$, and $w_a = \SIMwa$.

\subsection{Cosmology Analysis}
\label{subsec:CosmoAna}

Here we describe the general analysis strategy for measuring cosmological parameters and FoM.
After the simulation, selection cuts are applied (Section \ref{results}) and each light curve is fit with the SALT2 model
to determine three parameters: stretch ($x_1$), color ($c$), and amplitude ($x_0$).
In the limit of no intrinsic scatter and no measurement noise, $x_0 = X_0 10^{-0.4\mu}$.
The distance modulus for each event is obtained from the \citet{Tripp98} formulation,
\begin{equation}
  \mu = \mB  - M + \alpha \cdot x_{1} - \beta \cdot c, 
  \label{eqn:Tripp}
\end{equation}
where $\mB \equiv -2.5\log_{10}(x_0)$.
$M$, $\alpha$, and $\beta$ are global nuisance parameters determined from a fit
to minimize the Hubble residuals for an entire sample. 
Since $\mB$ and $M$ are degenerate, $M$ is often quoted to be around $-19.35$ mag with a corresponding $\mB$-offset  
such that $\mB$ appears to be a Bessell $B$-band magnitude. Here we leave out the $\mB$ offset to make clear
that $\mB$ is related to the amplitude ($x_0$) and is not related to the $B$ band.

For the WFC, the light-curve fit includes the imaging filters. 
For the IFC, the light-curve fit includes the 52 synthetic filters.
Typical WFC light-curve fits are shown in Figure~\ref{RICK} for a range of redshifts. 

We use the ``BEAMS with Bias Correction" ({\BBC}) method \citep{Kessler17} to determine distances from the
Tripp equation (Eq.~\ref{eqn:Tripp}). 
\BBC\ determines the bias on each fitted parameter ($c$,$x_{1}$,$\mB$) for each event, and minimizes the
Hubble residuals in a global fit to determine 
$\alpha$, 
$\beta$, 
a scale parameter ($\SCC$) for the CC probabilities, 
and an average distance modulus in approximately 40 log-spaced redshift bins. 
The output of \BBC\ is a redshift-binned Hubble diagram that is corrected for biases from 
selection effects and from CC contamination. 
The \BBC\  distance modulus uncertainties include 
statistical uncertainties on the fitted parameters ($c$,$x_{1}$,$\mB$),
Gaussian lensing scatter ($\sigma_{\rm {\red \mu}} = 0.055\times z$),
peculiar velocities ($\sigma_v = 150$~km/s), and 
intrinsic scatter.
In this analysis we have not trained a photometric classifier to determine CC probabilities, {\red and therefore we do not include the CC SN likelihood, but fix $\SCC=0$. This makes our estimate of the systematic uncertainty from CC contamination conservative, because the analysis does not take advantage of the BEAMS approach to account for contaminants.} 

Following the SDT report, we often plot the ``fractional distance uncertainty (FDU)''
defined as
\begin{equation}
  {\rm FDU} \equiv 10^{0.2\sigma_{\mu}} - 1,
    \label{eq:fdu}
\end{equation}
where  $\sigma_{\mu}$ is the {\BBC}-fitted uncertainty in the distance modulus. 

For each systematic uncertainty (Section~\ref{sysun}), the SALT2 light-curve fits and \BBC\  fit are repeated.
The resultant set of \BBC\  Hubble diagrams are used to compute a
total covariance matrix ($\Ctot$) that includes both statistical and systematic uncertainties. 
Compared to using individual SN distances, using redshift-binned distances greatly reduces the 
required computing time for constructing $\Ctot$.

The redshift-binned distance moduli, their respective uncertainties, and $\Ctot$ are passed to \CosmoMC\ \citep{Lewis13} to determine cosmological constraints.
FoMs are calculated corresponding to the inverse area of the 95\% confidence contours in the $w_{0}$--$w_{a}$ space \citep{Albrecht06}. 
For each FoM determination, we assume a flat Universe and marginalize over $H_{0}$ and $\Omega_{M}$. 
Furthermore, we include prior constraints from both baryon acoustic oscillation \citep[BAO;][]{Anderson14} 
and cosmic microwave background \citep[CMB;][]{Planck15} datasets. }

When assessing the effect of each individual systematic uncertainty, a modified version of \CosmoMC\ is used to reduce the computational complexity of determining the FoM. 
This version of \CosmoMC, which we call ``{\CosmoMCfast,}"  encodes the CMB information using the compressed 
Gaussian likelihood presented by the \citet[see their Table~4; the version which does not marginalize over $A_{L}$]{Planck15} 
and only accounts for the geometric effects of dark energy. Therefore \CosmoMCfast\ fixes $\tau$, the reionization optical depth, and $\log (A)$ (equivalent to $\ln (10^{10} A_{s}$), where $A_{s}$ is the inflation power spectrum amplitude). These changes significantly reduce the time to compute the FoM. We have done extensive checks to ensure that the \CosmoMCfast\ model produces accurate results relative to the original version that uses the full set of Planck likelihoods. Fluctuations of a few percent in the FoM value are expected {\red to arise from the sampling variance of the finite Markov Chain Monte Carlos (MCMC).}

\section{Simulated Strategies}
\label{results}

Here we describe simulations of several different strategies for the {\it WFIRST} SN survey. The survey variations examined here are summarized in Table~\ref{strat}. This table lists the strategy names, filters, imaging tiers, areas, and the resultant number of simulated SNe~Ia. 

We include strategies that use both the WFC imager and the IFC-S spectrograph (the SDT, SDT*, and SDT* Highz strategies) 
as well as strategies that employ imaging exclusively (Imaging, Imaging:Lowz, Imaging:Highz strategies).

For each strategy {\red we account for the 42 s slew-and-settle time per exposure, and we satisfy the constraint of 6 months total observing time.} 

For the imaging component of the survey, the exposure time per tier, filter zero-points, and sources of noise for each filter are specified in Tables~\ref{zpts}, \ref{exptim}, \ref{zpts2}, and \ref{RNval}. When the IFC-S is used, its wavelength range, redshift-dependent exposure times, and sources of noise remain set to the values given in Tables~\ref{ifuextend} and \ref{ifutime}. 

If an instrument (i.e., the IFC-S), tier (shallow or deep), or filter within a survey simulation is removed or added, 
the areas (listed in Table~\ref{filtttt}) of the remaining tiers are scaled evenly (except in the SDT* Highz {\red and SDT Imaging cases}; see Section~\ref{HIGHz-IFC+IMG}) to account for the loss or gain of time. 

Note that we have not changed the cadence, depth of a given tier, IFC-S strategy (epochs and number of SNe), or {\it WFIRST} filter bandpass for any strategy outlined within this paper. Such investigations/optimizations will be the focus of future papers.

For strategies that only have an imaging component, we consider the impact 
of {\red adding bluer filters that are not in the current {\WFIRST} design. We do not consider filters redder than $F$ band.
For any given strategy, we choose no more than six filters total.}
 
For simplicity, we assume that the additional filters are similar to those from {\it HST's} WFC3, and as such we have used their throughputs and taken the average AB magnitudes of the two WFC3 chips to be our zero-points (see Table~\ref{wfc33}). The FWHM values for these filters are calculated in part via the use of the WebbPSF for {\it WFIRST}\footnote{https://pythonhosted.org/webbpsf/wfirst.html} tool. This tool allows the user to input appropriate SN spectra, and account for wavefront aberrations, in order to calculate binned and unbinned PSF data. WebbPSF, however, is not designed for filters bluer than the $Z$ band. We therefore modified this tool to calculate bluer wavefront aberrations via the extrapolation and application of higher order Zernike coefficients. Pixelation is then applied to these results along with an inter-pixel capacitance effect on the order of $\sim$2\%. As PSF FWHM values change slightly between each tier, resultant average values are presented in Table~\ref{wfc33}.

\begin{deluxetable}{ccc}
\tablecaption{{\it HST} WFC3 filters included within {\it WFIRST} SN simulations \label{wfc33}}
\tablecolumns{3}
\tablewidth{0pt}
\tablehead{
\colhead{Filter} & \colhead{Zero-Point} & \colhead{PSF FWHM}\\
\colhead{} & \colhead{(AB Mag)} & \colhead{(pixel)}
}
\startdata
$F425W$ ($B$) & 24.75 & 1.62\\
$F555W$ ($V$) & 25.72 & 1.62\\
$F814W$ ($I$) & 25.03 & 1.67\\
\enddata                                      
\end{deluxetable}

In the current SDT strategy, a set number of SNe in the $0.1 < z \leq 1.7$ range are followed-up with the IFC-S (2726 SNe). For imaging-only strategies, there is no need to fix the number of SNe or the redshift range. We therefore allow the redshift range to extend from 0.01 to 2.99.  However, additional selection criteria as described
in Section~\ref{simulation} are implemented, and when combined with typical cuts on the color and stretch of the SN light curves the photometric classification purity is $>$99\%.

This small CC SN contamination of the SN~Ia sample is included as a systematic uncertainty within our work. Note that host-galaxy redshifts in an imaging-only survey could be collected after the {\it WFIRST} survey is completed, since they are not needed to define the imaging
sequence (as is the case for the SDT survey).

The design of each survey strategy is discussed below.

\begin{deluxetable*}{l ccc ccc ccc ccc}
\tabletypesize{\scriptsize}
\tablecaption{Simulated strategies investigated for the {\it WFIRST} SN survey, including the strategy suggested within the SDT report. \label{strat}}
\tablewidth{0pt}
\tablehead{
\multirow{3}{*}{Name}& \multicolumn{3}{c}{Redshift Range} & \multicolumn{3}{c}{Filter Set Used}  & \multicolumn{3}{c}{Area (deg$^{2}$) } & \multicolumn{3}{c}{Number of SN~Ia Selected}\\ 
\colhead{}& \colhead{Shallow} & \colhead{Medium} & \colhead{Deep} & \colhead{Shallow} & \colhead{Medium} & \colhead{Deep} & \colhead{Shallow} & \colhead{Medium} & \colhead{Deep} & \colhead{Shallow} & \colhead{Medium} & \colhead{Deep} 
}
\startdata
SDT & 0.10--0.39 & 0.40--0.79 & 0.80--1.70 & IFC-S, $YJ$ & IFC-S, $JH$ & IFC-S, $JH$ &  27.44 & 8.96 &  5.04 & 27 & 300 &  1181\\
SDT* & 0.10--0.39 & 0.40--0.79 & 0.80--1.70 & IFC-S, $YJ$ & IFC-S, $JH$ & IFC-S, $JH$ &  27.44 & 8.96 & 5.04  & 149 & 598 &  1221 \\
SDT* Highz & \nodata & 0.10--0.79 & 0.80--1.70 & \nodata & IFC-S, $JH$ & IFC-S, $JH$ & \nodata & 22.80 & 5.04 & \nodata & 1214 & 1217\\
SDT Imaging & 0.01--2.99 & 0.01--2.99 & 0.01--2.99 & $YJ$ & $JH$ & $JH$ & 27.44 & 8.96 & 5.04 & 1 & 546 & 3046 \\
Imaging:Allz & 0.01--2.99 & 0.01--2.99 & 0.01--2.99 & $RZYJ$ & $RZYJ$ & $YJHF$ & 48.82 & 19.75 & 8.87 & 1225 & 5723 & 6640 \\	
Imaging:Lowz & 0.01--2.99 & 0.01--2.99 & \nodata & $YJ$ & $JH$ & \nodata & 142.30 & 66.91 & \nodata & 6 & 4560 & \nodata \\	
Imaging:Lowz* & 0.01--2.99 &  0.01--2.99 & \nodata & $RZYJ$ & $RZYJ$ & \nodata & 73.57 & 32.24 & \nodata & 1828 &  9396 & \nodata \\
Imaging:Lowz+ &	0.01--2.99 & 0.01--2.99 & \nodata & $RZYJHF$	& $RZYJHF$ & \nodata & 50.66 & 20.68 & \nodata & 1237 & 5990 & \nodata \\
Imaging:Lowz-Blue & 0.01--2.99 & 0.01--2.99 & \nodata &  $BVRIYJ$ & $BVRIYJ$ & \nodata & 50.66 & 20.68 & \nodata &  1169 & 5644 & \nodata \\
Imaging:Highz* & \nodata & 0.01--2.99 & 0.01--2.99 & \nodata & $RZYJ$ & $YJHF$ & \nodata & 32.06	& 13.24 & \nodata & 9354 &  9640 \\
Imaging:Highz+ & \nodata & 0.01--2.99 & 0.01--2.99 & \nodata &  $RZYJHF$ & $RZYJHF$ & \nodata & 20.50 &  9.14 & \nodata & 5965 & 6759 \\
\enddata         
\tabletypesize{}                                           
\end{deluxetable*}

\subsection{The SDT and SDT* Strategies}
\label{SDTSN}
Here we present the simulated SDT survey strategy (see Section~\ref{currentsn}) along with a slightly modified version of this strategy (called SDT*), in which the efficiency of SN~Ia detection and selection has been significantly improved. These strategies use both WFI channels: the WFC imager and IFC-S.

The number of generated SNe is set by the volumetric rates, survey area, depth, and duration; the numbers are reported in Table~\ref{snn} and do not include selection requirements. 
Within the appropriate redshift ranges a total of 21,094 SNe are generated: 3,608 are SNe~Ia, and the remaining 17,486 are CC SNe. 
The initial SDT SNR requirement (Section~\ref{classification}) reduces the total to 6,640 ``detectable" events (3,514 of which are SNe~Ia, see Table~\ref{cutss}). For these detectable events, 2,621 pass the photometric cuts specified within the SDT report (see list given in Section~\ref{classification}). A breakdown of the number of SNe passing each cut is given in Table~\ref{cutss}.

\begin{deluxetable}{ccccc}
\tabletypesize{\scriptsize}
\tablecaption{Number of SNe generated per SDT survey tier. \label{snn}}
\tablecolumns{5}
\tablewidth{0pt}
\tablehead{
\colhead{Survey} & \colhead{Redshift} & \colhead{Number of} & \colhead{Number of} & \colhead{Total SNe}\\
\colhead{Tier} & \colhead{Range} & \colhead{SNe~Ia} & \colhead{CC SNe} & \colhead{per Tier}
}
\startdata
 Shallow & $0.1 \leq z <     0.4$ & 520   & 2437   & 2957\\
 Medium & $0.4 \leq z <    0.8$ & 932  & 5080   & 6012\\
 Deep     & $0.8 \leq z \leq 1.7$ & 2156 & 9969 & 12125\\ \hline   
& {\bf SN Total:} & {\bf 3,608} & {\bf 17,486} & {\bf 21,094}\\
\enddata    
\tabletypesize{}                                                
\end{deluxetable}

\begin{deluxetable}{cllllllc}
\tablecaption{Number of SNe that satisfy the photometric cuts defined in the SDT report\tablenotemark{a}. \label{cutss}}
\tablewidth{0pt}
\tablehead {
\multirow{2}{*}{Cut} & \multicolumn{2}{c}{Shallow} &  \multicolumn{2}{c}{Medium}  &  \multicolumn{2}{c}{Deep} & \multirow{2}{*}{Total} \\ 
\colhead{} & \colhead{Ia} & \colhead{CC} & \colhead{Ia} & \colhead{CC} & \colhead{Ia} & \colhead{CC} & \colhead{}
}
\startdata
0 & 426 & 175 & 932 & 672 & 2156 & 2279  & 6640 \\
1 & 159 & 67  &  718  & 202 & 2073 &  782 & 4001\\
2 &  27  & 28   & 305  &  110 &  1748  & 403 & 2621\\
\enddata                                                       
\tablenotetext{a}{Photometric cuts are defined in Section~\ref{classification}.}
\end{deluxetable}

{\red Our analysis implements the photometric cuts by defining a range of acceptable color and rise values. Details on how these ranges were defined are provided in Appendix~\ref{ap1b}.}

{\red A short- and medium-exposure spectrum is obtained for each SN that satisfies photometric {\it Cuts 0} and {\it 1} respectively. A long-exposure spectrum is obtained once the SN passes both the photometric {\it Cut 2}, and spectroscopic {\it Cut 3}.}
Within SNID \citep[][see references therein for a list of spectra used]{Blondin07}, each spectrum is compared to a library of real SN spectra. The number of SNe passing these additional spectroscopic selection criteria (see Section~\ref{classification}) are reported in Table~\ref{cutss2}. {\red Analysis within \SNANA\ reduces the sample to 1,957 SNe~Ia, i.e., 56\% of the total number of SNe~Ia detected (3,514 SNe~Ia). As is done with all current cosmological analyses, we then apply additional color and light-curve shape requirements, which further reduces the number of SNe in the cosmological sample.}

While the number of SNe observable with IFC-S dominated surveys is limited by the observing time required to reach the desired SNR in each spectrum, our work has shown that the original SDT-defined selection effects are inefficient in obtaining the ~2726 SNe~Ia desired by the report. In addition, the distribution of SNe~Ia obtained in our simulations does not match that of the SDT, which results in larger fractional distance uncertainties at low $z$ (z $\leq$ 0.6). The dependence of the SDT strategy on a long-exposure spectrum for final classification is also inefficient in that for tens of SNe (see difference in SNe between {\it Cuts 3} and {\it 4} of Table.~\ref{cutss2}), this spectrum indicates that they are not SNe~Ia, resulting in several hours of exposure time spent on contamination.

\begin{deluxetable}{ccc}
\tablecaption{Number of SNe that pass the photometric and spectroscopic cuts defined by the SDT report, and the number of SNe~Ia within this sample which then undergo further cuts based on color and stretch parameters. SNe that pass these final cuts are then used within our analysis. \label{cutss2}}
\tablecolumns{5}
\tablewidth{0pt}
\tablehead{
\colhead{Cut} & \colhead{Number of SNe} & \colhead{Number of SN~Ia}\\
\colhead{}& \colhead{}& \colhead{analyzed}
}
\startdata
 3 & 2109 & 1958\\
 4 & 2013 & 1957\\
\enddata                                                   
\end{deluxetable}

\begin{figure*}
\centering
\includegraphics[keepaspectratio=true,  width=0.8\textwidth]{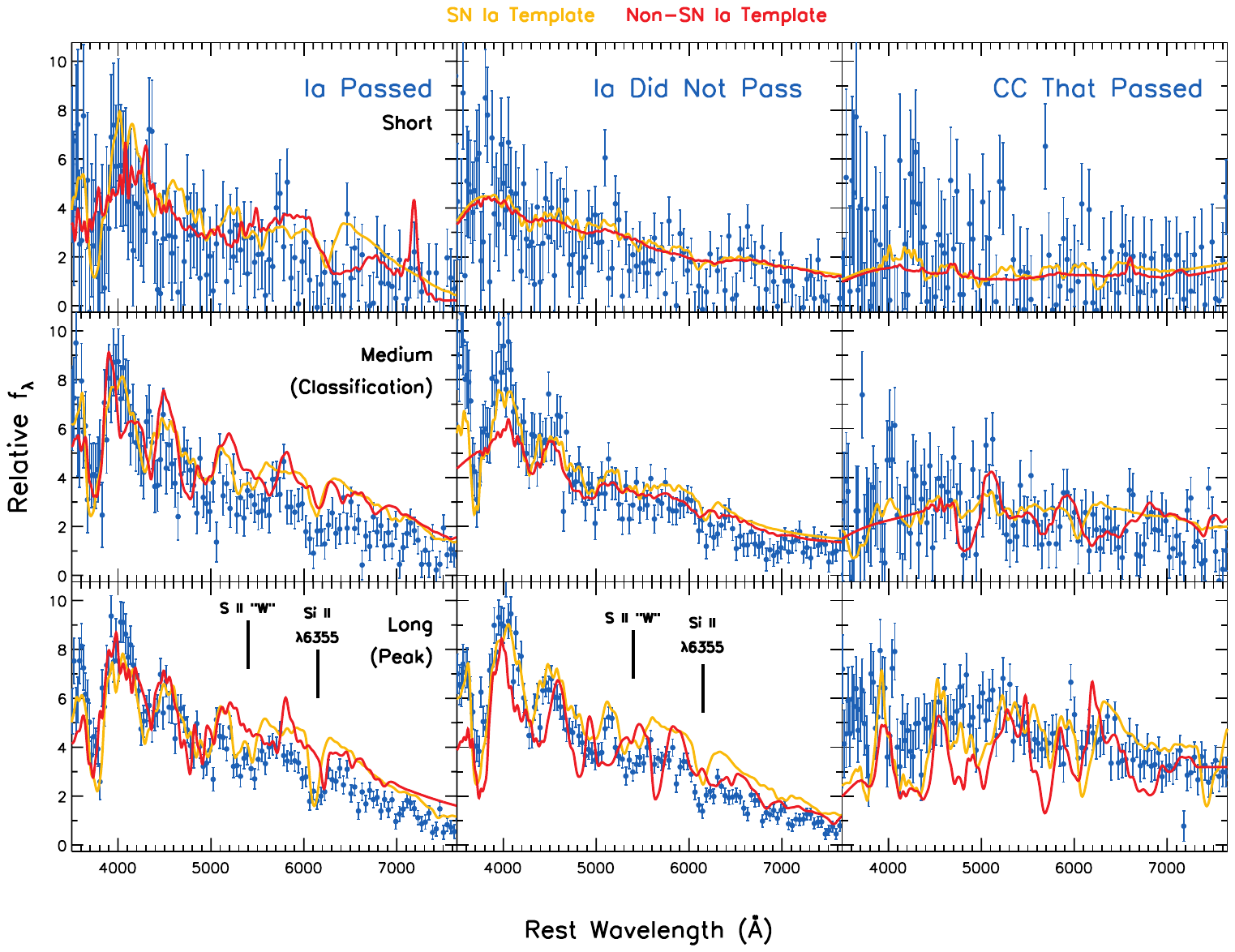}
\caption{Simulated rest-frame {\it WFIRST} IFC-S spectra of $z = 1$ SNe. The left panels correspond to a SN~Ia that passes all cuts and for which full follow-up observations would be obtained. The middle panels correspond to a SN~Ia that is not identified as SN~Ia based on its long-exposure spectrum and is thus rejected. 
The right panels correspond to a CC SN that passes all requirements, including photometric cuts, and would undergo the full set of follow-up observations. The top, middle, and bottom rows correspond to short-, medium-, and long-exposure spectra, respectively, for each SN. The {\it WFIRST} spectra are plotted as blue points with error bars. Note the changing resolution with wavelength. The best-matching SN~Ia and non-SN~Ia spectrum are plotted as gold and red, respectively.}
\label{spec}
\end{figure*}

When considering the purity of the SN~Ia sample, the photometric selection criteria ({\it Cuts 0}, {\it 1}, \& {\it 2}) result in a purity of $\sim$79\% {\red (2080 of the 2621 SNe within {\it Cut} 2 are SNe~Ia)}. Of the $\sim$21\% CC SN contaminants, $\sim$67\% are SNe~Ib/c and $\sim$33\% are SNe~II. The SNe~Ib/c that make up the majority of the contaminants are also the objects that are most spectroscopically similar to SNe~Ia, and therefore the most difficult to remove with low-SNR spectra. Example spectra of a SN~Ia that passes all cuts, a SN~Ia that is excluded based on its long-exposure spectrum, and a CC~SN (SN~Ic) that passes all cuts and is included in the cosmology sample are illustrated in Figure~\ref{spec}. This figure demonstrates the difficulty of classification using the SDT report strategy. {\red The middle panel of Figure~\ref{spec} illustrates a SN~Ia with spectral features that cannot be identified given its SNR and resolution within each of the different IFC-S exposures. In particular, the sulfur ``W," which the SDT report uses as a clear example of a SN~Ia feature, is not detected in the long-exposure SN~Ia spectrum (bottom row, center column of of Figure~\ref{spec}), and thus this object is rejected.

All spectral classification is conducted using SNID, and a SN is classified as a SN~Ia if $\>80\%$ of the matches and the top match are a SN~Ia at the correct redshift (see Section~\ref{simulation} for further details)}. 
{\red The SDT report does not indicate any use of the photometric light-curve data beyond that of the first three epochs after detection. 
If all epochs for a given SN light curve were used in the classification, and SNR cuts on the light curves are applied (as 
discussed below), we would likely reduce CC~SN contamination to a negligible level (see Section.~\ref{cont}).}

The current number of correct spectral classifications for the SDT strategy is likely optimistic. 
While correlated template sky noise is included in the simulations, we have optimistically ignored flux and noise from un-subtracted host galaxy light that will be there as a consequence of not having a galaxy template at the time of classification.
Even if a spectrum of the host-galaxy does exist (e.g., from a ground-based spectrograph), the exact galaxy SED at the position of the SN will not be accurately measured.

\begin{figure*}
\centering
\includegraphics[keepaspectratio=true,  width=\textwidth]{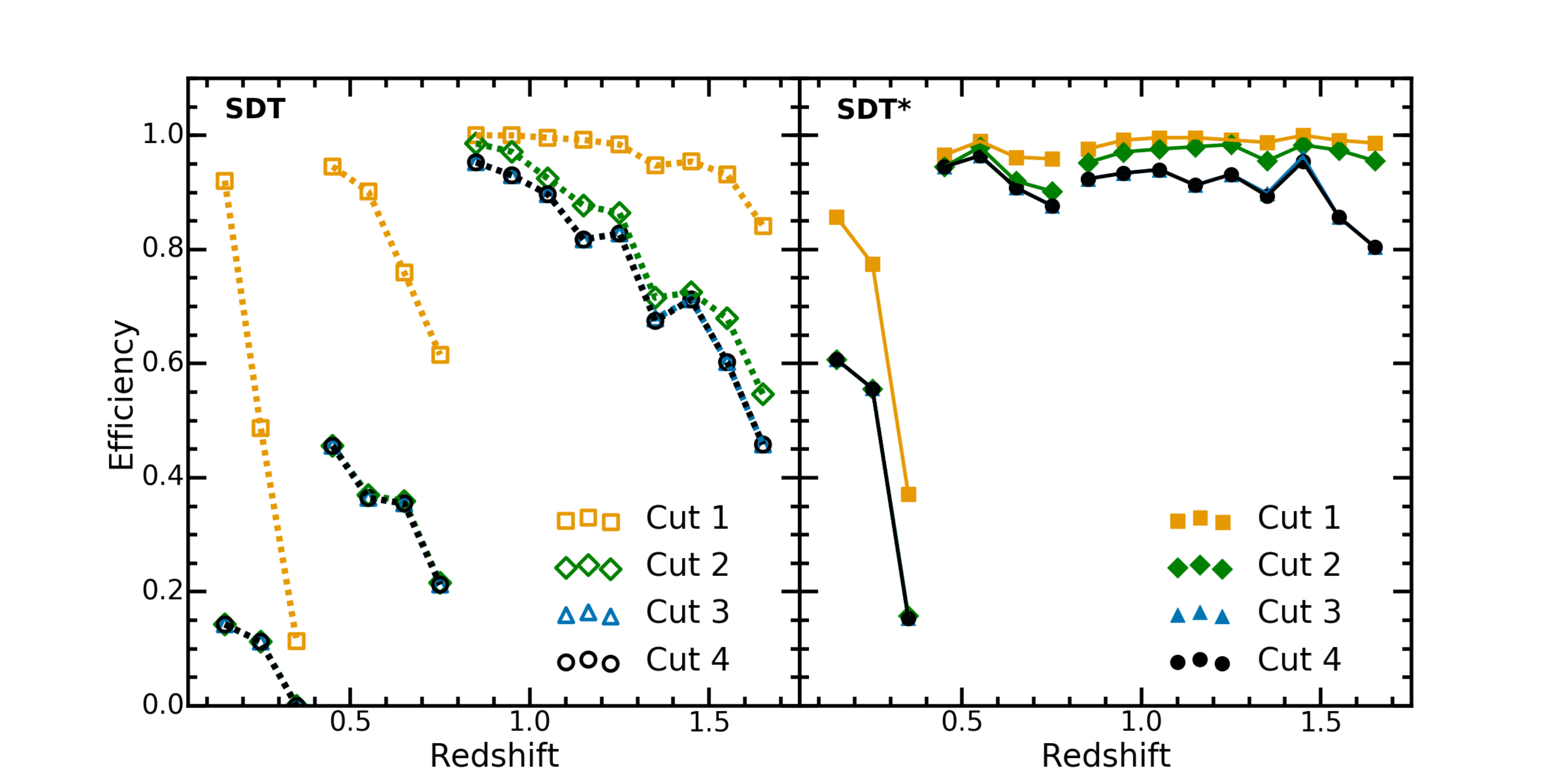}
\caption{{\red SDT and SDT* (left and right panels, respectively) SN~Ia selection efficiency as a function of redshift. 
{\red Efficiency is defined as the number of SNe correctly identified as a Type~Ia within each cut, divided by the number of SNe~Ia that pass {\it Cut 0} (see Section~\ref{classification}). The efficiency is calculated  per 0.1 redshift bin.} The gold squares, green diamonds, blue triangles, and black circles represent the efficiency of SNe~Ia that pass {\it Cuts} 1, 2, 3, and 4 respectively. Lines connect data from the same tier of the survey. The large drop in efficiency from the first to second cut at $z < 0.8$ for the SDT strategy is caused primarily by the SDT requirement that a SN flux rises between each epoch. Comparison of the two strategies suggests that the looser selection criteria of the SDT* strategy is significantly more efficient than that of the SDT.}}
\label{eff}
\end{figure*}

{\red The SN~Ia efficiency for a given selection cut (see Section~\ref{classification}) is shown in Figure \ref{eff}. Efficiency is defined as the number of SNe correctly identified as Type~Ia and passing each cut, divided by the number of SNe~Ia that pass {\it Cut 0}. This calculation is per 0.1 redshift bin. SNe that pass {\it Cut} 4 have short, medium, and deep exposure IFC-S spectra, in addition to six-short follow-up spectra.}

The efficiency is low at particular redshifts, $0.3 < z \leq 0.4$ and $0.7 < z \leq 0.8$. 
This efficiency gap is partially the result of the survey design producing insufficient SN discoveries at the high-$z$ end of each tier. However, photometric selection criteria that require the color to be
consistent with a SN~Ia at their host-galaxy redshift, and that the flux rise between epochs, are the main contributors for the low efficiency. For the shallow imaging tier, which covers $0.1 \leq z < 0.4$, these criteria are problematic due to the tier's short 13-s exposure, which results in noisy light curves and an undetectable rise value.

The large reduction in the number of SNe between {\red {\it Cuts} 1 and 2} is also caused by the required increase in flux between epochs. In many cases statistical noise causes a candidate to appear to fade between two successive epochs. 
To reduce this selection artifact, this criterion is loosened via the iterative examination of a range of measured rise values (including negative values) 
for the simulated SNe~Ia as a function of redshift for each tier, and a looser cut is applied.

{\red The requirements on the discovery colors ($Y-J$ for shallow, $J-H$ for medium and deep)} are also tightened, excluding some of the most extreme SNe~Ia and significantly reducing the number of CC SNe at each step. The effect of these improved selection criteria is {\red to increase the SN~Ia acceptance from $56\%$ to $\sim 81\%$} (2909 SNe~Ia make it to the final sample), and to decrease the number of misclassified CC SNe, all with minimal SN~Ia losses. Hereafter we refer to this sample as SDT*, a simulated SDT strategy where the selection criteria have been modified.

Using the results of both the SDT and SDT* selection procedure, the efficiency of each strategy is shown in Figure~\ref{eff}. {\red The SDT methodology results in a rapidly falling efficiency at high $z$, while the STD* strategy has a relatively flat efficiency versus redshift.}

Spectroscopic classification for the SDT* strategy, however, suffers from the same issues as the SDT strategy, reducing the efficiency to $\sim$82\% (average of efficiency measurement for each 0.1 redshift bin). While we have not yet examined potential biases related to the spectroscopic selection, previous experience with spectroscopically confirmed SN samples show that this selection will introduce a distance bias that must be corrected. 
{\red The resultant} {\red combined photometric and spectroscopic selection procedure} has a $\sim$99\% purity, which will further increase when considering full light curves and all spectral data.

To accurately match the SDT description of their survey strategy 
we select SN redshifts only from their corresponding tiers. This is particularly important for $z < 0.4$, where the shallow tier is conducted in the $Y$ and $J$ filters, while deeper tiers use the $J$ and $H$ filters. 
To be specific, SNe with $z < 0.4$, $0.4 \leq z < 0.8$, and $0.8 \leq z \leq 1.7$ are selected exclusively from the shallow, medium, and deep tiers, respectively. The overall efficiency is not strongly affected by this decision since the IFC-S exposure time is only a function of redshift and not, for instance, brightness.

\begin{figure*}
\centering
\includegraphics[keepaspectratio=true,  scale=0.25]{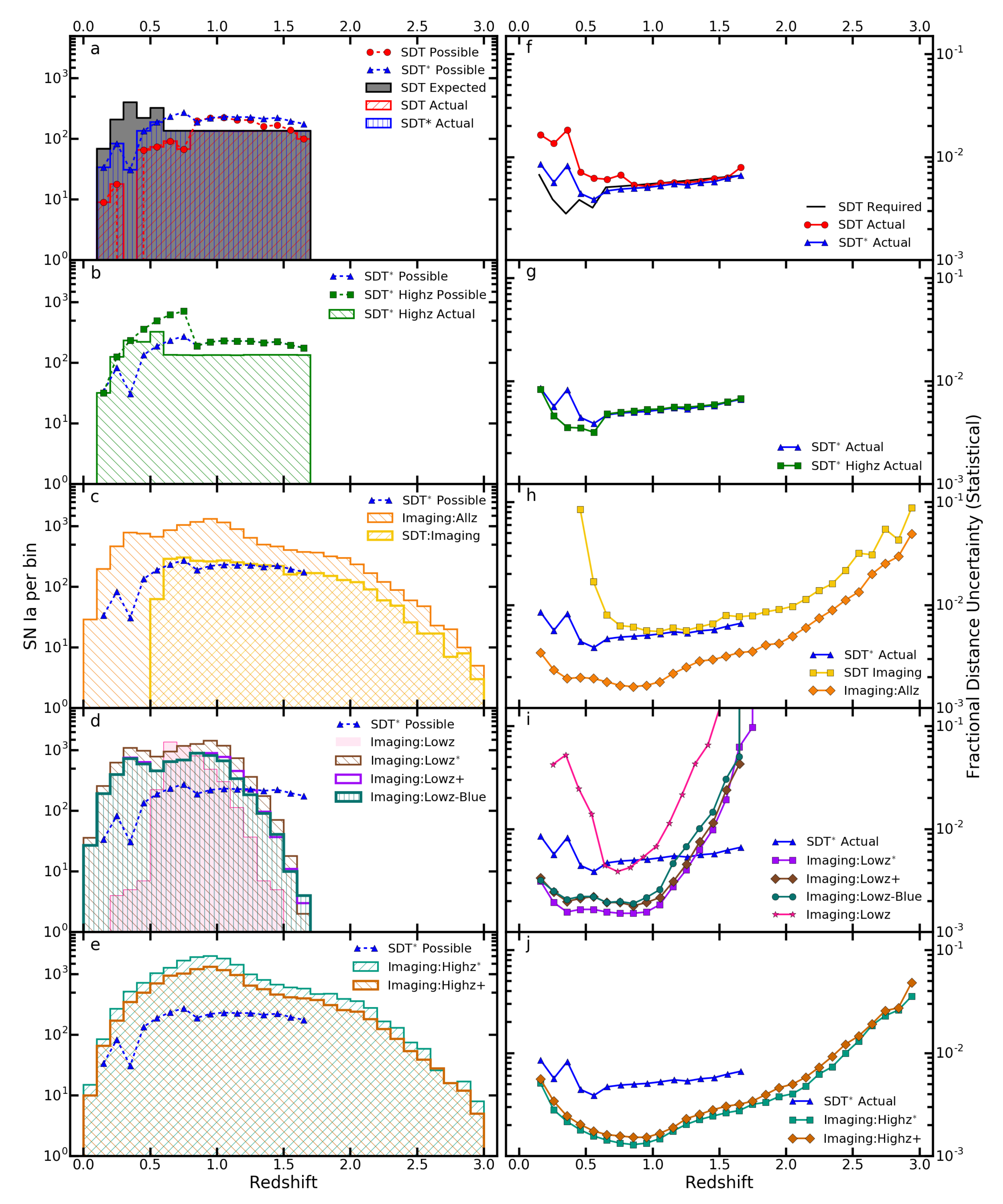}
\caption{{\it Left panels}: Redshift distribution for each simulated {\it WFIRST} SN survey examined. 
For comparison, the SDT-required distribution 
is presented as a grey histogram in panel (a)
(equivalent to Fig.~\ref{hist22}). 
Panel (a) also shows the ``possible'' SDT distribution
for discovered SNe~Ia that pass all SDT cuts (red circles).
The red histogram represents the ``actual'' SDT  distribution
as described in the text.
A similar curve and histogram for the SDT* strategy are shown as blue triangles and a blue histogram. 
The remaining left panels present the redshift distributions of the other strategies examined.   
{\it Right panels}: Fractional statistical {\red distance uncertainties}  for each simulated {\it WFIRST} SN survey as a function of redshift. 
The assumed SDT uncertainties are plotted as the thick black line in the top-right panel (see Figure~\ref{hist22}), with the measured uncertainties 
for the SDT (SDT*) strategies represented by red circles (blue triangles).  The remaining right panels present the fractional statistical {\red  distance uncertainties} of the other strategies, with the left and right panels of a given row corresponding to the same strategy.  
For comparison, the ``actual'' SDT* {\red  fractional distance  uncertainties} are presented as blue triangles in each panel.}
\label{allhistres}
\end{figure*}

The choice to only select SNe within their corresponding redshift tiers, however, does reduce the number of SNe available for follow-up observations. For the SDT* simulation, the final redshift distribution (Figure~\ref{allhistres}a) has only 475 SNe, compared to the 1230 $z < 0.6$  SNe~Ia expected in the SDT report (i.e., only 39\%; see Figure~\ref{allhistres}a). 
This deficit is caused primarily by the low SNR for objects in the shallow survey tier and the choice not to obtain  follow-up observations of low $z$ SNe from the medium and deep tiers.  

If SNe were to be selected from any of the three imaging tiers regardless of their of redshift, then the number of SNe per redshift range would likely increase (see results of SDT* Highz, Section~\ref{HIGHz-IFC+IMG}). The total number of SNe discovered, however, will be similar to that in the SDT report, and additional selection criteria would likely reduce the final number below that desired.

For the simulated SDT and SDT* samples, the statistical uncertainties on the fractional distance (Eq.~\ref{eq:fdu})
as a function of redshift are presented in Figure~\ref{allhistres}f. 

The significant disagreement between statistical uncertainties 
{\red  forecast in the SDT report (henceforth SDT-required)} and SDT* surveys at {\red $z < 0.6$} is due to the lack of low-$z$ SNe~Ia in the SDT* sample.

\subsection{SDT* Highz} 
\label{HIGHz-IFC+IMG}
As the shallow tier did not yield many SNe for the simulated SDT* survey (Section~\ref{SDTSN}), we examined the effects of removing that component and reallocating the time to the medium tier. This IFC+imaging based strategy therefore consists of only two tiers: medium and deep. This simulation allows the medium tier to sample SNe within a greater redshift range, $0.1 \leq z < 0.8$ (instead of the previously defined $0.4 \leq z < 0.8$ range). The deep tier is unchanged from its description in Section~\ref{SDTSN}. The area of the medium WFC imaging component is therefore increased by a factor of $\sim 2.3$ by using the survey time from the shallow tier. 
The numbers presented in Table~\ref{strat} are limited (where applicable) to the maximum number of SNe per 0.1 redshift bin as outlined in the SDT report. In addition, the modified selection criteria (Section~\ref{SDTSN}) are implemented.

The SDT versions of selection criteria were also applied to this simulation, but as in the previous SDT* scenario our modified selection yields a larger statistical sample. The results of this strategy can be seen in Figure~\ref{allhistres}b and \ref{allhistres}g. 
The redistribution of time to the medium tier 
results in 24\% more SNe~Ia within the final classified sample in comparison to the SDT* {\red actual} results.

\subsection{SDT Imaging} 
\label{SDT-IMG-ONLY} 
This simulation is based on a worst-case scenario where the SDT strategy is executed, but after the fact it is determined that the IFC-S data are unusable, resulting in exclusive use of the existing WFC imaging data. Presumably this analysis can happen even if the IFC-S works perfectly. There is no increase in the areas of this strategy as it is exploring the idea of data obtained when an instrument is ``faulty." There are also no SN selection criteria as outlined in Section~\ref{classification} as there are no spectra. Purity of the resulting SN~Ia sample is implemented via the aforementioned SNR requirements made on fitting (see Section~\ref{simulation}). The results of this simulated survey are presented in Figures~\ref{allhistres}c and \ref{allhistres}h. 

While number of SNe~Ia obtained within the simulation is a factor of $\sim$1.24 {\blue more than} the possible {\red SDT* strategy sample}; only $\sim$76\% of these have $0.1\leq z \leq1.7$, the rest are spread over higher $z$. In addition, only two SNe~Ia are detected at $z < 0.5$. This issue can again be attributed to the insufficient SNR of the low-$z$ SNe in the shallow tier of the survey.  

{\red As stated above, this SDT Imaging strategy is a worst-case scenario and unlikely to happen. The strategy does, however, indicate the usefulness of limited imaging-only data. We do not consider the case where discovery filters in the WFC were to fail, as the mission would no longer be self-reliant for SN discovery.}

\subsection{Imaging:Allz} 
\label{3TIER-IMG-4FILT} 
This simulated WFC imaging-only survey uses all three SDT tiers, but four broadband filters instead of two. 
The four filters used are $RZYJ$
for the shallow and medium tiers, and $YJHF$ for the deep tier. 
{\red To account for removing the IFC-S component and increasing the number of filters,
the areas have been adjusted by factors of $\sim$1.8, 2.2, and 1.8 for the shallow, medium, and deep tiers, respectively.}
These filters are chosen to span the rest-frame optical and NIR wavelength range, where our spectral models are defined.

The redistribution of IFC-S time, lack of IFC selection criteria, and additional filter selection, results in a factor of $\sim$4.7 increase in the final SN sample over the possible SDT*. See Figures~\ref{allhistres}c and \ref{allhistres}h for the results of this strategy. This is the first scenario for which the number of SNe~Ia in each 0.1 redshift bin has exceeded the requirements set by the  SDT.

\begin{figure}
\includegraphics[keepaspectratio=true, width=0.5\textwidth]{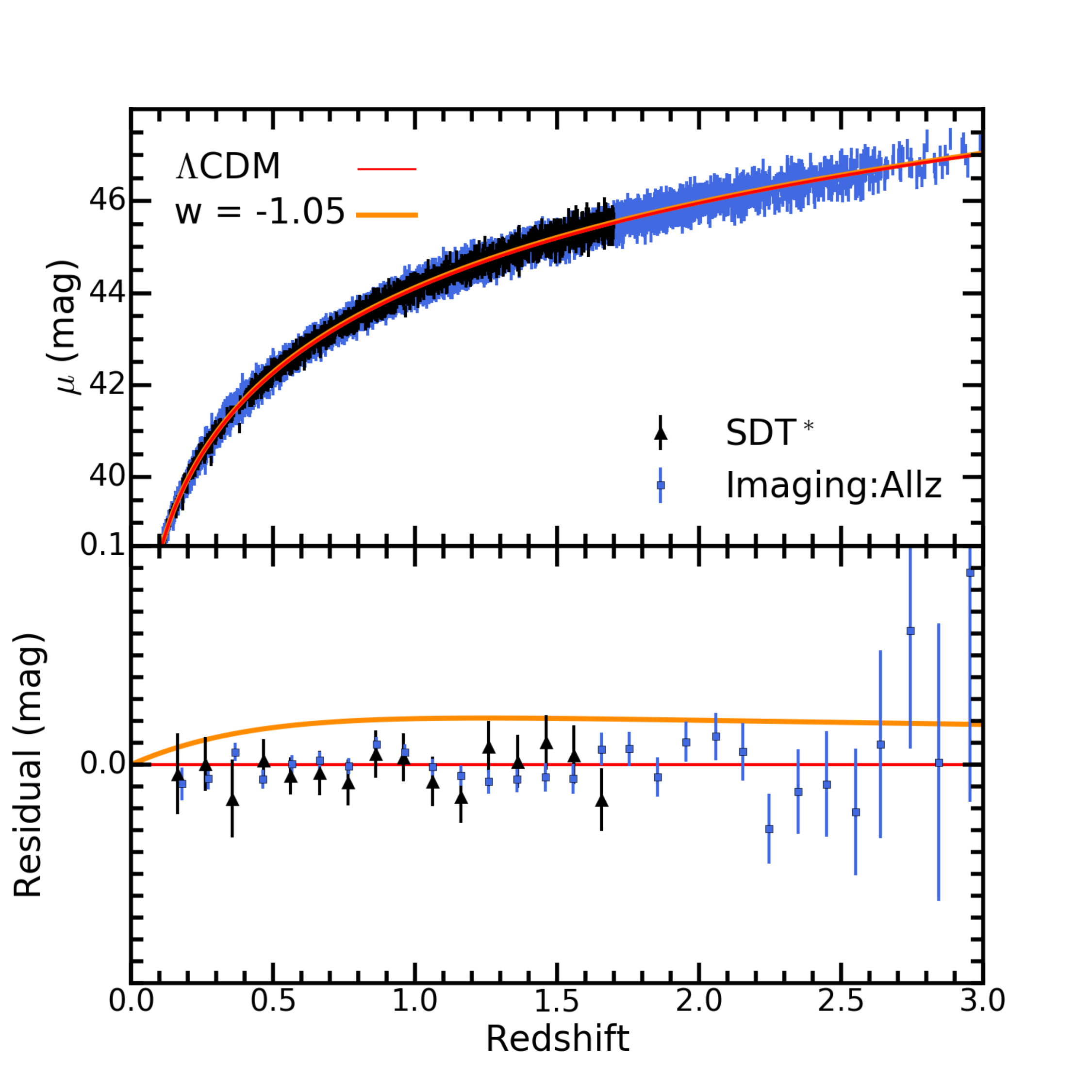}
\caption{{\red{\BBC}-fitted Hubble diagram of the simulated {\it WFIRST} SDT* sample (black points), and 
Imaging:Allz sample (blue points). 
The red lines represent the $\Lambda$CDM model used to generate the simulation. Gold lines show a $w$CDM model with $w = -1.05$. 
The bottom panel displays binned Hubble residuals, relative to the $\Lambda$CDM model, for SDT* (black triangles) and Imaging:Allz (blue squares).}}
\label{hubble}
\end{figure}

After selection requirements and the \BBC\ fit, we present a Hubble diagram for the Imaging:Allz and SDT* SN surveys in Figure~\ref{hubble}. 

The SDT* survey covers $0.1 \leq z \leq 1.7$, while the Imaging:Allz dataset covers $0.0 \leq z \leq 3.0$. Within the {\red $0.1 \leq z \leq 1.7$ range, the SDT* distance uncertainties are on average a factor of $\sim 2.4$ worse than those from the Imaging:Allz. 
By contrast, the uncertainties within the high-$z$ end ($z > 2.0 $) of the Imaging:Allz simulation are on average $\sim 3.2$ times worse than the average uncertainty within the  $0.1 \leq z \leq 1.7$ range for the SDT* survey. To illustrate the sensitivity of \WFIRST, the gold lines show a $w$CDM model with $w=-1.05$.} 

\subsection{Imaging:Lowz} 
\label{LOWz-IMG} 
We simulated a WFC imaging-only survey that consists of the shallow and medium tiers only, and the use of the two SDT discovery filters ($Y$+$J$ and $J$+$H$). The SDT time from the deep imaging and IFC-S are used to increase the area of the shallow tier by a factor of $\sim 5.2$ and the medium tier by $\sim 7.5$. 

This scenario results in a factor of  $\sim 1.6$ increase in the number of SNe~Ia compared to the possible sample in the SDT* scenario (see Figures~\ref{allhistres}d and \ref{allhistres}i). 
There are no SNe~Ia at $z < 0.2$, and only $\sim 5\%$ of the total sample have $z < 0.6$. The fraction of SNe~Ia with $z \geq 1.2$ is only $\sim 1\%$.

\subsection{Imaging:Lowz*} 
\label{LOWz-IMG-4FILT} 
Here is another imaging-only simulation that consists of the shallow and medium tiers, where the area of each respective tier has been increased by a factor of $\sim 2.7$ and $\sim 3.6$ to account for not using the IFC-S and deep-tier components. $RZYJ$ filters are used in both tiers, maximizing our coverage of the rest-frame optical and extending to the rest-frame NIR.
Comparing to the Imaging:Lowz strategy, we have included two additional filters, $R$+$Z$, and thus decreased the observed areas. This Imaging:Lowz* strategy
results in a factor of $\sim 3.9$ increase in the number of SNe~Ia in the final sample. Of the total sample, only $\sim 6\%$ have $z \geq 1.2$ (see Figures~\ref{allhistres}d and \ref{allhistres}i).

\subsection{Imaging:Lowz+}
\label{LOWz-IMG-6FILT}
Six filters, $RZYJHF$, are used in this imaging-only strategy with two tiers (each tier uses six filters). The areas of the shallow and medium tiers are increased by factors of $\sim 1.8$ and 2.3, respectively, with respect to SDT*. This 6-filter strategy leads to a factor of $\sim 2.5$ increase in the number of SNe~Ia compared to SDT*. Only $\sim 5\%$ of the sample has $z \geq 1.2$ (see Figures~\ref{allhistres}d and \ref{allhistres}i).

\subsection{Imaging:Lowz-Blue} 
\label{LOWz-IMG-BLUE} 
The number of tiers and their area on-sky within this strategy are the same as for Imaging:Lowz+; however,
bluer filters have been selected, which include the WFC3 $F425W$ ($B$), $F555W$ ($V$), and $F814W$ ($I$) filters in combination with the {\it WFIRST} SDT discovery filters, $Y$+$J$. The number of SNe~Ia found by this strategy is $\sim$2.3 times greater than in the SDT* survey. Only $\sim$5\% of the SN sample has $z \geq 1.2$  (see Figures~\ref{allhistres}d and \ref{allhistres}i).

\subsection{Imaging:Highz*} 
\label{HIGHz-IMG-4FILT} 
This strategy is similar to the Imaging:Lowz* strategy, but here the medium and deep tiers are used rather than the shallow and medium tiers. Time from the IFC-S and shallow components is used to increase tier areas by factors of $\sim 3.6$ and 2.6, respectively. Filters selected for the medium tier are $RZYJ$, and $YJHF$ for the deep. The number of SNe~Ia found by this strategy is $\sim 6.5$ times greater than in the SDT*.
This two-tier survey results in a more complete sample of SNe~Ia across the required redshift range. This is the second scenario for which the number of SNe~Ia per 0.1 redshift bin exceeds the requirements set by the SDT report 
(see Figures~\ref{allhistres}e and \ref{allhistres}j)

\subsection{Imaging:Highz+} 
\label{HIGHz-IMG-6FILT}
This strategy is similar to the Imaging:Lowz+ strategy, but here the medium and deep tiers are used rather than the shallow and medium tiers, and 6 bands are used: $RZYJHF$ for medium, and $RZYJHF$ for deep.
The areas of the two tiers have been increased by factors of $\sim 2.3$ and 1.8, respectively, accounting for not using the IFC-S and shallow tier. The number of SNe~Ia found by this strategy is a factor of $\sim 4.4$ times greater than in the {\red possible SDT* survey sample}. There is a reduction of SNe~Ia found here in comparison to the Imaging:Highz* strategy due to the addition of two filters ($HF$ in  medium and $RZ$ in deep), and thus comparative reduction in area size. See Figures~\ref{allhistres}e and \ref{allhistres}j for the results of this strategy.

\subsection{Summary of Simulated Surveys}
\label{sum1}
Redshift distributions of SNe~Ia and their associated {\red fractional distance uncertainties (Eq.~\ref{eq:fdu})} are shown in Figure~\ref{allhistres}. For the Imaging:Highz and Imaging:Allz strategies, the number of SN~Ia detected per 0.1 redshift bin
is significantly greater compared to the SDT* results. However, for each low-$z$ strategy
the number of SNe~Ia at $z \geq 1.3$ drops dramatically owing to the loss of the long-exposure deep-tier component with redder bands. 

The SDT Imaging and Imaging:Lowz strategies 
indicate how ineffective the shallow tier is. 
The dearth of SNe~Ia at $z < 0.5$ is driven by a combination of the shallow exposure time resulting in objects with a low SNR, and poor filter selection which provides minimal rest-frame coverage. {\blue The slew-and-settle time has a drastic impact on the shallow imaging tier, explaining the much lower utility we find for this tier relative to the SDT report.}

In many of the imaging-only scenarios the redshift range extends beyond
$z > 2$, but note that the SN~Ia rate 
has a larger uncertainty for $z > 2$. The increased number of SNe~Ia for particular redshift ranges leads to an increase in the statistical precision per redshift bin, as much as a factor of $\sim 2.4$ better than  SDT* for the Imaging:Highz* strategy.

\section{Systematic Uncertainties}
\label{sysun}
In addition to the statistical uncertainties included in our analysis, 
several sources of systematic uncertainty have been investigated. These investigations are the first attempt to quantify the systematic uncertainties of the {\it WFIRST} SN survey without the use of ad hoc functions such as Equation~\ref{sys}.

We compute a covariance matrix to describe the {\red distance modulus uncertainties} such that {\bf $\Ctot = D_{\rm stat} + C_{\rm sys}$} \citep{Conley11}.

D$_{\rm stat}$ is the purely diagonal matrix, where the diagonal elements correspond to the individual SN {\red distance modulus uncertainties} given by Equation~\ref{stat}. The systematic component, $C_{\rm sys}$, can be described as the summation over each systematic uncertainty such that
\begin{equation}
  C_{{\rm sys}, ij} = \sum_{k} \left ( \frac{\delta \mu_{i}}{\delta S_{k}} \right ) \left ( \frac{\delta \mu_{j}}{\delta S_{k}} \right ) \left ( \Delta S_{k} \right )^{2}, \label{syseq}
\end{equation}
{\red where $\delta \mu_{i} / \delta S_{k}$ expresses the dependence of distance modulus on the $k^{\rm th}$ systematic uncertainty for the $i^{\rm th}$ SN, and the magnitude of the $k^{\rm th}$ systematic uncertainty is given as $\Delta S_{k}$}.  
To calculate C$_{\rm sys}$, 
we change each systematic effect by $1\sigma$ and
determine the distance modulus difference from the {\BBC} fit.
  
During this process, we fix $\alpha$ and $\beta$ from Equation~\ref{eqn:Tripp} to the values found 
from the \BBC\ fit using
only the statistical uncertainties.

To characterize the dependence of the FoM on the value of each systematic uncertainty, we introduce a bias in our measurements that mimics the effect of each systematic uncertainty. We vary each systematic effect with multiplicative scaling from 0 (no effect) to 12 times the value of our current constraints for that uncertainty.  For each case, we compare the 
{\BBC}-fitted distance moduli determined with the included uncertainty to that determined without the effect. We display the absolute median distance modulus bias as a function of redshift for the nominal case (multiplicative factor of 1) in Figures~\ref{syscalphys} and \ref{zptall}. The $\mu$-differences are used to compute the derivative term in Equation~\ref{syseq}. Note that although absolute values are presented in Figures~\ref{syscalphys} and ~\ref{zptall}, the signs of the differences are used in the computation of the derivative. 
{\red In addition, the values presented in these figures are defined to have zero residual at $z=0$.}

To determine a FoM, we input the derived distances and the associated covariance matrix to {\CosmoMCfast}.  Additional constraints from CMB and BAO measurements are included in the fitting (as discussed in Section~\ref{simulation}). 
{\red For a subset of systematics, Figures~\ref{syscalphys}-\ref{zptall} show the $\mu$-shift versus redshift (left panels) and the relative FoM change with respect to the statistical-only FoM, $\FoMstat$ (right panels).}
  
The points marked as ``current'' represent the FoM calculated with our present understanding of the systematic uncertainty (i.e., a multiplicative scaling of 1), $\FoMcurr$. Points marked ``optimistic'' represent the FoM values calculated with assumptions for improved systematic uncertainties, $\FoMopt$. These optimistic systematic uncertainties are values which we hope will be achievable
by launch, and are based on reasonable assumptions.

The limited precision of {\CosmoMCfast} 
(Section~\ref{simulation}) and the artifacts of light-curve fitting 
add small numerical noise to 
FoM measurements, making their values deviate, on order of a few percent, from a smooth interpolation. However, all of our main findings are robust against these small variations.

\subsection{Calibration} 
{\it Calibration uncertainty} is currently the largest systematic uncertainty of all recent ground-based SN cosmology analyses \citep[e.g.,][]{Scolnic14:ps1}. The primary sources of calibration uncertainty can be split into three separate components, which are listed and discussed below. The nominal size of each component is set to match the current values determined for the {\it HST} system. This is likely a conservative assumption, and is varied within the present analysis.\\

\begin{figure*}
\centering
\includegraphics[keepaspectratio=true, width=\textwidth]{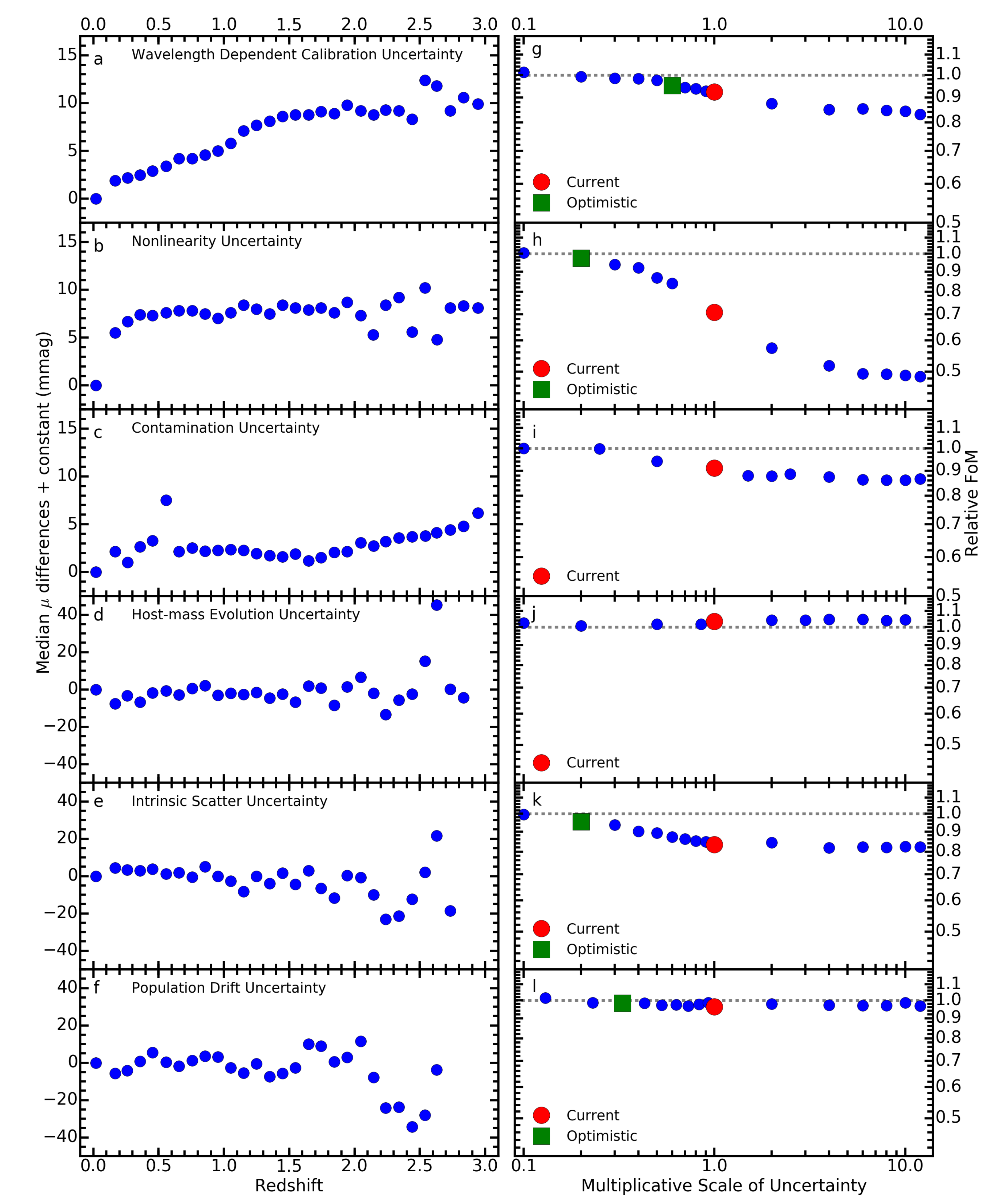}
\caption{{\it Left panels}: {\red For the Imaging:Allz strategy, median distance modulus difference vs. redshift for a current systematic effect labeled on each panel. Note that the values presented in these figures are defined to have zero residual at $z=0$.
{\it Right panels}: FoM/FoM$_{\rm stat}$} for different values of  each systematic uncertainty (with the scaling relative to the current value). 
The dashed line represents the statistical FoM. Red circles and green squares represent the current and optimistic values of each uncertainty, respectively.}
\label{syscalphys}
\end{figure*}

{\bf 1.} The {\it wavelength-dependent flux uncertainty:} The accuracy of the {\it HST} Calspec system \citep{Bohlin14} is described as a linear function with a slope of roughly 5~mmag per 7000~\AA\ \citep{Bohlin07}. Assuming the functional form of the calibration of {\it WFIRST} is similar to that of {\it HST}, we use the magnitude of the {\it HST} systematic uncertainty as the nominal uncertainty for the WFC (see Figure~\ref{syscalphys}a and g). 
{\red For the IFC-S we assume this uncertainty
to be 50~mmag per 7000~\AA, as there are many unknown calibration issues for this instrument. This higher uncertainty for the IFC-S is based on
work conducted in Section 4.6 of \citet{Bacon15}, which compares synthesized broadband magnitudes from the Multi Unit Spectroscopic Explorer (MUSE -- a panoramic integral-field spectrograph) to that of {\it HST}, and finds a mean bias of 50~mmag, with a statistical uncertainty of 40~mmag. In addition, similar values were also found by \citet{Childress16} in which data from a SN survey using the Wide Field Spectrograph on the Australian National University Telescope, enabled the determination of a color variation ranging from 40~mmag in the red to 90~mmag in the blue.}

For both the IFC-S and WFC we assume an optimistic {\red {\it wavelength-dependent flux uncertainty}} of 3~mmag per 7000~\AA, a factor of $\sim$1.7 better than {\it HST}. This is the main calibration systematic uncertainty for the IFC-S, and also a significant uncertainty for the WFC filters.\\

\begin{figure*}
\centering
\includegraphics[keepaspectratio=true, width=\textwidth]{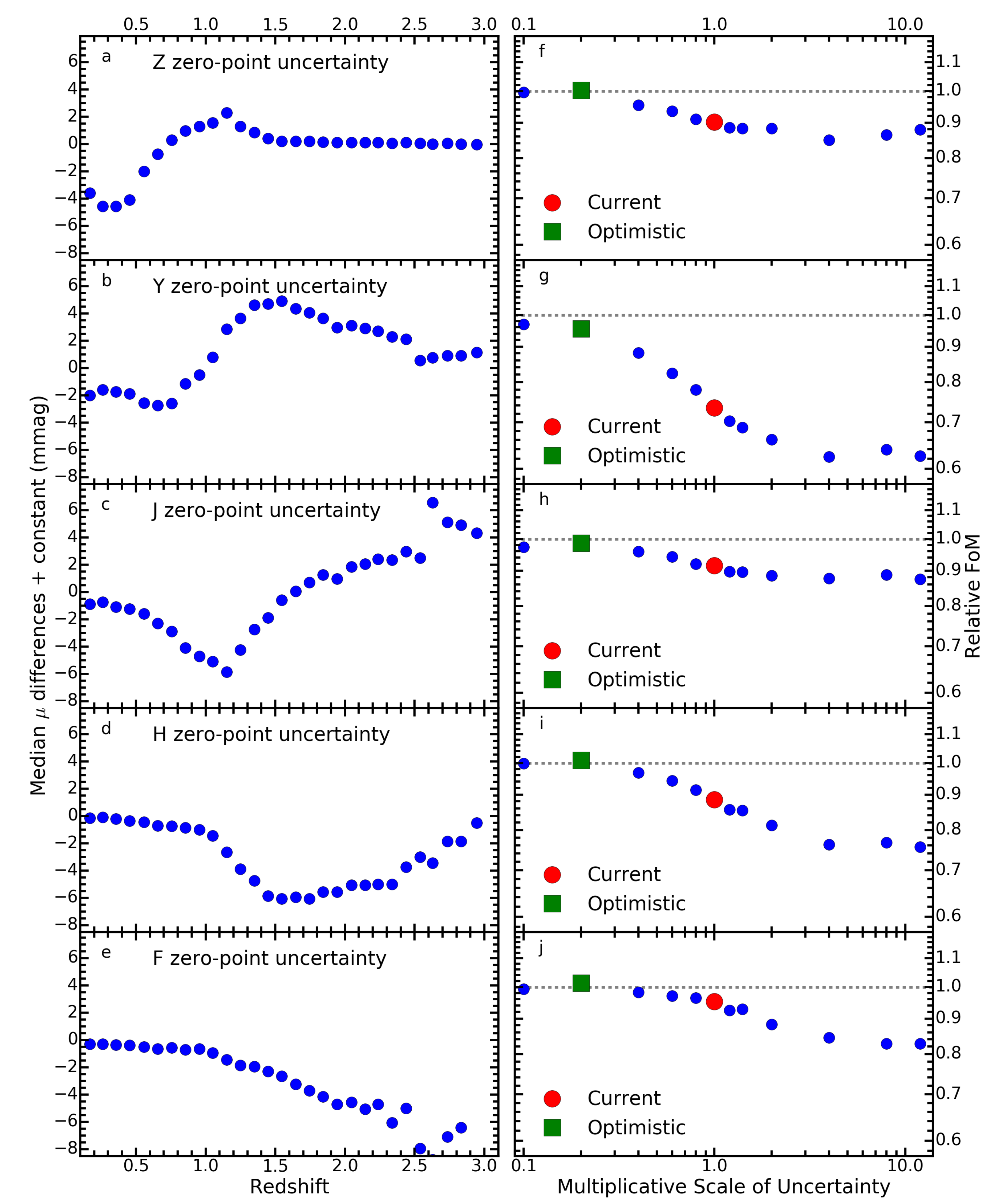}
\caption{Same as Figure~\ref{syscalphys}, but for additional systematic uncertainties.}
\label{zptall}
\end{figure*}

{\bf 2.} {\it Nonlinearity of the detector:} Detector response nonlinearity can impact photometric precision in astronomical observations. Recent work \citep[e.g.,][]{Riess10} suggests
that a count-rate dependent nonlinearity is common in the HgCdTe detectors that will be used by {\it WFIRSTs} WFI. 
\citet{Riess10} measured a WFC3-IR non-linearity of $\sim 1\%$ per dex over a range of 10~mag (4 dex), which was independent of wavelength. We take this  value as our baseline systematic\footnote{Note added in proof. For the IFC-S, we assumed an optimistic nonlinearity for both the current and optimistic case; we realized this only after the analysis, and thus all quoted IFC-S FoM values presented within the paper use an optimistic nonlinearity value. If a nominal nonlinearity is applied, the effect is to reduce the current FoM listed in Table 13 by a maximum of 13\%.}
(see Figures~\ref{syscalphys}b and h).

Our optimistic {\it nonlinearity systematic uncertainty} is assumed to be a factor of 5 better than the values obtained from {\it HST} studies. This is reasonable given current and future improvements of detectors.\\

{\bf 3.} {\it Zero-point uncertainty:} Recent ground-based imaging surveys have achieved 5~mmag
uncertainties in their filter zero-points 
\citep{Betoule13, Scolnic15}. Because of the color term in the SN distance equation, the $\mu$-bias can be $\beta \simeq 3$ times larger than the zero-point bias.
 
A space-based observatory, being above the atmosphere, should have zero-point uncertainties no larger than those from
state-of-the-art ground-based surveys. We have therefore  included 
this uncertainty with
a 5~mmag shift to each filter {\it zero-point} (see Figure~\ref{zptall}). Optimistic WFC imaging zero-point uncertainties are assumed to be 1~mmag.\\

{\red As stated in Section~\ref{simulation}, we include a sample of 800 low-$z$ SNe simulated based on predictions of the Foundation survey from \citet{Foley17}. These low-$z$ SNe 
provide an anchor to the SN~Ia Hubble diagram.

Here we consider two main sources of calibration uncertainties that impact the low-$z$ sample.
First is how well the photometric system of the low-$z$ sample is calibrated and tied to {\it HST} Calspec standards. Since the Foundation survey uses the Pan-STARRS photometric system \citep{Chambers17}, 
we adopt  $g'$, $r'$, and $i'$-band zero-point uncertainties of 5 mmag \citep{Scolnic15}, and assume that these uncertainties will improve with a dedicated standard-star observing program to more optimistic values of  3~mmag. The second source
of calibration uncertainty is that of the {\it HST} Calspec standards themselves (as discussed earlier), which affects both the low-$z$ and \WFIRST\ samples.

Given that we aim to obtain a relative calibration accuracy of 1-2 mmag between ground-based SN surveys (e.g., LSST/PS1/DES) and {\it WFIRST}, it might be appropriate to drop the low-$z$ (shallow) tier of the {\it WFIRST} SN survey. This shallow tier 
has proved to be inefficient in obtaining SNe with the SNR required for detection and accurate classification within the {\it WFIRST} cosmology sample. The redistribution of the shallow-tier time into that of the medium or deep tiers serves to benefit the overall efficiency of the survey (see, for example, the SDT* Highz or Imaging:Highz* surveys).

If, however, the relative calibration between ground and space-based data is suboptimal, then a low-$z$ {\it WFIRST} SN survey may be required. Such a survey would require a different strategy than those proposed within this paper, and will be the focus of future investigations.}\\

{\red Within this section we have addressed some of the larger systematic uncertainties affecting the calibration of the {\it WFIRST} instruments. We have not, however, considered calibration uncertainties at the pixel level.
For these systematics the implicit assumption is that they average out due to the distribution of observations of SNe over the detector, resulting in a random measurement uncertainty, and thus their effect is considered minimal compared to the more dominant uncertainties listed within this section. 
In addition, we ignore sub-pixel sensitivity variations and persistence uncertainties, which may introduce redshift-dependent effects owing to the reduced resolution of host galaxies at high redshift.}

\subsection{Core-Collapse SN Contamination} 
\label{cont}
For surveys that use the IFC-S we assume that there are no CC SNe in the cosmological sample; i.e., we set the current and optimistic {\it contamination} to be $0\%$. Although some CC SN {\it contamination} was present within the SDT/SDT* results (see Section~\ref{SDTSN}), we expect that by using all spectra and light-curve data
(not just the first 5 imaging data points and 3 IFC-S spectra), {\it contamination} will drop significantly. 
In addition, we expect that
improved classification techniques in the future will aid in CC SN contamination reduction.

For each imaging-only scenario (Sections~\ref{SDT-IMG-ONLY} through \ref{HIGHz-IMG-6FILT}), however, CC SN contamination 
must be considered. 
Our analysis results in a photometric classification purity of $>$99\% (for the Imaging:Allz survey). To account for the systematic uncertainty introduced by any remaining 
{\it contamination}, {\BBC}-fitted Hubble residuals are obtained
with and without CC SNe, and differences in distance vs. redshift are used as a systematic uncertainty. 
{\red We further reduce the {\red{\it contamination} by a} factor of 5 based on the assumed improvements from photometric classifiers using machine learning (see Figures~\ref{syscalphys}c and i). This assumption is based on a factor of 3.6 contamination reduction for DES simulations \citep{Kessler16} using a simple nearest-neighbor (NN) method on only three variables (redshift, color, stretch). Using a wider range of light-curve features, better machine-learning methods have been demonstrated, such as by \citet{Lochner16}. The addition of NIR data in {\WFIRST} can also be used to reduce contamination.
Training a photometric classifier is beyond the scope of this work,} but will be considered in future investigations.

Our optimistic {\it core-collapse SN contamination uncertainty} for imaging-only simulations is assumed to be negligible, as we expect classification methods to have improved substantially by launch and to be able to take advantage of the rest-frame NIR data.\\

\subsection{SN Physics} 
\label{phys}

Our analysis includes
five systematic uncertainties related to SN physics.\\

{\bf 1.} {\it The host-galaxy -- SN luminosity relation:} After correcting for SN light-curve shape and color, SN~Ia Hubble residuals still correlate with host-galaxy properties \citep[e.g.,][]{Kelly10, Lampeitl10:host, Sullivan10}. Although the cause of this effect is still unknown, it is possible that it is related to different progenitor properties, such as metallicity or age, that correlate with environment.

Currently, cosmology analyses \citep[e.g.,][]{Betoule14} correct SN luminosities based on the mass of the SN host-galaxy relative to a central split value. The exact functional form of this correction is still poorly constrained, but most assume a binary population split at 10$^{10}$~M$_{\odot}$ \citep{Sullivan10}. It is possible that the magnitude of this correction and the form could change with redshift \citep[e.g.,][]{Rigault13, Rigault15, Childress14}. However, the size of the systematic uncertainty due to the mass-dependent evolution can be mitigated by measuring the relation between distance residuals and mass at different redshifts. This method is similar to ideas presented by \citet{Shafer14}. Therefore, the size of the systematic uncertainty is actually dependent on how well the evolution of the relation is measured. 

To study
this effect, we label half of our
SN~Ia sample to be ``high-mass''
and introduce a redshift-dependent offset in the fitted $\mB$,
\begin{equation}
m_{B, \rm shift} = m_{\rm B} + 0.06 - \left [ 0.06 \times (1 - z) \right ].
\label{shift_mB}
\end{equation}
We then determine the redshift dependence of the difference in Hubble residuals between our altered ``high-mass" sample and the unaltered ``low-mass" sample. The difference in our recovered dependence and our input dependence, given in Equation~\ref{shift_mB}, is used as the size of our nominal systematic uncertainty (see Figures~\ref{syscalphys}d and j). For this analysis, we assume that the uncertainty in the difference of the Hubble residuals for the two host-mass bins is dominated by the distance uncertainty, rather than by uncertainties in the mass estimates of the host galaxies. There may be a larger systematic uncertainty related to a population drift of the host-galaxies; however, we choose this particular kind of systematic for the {\it host-galaxy-SN luminosity relation} because it can be reduced with increasing statistics. As such, there is no optimistic value for this bias because it is based purely upon the statistics of the survey.

{\red We note that within Figure~\ref{syscalphys}j the FoM appears to increase with multiplicative scale factor. Further analysis of the host-mass systematic indicates that within in the 0.1--10 range of the fiducial value, $w_{a}$ appears to have shifted by approximate 1/2 sigma. Because $w_{a}$ correlates with other cosmological parameters that are fixed by the CMB, this bias spuriously increase the FoM. For a more detailed discussion of similar issues see \citet{Scolnic14:ps1}.}\\

{\bf 2.} {\it Intrinsic scatter uncertainty:} There is still uncertainty in the relative proportion of color variation and luminosity variation in the intrinsic scatter model for SNe~Ia \citep[see][for a review]{Scolnic16}. The distance bias corrections applied depend on the assumption of the intrinsic scatter model. The differences between the bias corrections are typically largest where selection effects are strongest because the intrinsic scatter model will determine whether predominantly bluer objects are selected or predominantly brighter objects are selected. To determine the impact on our cosmological measurements from this uncertainty, we first simulated our samples with two different intrinsic scatter models: G10, a model from \citet{Guy10} which has 70\% luminosity variation and 30\% color variation, and a model from \citet[][hereafter C11]{Chotard11}, which has 25\% luminosity variation and 75\% color variation. Following \citet{Kessler13}, we converted the G10 and C11 models into spectral-variation models for {\SNANA}. The difference between the recovered distances from these two models is the systematic uncertainty, shown in Figure~\ref{syscalphys}e. The structure of the distance differences with redshift shown in Figure~\ref{syscalphys}e is due to the impact of various selection effects (from the tiered surveys) on the different scatter models.

The optimistic {\it intrinsic scatter uncertainty} is assumed to be a factor of 5 better than current estimates due to improved models in the IR \citep[e.g.,][]{Mandel11}, {\red and improved spectra to empirically model intrinsic SED variations}.\\

{\bf 3.} {\it Population drift:} Related to uncertainty in the intrinsic scatter model, there is uncertainty in whether this form of the scatter could evolve with redshift. This issue is conflated in past analyses with the possibility that the color of the SN population could evolve with redshift \citep{Mandel16, Scolnic16}, and this evolution is not accounted for in the analysis. To determine the impact on our cosmological measurements from this uncertainty, we introduced a SN color population drift of $0.01\times z$~mag, keeping the defined color range and Bifurcated Gaussian $\sigma$ identical to previous simulations. {\red This means that the center of the color ($c$) population for a sample of SN~Ia, at fixed absolute magnitude, $M$, increases as a function of redshift by 0.01 $\times$ z.}

While there is evidence for an $x_{1}$ population drift 
\citep[see][]{Scolnic16,Scolnic17}, 
it will have less impact on possible distance biases than a $c$ population drift because of the different correlations between $c$ and $x_{1}$ with luminosity. Therefore in this analysis, we do not include an uncertainty from
$x_{1}$ population drift. The difference between the recovered distances from the color shift and the nominal simulation are shown in  Figure~\ref{syscalphys}f, with relative FoM values given in Figure~\ref{syscalphys}l.  

It is possible that with the IFC-S, evolution of the intrinsic color can be constrained by measuring the SN ejecta velocities \citep{Foley11:vel, Foley11:vgrad, Foley12:vel, Mandel14}. This claim is analyzed further in Appendix~\ref{ap2}, though for our nominal systematic uncertainty, we do not assume any improvement in the constraint on intrinsic color evolution or population drift from the IFC-S.

The optimistic {\red{\it color-vs.-redshift drift is taken to be $0.0033\times z$.}}

This is an estimate, as it is unclear from recent studies \citep[e.g.,][]{Rubin15, Scolnic16, Rubin16} what an optimistic constraint should be.\\

\noindent The following systematic uncertainties are not included in our FoM$_{\rm tot}$ predictions, but their effects have been considered.\\

{\bf 4.} {\it Beta evolution:} The properties of interstellar dust may change with redshift, affecting the ratio of total to selective extinction. This evolution would manifest itself in a change in the recovered value of $\beta$ \citep{Scolnic14:col} with redshift \citep{Conley11}. Furthermore, as shown in \citet{Mandel16}, the color law (CL in Equation~\ref{eq:salt2flux})
may be composed of a reddening law as well as a separate relation between SN intrinsic color with luminosity, and the relative components of the two  may change with redshift. Similar to the correlation of Hubble residuals with host mass, $\beta$ evolution can be included with additional {\BBC} fit parameters and their uncertainties will decrease with increasing sample size.

Therefore, the {\it $\beta$-evolution uncertainty} is expected to be small compared to the systematic uncertainty from the {\it intrinsic uncertainty} or {\it population drift}, so it is not included here.\\

{\bf 5.} {\it K-corrections:} The SDT report lists {\it K-corrections} as a top systematic uncertainty and a large motivation for the use of the IFC-S over broadband imaging. However, since modern distance-fitting algorithms employ spectral models to fit SN data in the observer frame, no true {\it K-correction} is ever applied. Instead, a {\it K-correction} uncertainty should be described as an imperfect knowledge of a SN SED. Since most of our SN~Ia training set is at $z \approx 0$, certain regions of the spectral model (near the effective wavelength of certain filters) are better constrained than others. If the deredshifted observer-frame and rest-frame filters are not well aligned, the diversity of spectral features could cause an additional statistical uncertainty of up to 0.04~mag \citep[e.g.,][]{Saunders15}. With IFC-S measurements, one can synthesize photometry over any wavelength range, largely eliminating this uncertainty.

{\red It has been argued \citep[e.g.,][]{Aldering02, Perlmutter03} that this uncertainty will be dominant for a space-based SN mission. However, even with the most pessimistic scenario \citep{Saunders15}, the additional $0.04$ mag of scatter is still negligible compared to the 0.13~mag intrinsic scatter of SNe~Ia \citep{Scolnic14:ps1, Betoule14}. \citet{Saunders15} also predict systematic biases on the $0.01$ mag level which follow oscillatory behavior with a period of $\Delta z \approx 0.1$.  For a redshift range of a SN sample with $\Delta z \ge 1.0$, the impact of this $0.01$ mag bias should be reduced to the sub-mmag level.  

To further examine the impact on the measurement uncertainties using the IFC-S, the IFC-S spectra were rebinned, maintaining the overall SNR, mimicking progressively lower resolution spectra (or wider filters). }
Note that intrinsic spectral variations are included in the simulated spectra, but the analysis does not account for these variations except for a global intrinsic scatter term in the distance modulus uncertainty.
No systematic bias is found and {\red distance modulus uncertainties} do not increase, confirming that this uncertainty will be subdominant. Therefore, {\it K-corrections} are not included as an additional systematic uncertainty.\\

{\bf 6.} {\it Milky Way extinction:} Systematic uncertainties in the amount of Milky Way (MW) extinction along the line-of-sight to the SNe will propagate to systematic uncertainties in the recovery of the cosmological parameters. The {\it WFIRST} SN fields have not yet been chosen, but it is likely they will be picked to minimize the amount of MW extinction: MW $E(B-V) < 0.02$~mag.  As discussed by \citet{Scolnic14:col}, systematic uncertainties in the MW extinction take the form of a multiplicative component and additive component.  Assuming a $10\%$ multiplicative uncertainty and a separate 3~mmag additive uncertainty, we find that the impact on the FoM is small ($<$ 10\%), relative to the other systematic uncertainties discussed above. Therefore, it is not included in our analysis.\\

{\red The list of systematic uncertainties presented within this section, although not complete, covers some of the most dominant uncertainties found today in ground-based (and space-based) SN surveys. It is important to note, however, that some of the uncertainty descriptions
may be oversimplifications of a more complex problem. Future studies into these systematic uncertainties as well as additional biases could include the consideration of population drift outside of the SALT2 framework; the redshift-dependent efficiency and bias
of imaging-only surveys; selection biases in the IFC-focused surveys, including the effect of SN classifications with photo-redshifts instead of spectroscopically determined redshifts; and host-subtraction uncertainty, i.e., random noise or redshift dependency. See Section~\ref{diss} for a more detailed discussion of these biases.}

\subsection{Statistical and Systematic Uncertainty Summary} 
\label{syssum}

A summary
of the current and optimistic systematic uncertainties investigated by our various simulations is presented in Table~\ref{optsystab}.

\begin{table*}
\centering
\caption{Current and optimistic systematic uncertainties investigated for both the WFC and IFC-S.
}
\scriptsize\addtolength{\tabcolsep}{+2pt}
\begin{tabular}{lllll}
\hline\hline
\centering  
Systematic & \multicolumn{2}{c}{Current} &  \multicolumn{2}{c}{Optimistic }\\ 
Uncertainty & WFC & IFC-S & WFC & IFC-S \\ \hline
\rule{0pt}{3ex}Wavelength dependent calibration & 5 mmag per 7000~\AA\ & 50 mmag per 7000~\AA\  & 3 mmag per 7000~\AA\ &  3 mmag per 7000~\AA\\ 
Nonlinearity & 1\% per dex over 10 mag & 0.2\% per dex over 10 mag & 0.2\% per dex over 10 mag &  0.2\% per dex over 10 mag \\
Zero-point offsets & 5 mmag &  \nodata  & 1 mmag & \nodata \\
CC contamination & 1/5th of derived systematic$^{e}$ &  0\%  &  0\% & 0\%\\
Population drift & $10~{\rm mmag} \times z$ & $10~{\rm mmag} \times z$ & $3.3~{\rm mmag} \times z$ & $3.3~{\rm mmag} \times z$\\
Intrinsic scatter & \multicolumn{2}{c}{The difference between the G10 and C11 models} & 1/5th that of current & 1/5th that of current\\
Host-mass evolution & \multicolumn{4}{c}{Calculated for each strategy$^{f}$}\\ \hline
\rule{0pt}{3ex}Beta evolution &  \multicolumn{4}{c}{Considered Negligible$^{g}$}\\
K-corrections & \multicolumn{4}{c}{Considered Negligible$^{g}$}\\
MW extinction & \multicolumn{4}{c}{Considered Negligible$^{g}$}\\
\hline                                                   
\end{tabular}
\\
$^{e}$ See Section~\ref{cont} for details on this systematic uncertainty.\\
$^{f}$ For each simulated survey strategy the host-mass systematic uncertainty was calculated as described within Section~\ref{phys}.\\
$^{g}$ As the effect of this systematic uncertainty is considered negligible (see Section~\ref{phys}) we have not included it within our final analysis.
\label{optsystab}
\end{table*}

\begin{figure*}
\centering
\includegraphics[keepaspectratio=true, scale=0.24]{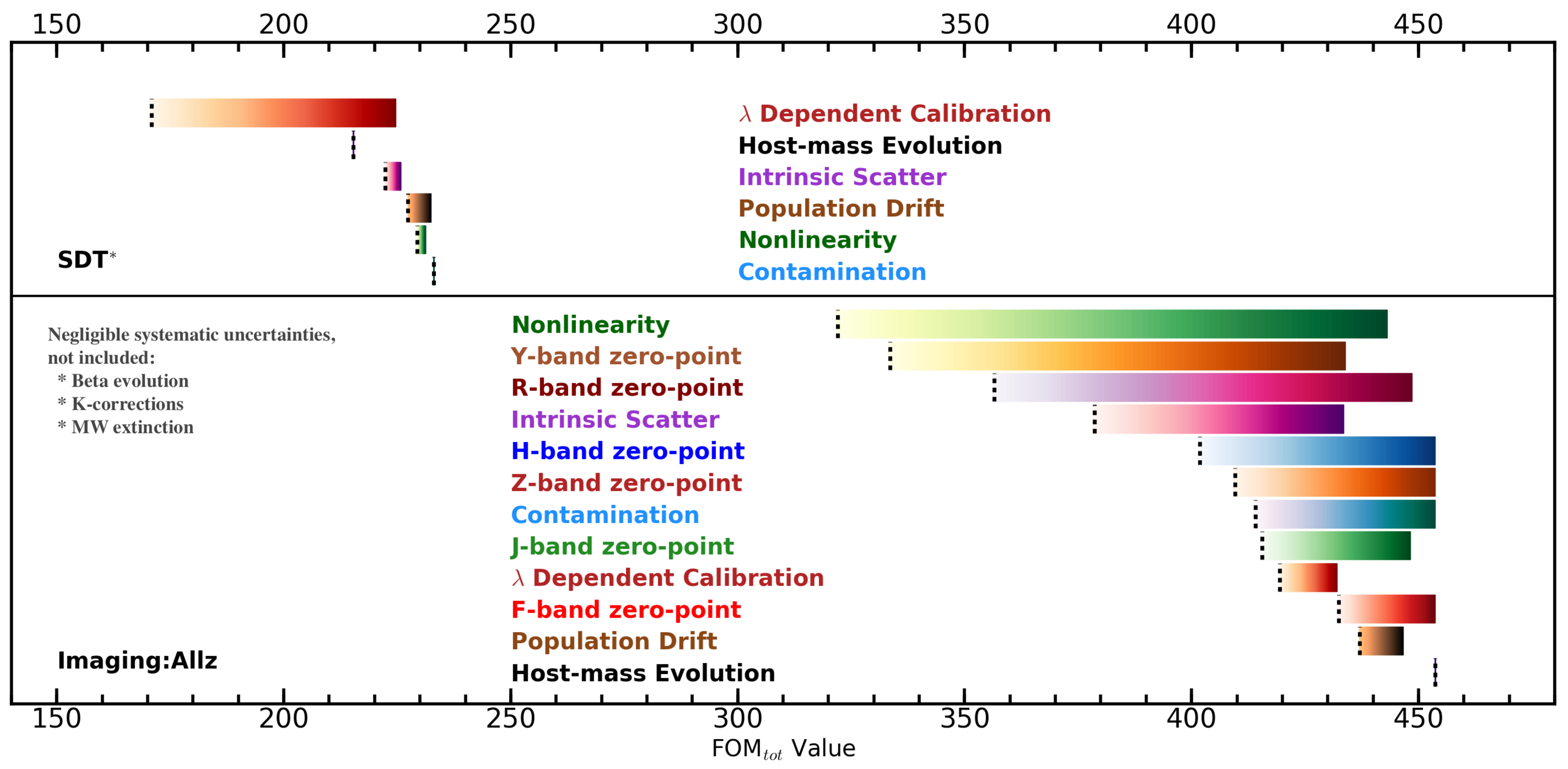}
\caption{$\FoMtot$ when 
only a single systematic uncertainty is included, as indicated in the figure.
The range for each $\FoMtot$ spans current (dotted line) to optimistic.
The top and bottom panels include
systematic uncertainties 
for the SDT$^{*}$ (see Section~\ref{SDTSN}) and Imaging:Allz strategies (see Section~\ref{3TIER-IMG-4FILT}), respectively.  In cases where noise fluctuations makes $\FoMtot$ 
slightly greater than $\FoMstat$, the FoM is set as $\FoMstat$.
Note that the SDT$^{*}$ results do not include the effect of {\it zero-point} uncertainties as this strategy does not use any imaging to measure distances.  
Negligible uncertainties 
(such as {\it beta evolution}, {\it K-corrections}, {\it MW extinction}) are not included.}
\label{diffsys}
\end{figure*}

The effect of each individual systematic uncertainty, both current and optimistic, is presented within Figure~\ref{diffsys}. The values plotted here are produced using {\CosmoMCfast}. For the SDT$^{*}$ survey simulation, the largest uncertainty is the {\it wavelength dependent calibration uncertainty}. Our current estimate for this uncertainty is 50~mmag per 7000~\AA. We hope that this value will decrease by over a factor of 10 by launch, leading to the much larger relative FoM$_{{\rm tot}, {\rm opt}}$. For the imaging-only scenario the largest systematic uncertainties are the {\red{\it nonlinearity}, {\it intrinsic scatter}, and {\it zero-point offsets}, specifically for the $Y$ and $R$ bands}. Further evaluation of these uncertainties is required in order to fully understand their effects and enable optimization of survey strategies. 

\section{Comparison of Simulated Survey Strategies}
\label{compss}

Within this paper we have simulated a total of 
\NSTRATEGY\  different SN survey strategies for the {\it WFIRST} mission. Here we compare each strategy, assessing how successful they are at constraining dark energy models, via their FoM values. We also examine the details of these strategies, such as redshift distribution of SNe~Ia, and suggest how they may be improved. 

\begin{figure*}
\centering
\includegraphics[keepaspectratio=true, width=\textwidth]{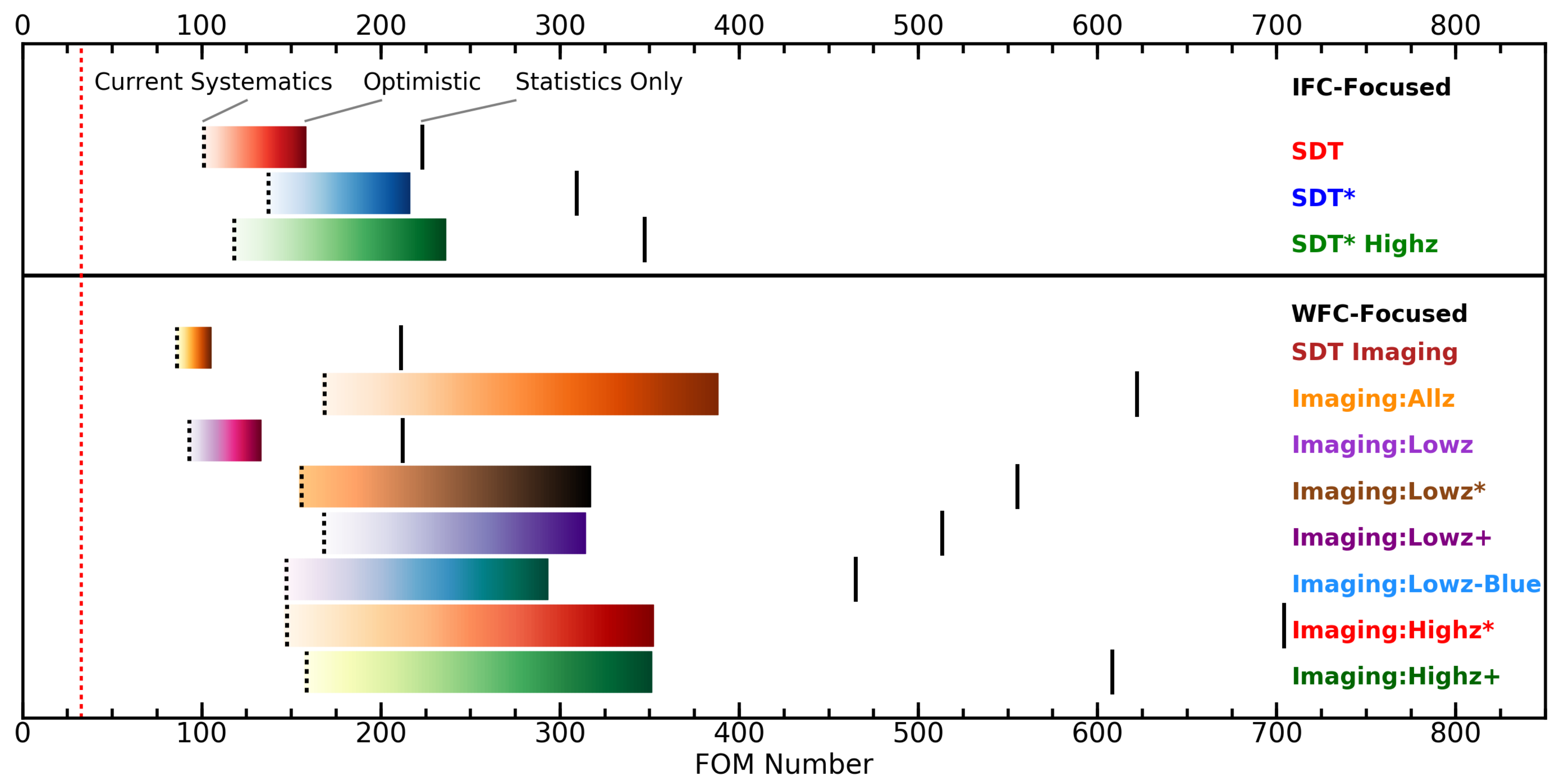}
\caption{Predicted dark energy FoMs for the simulated {\it WFIRST} SN survey strategies outlined in Section~\ref{results}. IFC-focused and WFC-focused strategies are presented in the top and bottom panels, respectively.  The gradients for each strategy represent the range of FoMs from $\FoMcurr$ (dotted lines) to $\FoMopt$. The thick black lines represent $\FoMstat$. 
The red dashed vertical line indicates the current FoM value of 32.6 \citep{Alam16}.}
\label{allsys}
\end{figure*}

Using the {\blue``optimistic"} systematic uncertainties described above we have evaluated the impact of each uncertainty on each strategy, 
the results of which are presented in Figure~\ref{allsys}, with Table~\ref{fomvalss} listing the 
$\FoMstat$, 
$\FoMcurr$, and 
$\FoMopt$
values determined for each case. For completeness the FoM values presented here are calculated using the original version of \CosmoMC, where the full set of Planck likelihoods is used.
 
Figure~\ref{allsys} shows that these strategies result in a wide range of
{\red $\FoMstat$ (211 -- 704)} values compared to a much narrower range of 
{\red $\FoMcurr$ (86 -- 169) values}. 

\begin{deluxetable*}{ l  c  c  c  c  c  c  c  c  c }
\tablecaption{{\red FoM values and cosmological parameter uncertainties for each strategy.}}
\label{fomvalss}
\tablewidth{0pt}
\tablehead{
\multirow{3}{*}{Strategy\tablenotemark{a}} & \multicolumn{3}{c}{Statistical} & \multicolumn{3}{c}{Current} & \multicolumn{3}{c}{Optimistic}\\
\colhead{} & \colhead{FoM} & \colhead{$\sigma(w_{0})$} & \colhead{$\sigma(w_{a})$} & \colhead{FoM} & \colhead{$\sigma(w_{0})$} & \colhead{$\sigma(w_{a})$} & \colhead{FoM} & \colhead{$\sigma(w_{0})$} & \colhead{$\sigma(w_{a})$}
}
\startdata
SDT				&	223	&	0.053	&	0.26	&	101	&	0.067	&	0.29	&	158	&	0.061	&	0.28\\
SDT* 			&	309	&	0.042	&	0.20	&	137	&	0.058	&	0.26	&	216	&	0.050	&	0.22\\
SDT* Highz		&	347	&	0.037	&	0.18	&	118	&	0.060	&	0.28	&	236	&	0.048	&	0.20\\
SDT Imaging		&	211	&	0.059	&	0.29	&	86	&	0.071	&	0.27	&	105	&	0.068	&	0.25\\
\bf{Imaging:Allz}	&	622	&	0.024	&	0.12	&	169	&	0.061	&	0.25	&	388	&	0.035	&	0.16\\
Imaging:Lowz		&	212	&	0.057	&	0.29	&	93	&	0.078	&	0.33	&	133	&	0.067	&	0.31\\
Imaging:Lowz*		&	555	&	0.027	&	0.13	&	156	&	0.066	&	0.28	&	317	&	0.042	&	0.19\\
Imaging:Lowz+		&	513	&	0.029	&	0.14	&	168	&	0.063	&	0.27	&	314	&	0.043	&	0.19\\
Imaging:Lowz-Blue	&	465	&	0.031	&	0.15	&	147	&	0.067	&	0.27	&	293	&	0.044	&	0.19\\
\bf{Imaging:Highz*}	&	704	&	0.022	&	0.11	&	148	&	0.069	&	0.29	&	352	&	0.039	&	0.17\\
\bf{Imaging:Highz+}	&	608	&	0.024	&	0.12	&	159	&	0.063	&	0.27	&	351	&	0.035	&	0.16\\
\enddata                                                    
\tablenotetext{a}{The ordering of this table follows the ordering in
Section~\ref{results}. Strategy names in bold have the highest $\FoMopt$ values.}  
\end{deluxetable*}

Examination of strategies that use both the IFC-S and WFC (e.g., SDT, SDT* and SDT* Highz) 
lead to several important conclusions. The SDT strategy as outlined by \citet{Spergel15} results in a lower than expected number of SNe~Ia at $z < 0.6$. This decrease in low-$z$ SNe is a result of the short exposure time within the shallow tier of the imaging survey, and the strict SDT spectrophotometric selection criteria. 

Slight modification of these selection criteria, 
as implemented in SDT* (see Figure~\ref{allhistres}a), increases the total number of low-$z$ ($z < 0.6$) SN~Ia by $\sim 300\%$, with an increase of $\sim 31\%$ for $0.1 \leq z \leq 1.7$. The short exposure time of the shallow imaging tier leads to low-SNR SNe, and thus even with modified selection criteria this shallow survey still hinders the number of  $z < 0.6$ SNe~Ia obtained, reaching only $\sim 39\%$ of the fraction stated within the SDT report. 

Based on these results we conclude that any selection criteria implemented must be very carefully chosen to maximize both efficiency and purity, and that the shallow tier of the imaging survey (with current overhead estimates) is a suboptimal use of survey time. Shifting exposure time from the shallow imaging tier to the medium tier, and applying the modified selection criteria, significantly increases the SN~Ia sample size
(460 and 923 more SNe~Ia compared to SDT*, and SDT surveys, respectively) as indicated in our SDT* Highz survey simulation (see Figure~\ref{allhistres}b). 
This SDT* Highz strategy results in a much higher 
$\FoMopt$ value of 236 in comparison to the 
$\FoMopt = 158$ value of the SDT.

\begin{figure*}
\centering
\includegraphics[width=3in]{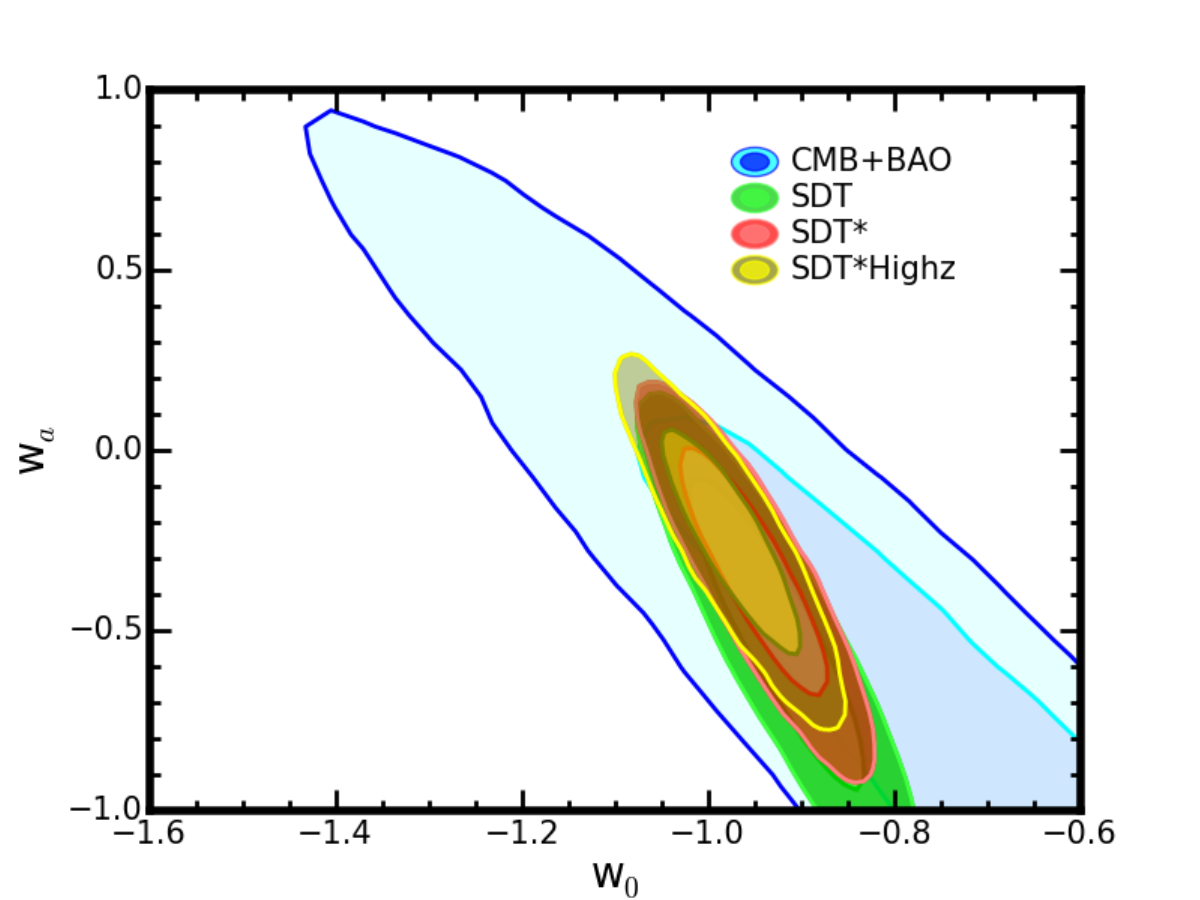}
\includegraphics[width=3in]{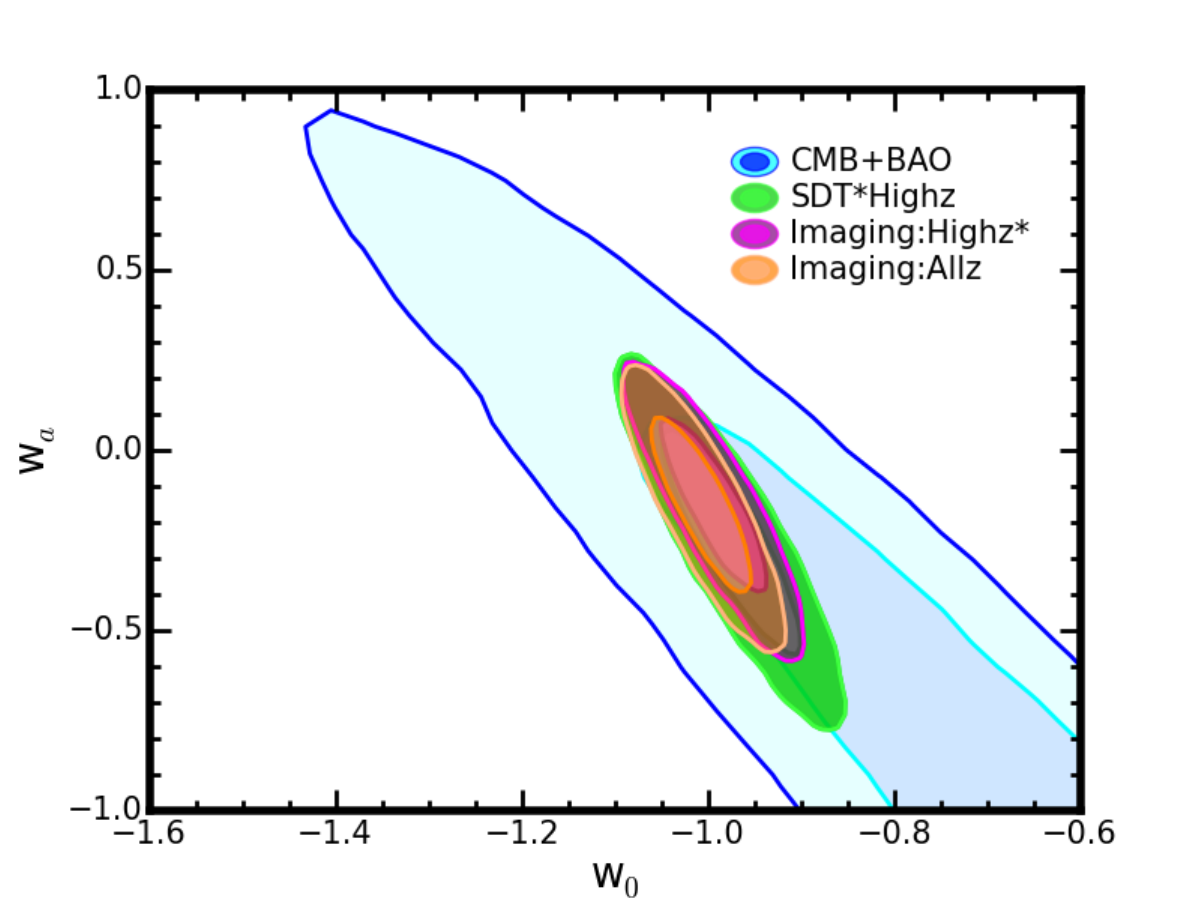}
\caption{The $w_{0}$--$w_{a}$ 68\% and 95\% confidence contours for the simulated SDT, SDT*, and SDT* Highz {\it WFIRST} SN surveys (left panel) and the SDT, Imaging:Allz, and Imaging:Highz* {\it WFIRST} SN surveys (right panel). Each contour represents total (statistical plus optimistic systematic uncertainties) SN~Ia constraints combined with CMB and BAO constraints. For comparison we have included the confidence contours created using CMB+BAO data only.}
\label{pl12}
\end{figure*}

Figure~\ref{pl12} (left) presents the $w_{0}$--$w_{a}$ 68$\%$ and 95$\%$ confidence contours for the simulated SDT, SDT*, and SDT* Highz surveys. These contours illustrate how slight modification of the classification criteria presented by the SDT report (see Section~\ref{classification}) can lead to an increase in $\FoMopt$, whereas moving time to focus on the medium tier of the survey results in a more significant impact.  

As an informative worst-case scenario, the SDT Imaging simulation replicates a situation where all IFC-S data are determined to be unusable, but only after completion of the {\it WFIRST} mission. As a result, only the imaging data as part of the SDT survey would be used for cosmological analyses. Unsurprisingly, this SN survey produces too few SNe~Ia at $z < 0.6$ and delivers a low $\FoMopt = 105$.

The Imaging:Allz simulation has $\FoMopt = 388$, and is our most successful imaging-only strategy. It is a 3-tier imaging strategy that uses four broadband filters ($RZYJ$ or $YJHF$). This strategy results in
$>$5 times as many SNe~Ia as any IFC-S strategy and has $\FoMstat = 622$. {\it Zero-point uncertainties} are {\red some of the} largest systematic uncertainties for this strategy (see Figure~\ref{diffsys}), as with all imaging strategies. 

\begin{figure*}
\centering
\includegraphics[keepaspectratio=true, width=0.9\textwidth]{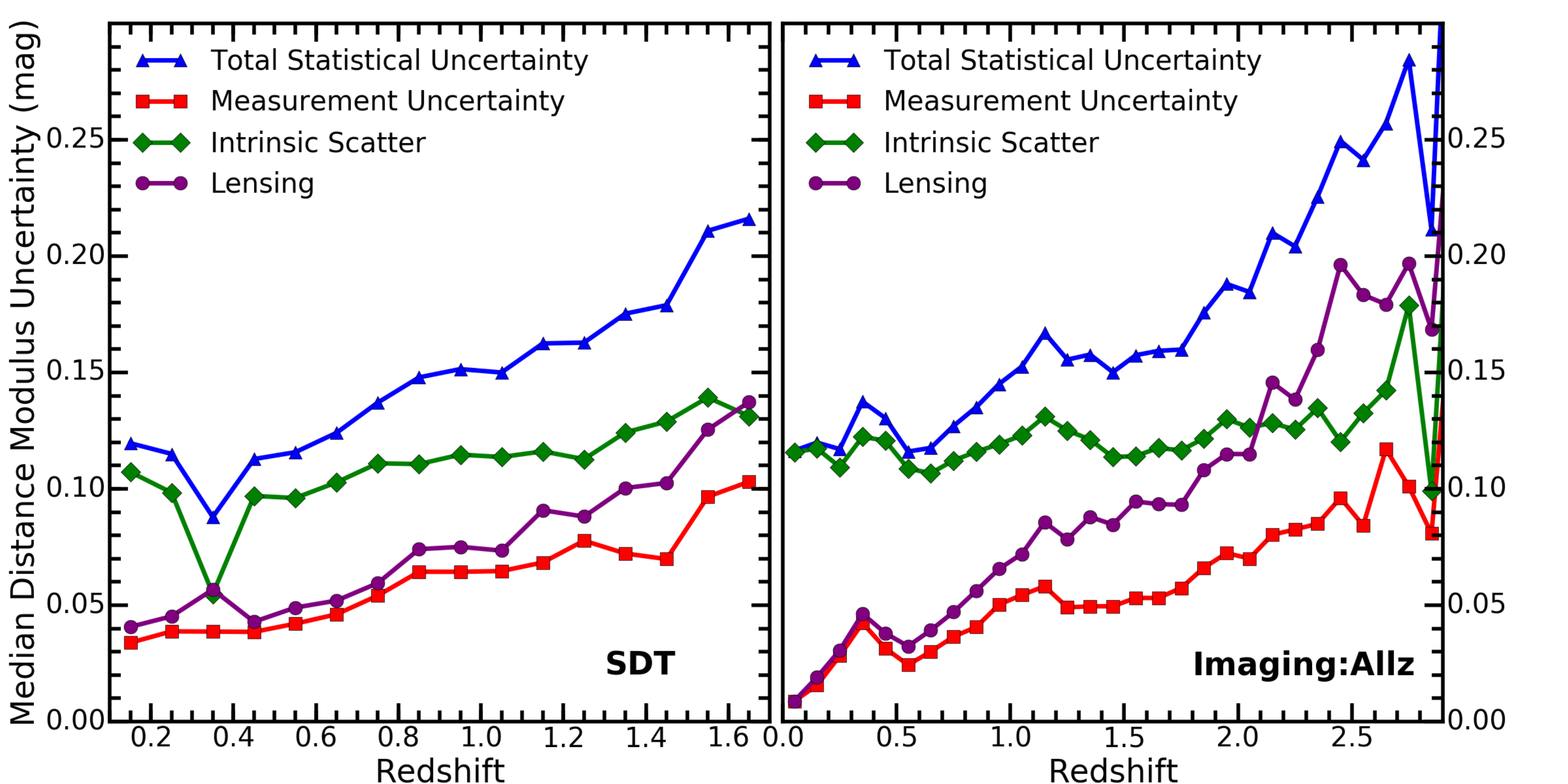}
\caption{The contribution to the median distance modulus uncertainty per {\red SN} within a 0.1 redshift bin, from the {\blue measurement uncertainty} (red squares), intrinsic scatter (green diamonds), and lensing uncertainty (purple circles) components for both the IFC-focused SDT (left) and Imaging:Allz (right) strategies.
The blue triangles show the combined {\blue statistical} uncertainties added in quadrature.}
\label{scat}
\end{figure*}

\begin{figure*}
\centering
\includegraphics[keepaspectratio=true, width=0.8\textwidth]{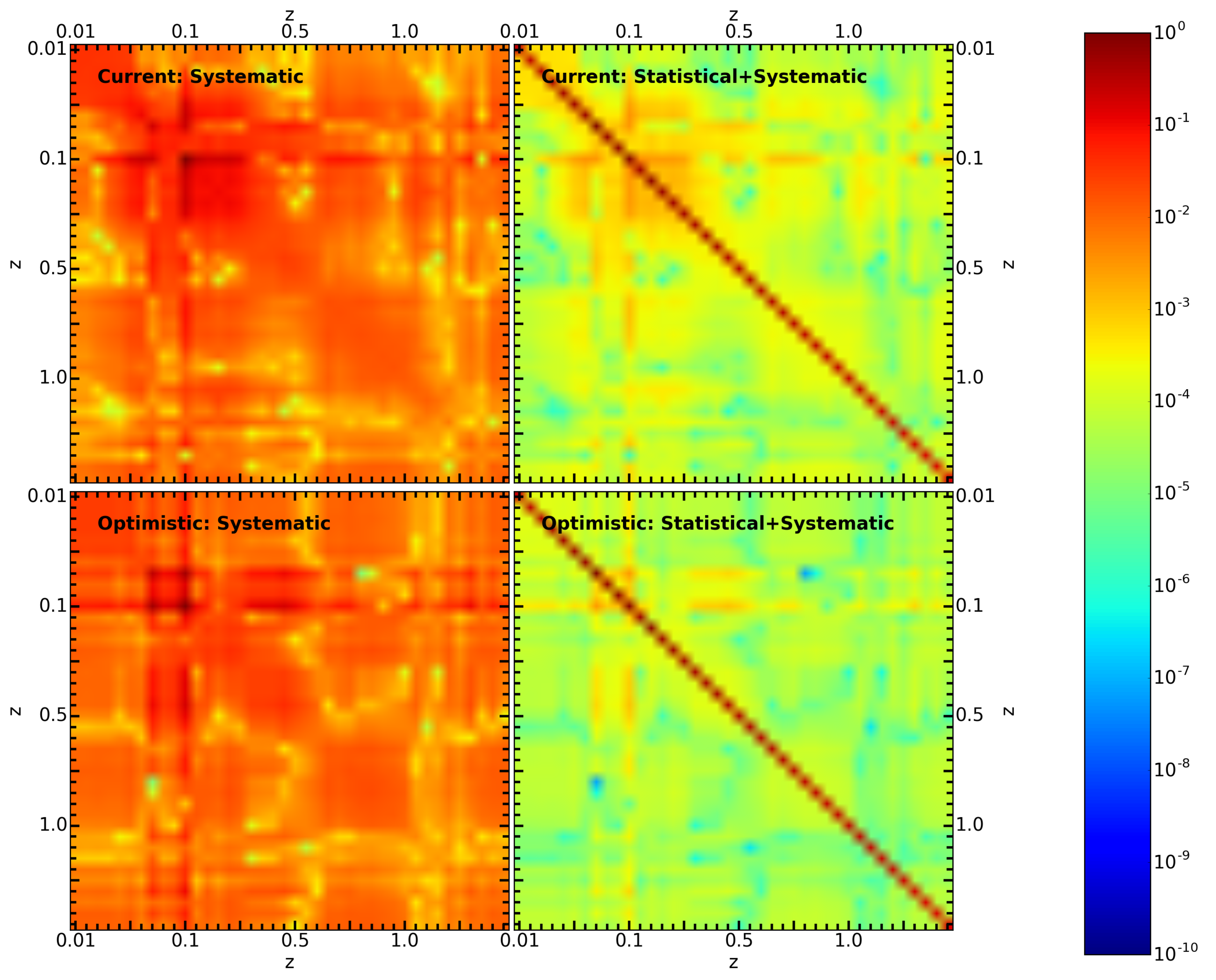}
 \caption{Redshift correlation matrices for the Imaging:Allz strategy. Within a given matrix the data have been normalized to the maximum array value (absolute values are displayed). A log color bar is used to show the relative distribution. Left panels show the correlation matrix for current (top), and optimistic (bottom) systematic uncertainties. Right panels show the full correlation matrix which includes statistical uncertainties.{\blue Note that although the off-diagonal elements from the systematic uncertainties appear to have a smaller magnitude than the diagonal elements from the statistical uncertainties, the cumulative impact of the off-diagonal elements can be greater on the FoM than the impact of the diagonal elements on the FoM.}}
\label{covmat}
\end{figure*}

{\red For both the SDT and Imaging:Allz simulations, Figure~\ref{scat} shows the contribution of {\blue measurement} (red line), lensing (purple line), and intrinsic scatter (green line) uncertainties on the total {\blue statistical} uncertainty. Within the SDT simulation the dominant uncertainty is that of intrinsic scatter, with the lensing uncertainty becoming comparable at $z \approx 1.7$. At higher redshift (i.e., $z\geq 2$) the Imaging:Allz simulation shows that lensing is the dominant source of uncertainty.

Although measurement uncertainties for high-redshift SNe are significant, Figure~\ref{scat} clearly illustrates that these uncertainties are not dominant.

Figure~\ref{covmat} shows the redshift correlation matrices for the Imaging:Allz strategy.
The statistical uncertainty dominates 
across all redshifts with some systematic contribution
around $0.1 < z < 0.3$. While the statistical uncertainty per redshift bin is larger than the systematic uncertainty, the impact of systematic uncertainties depends on the covariance between redshift bins. The effects of these systematic uncertainties are significantly reduced when considering their optimistic values.}

\begin{figure}
\centering
\includegraphics[keepaspectratio=true, width=0.5\textwidth]{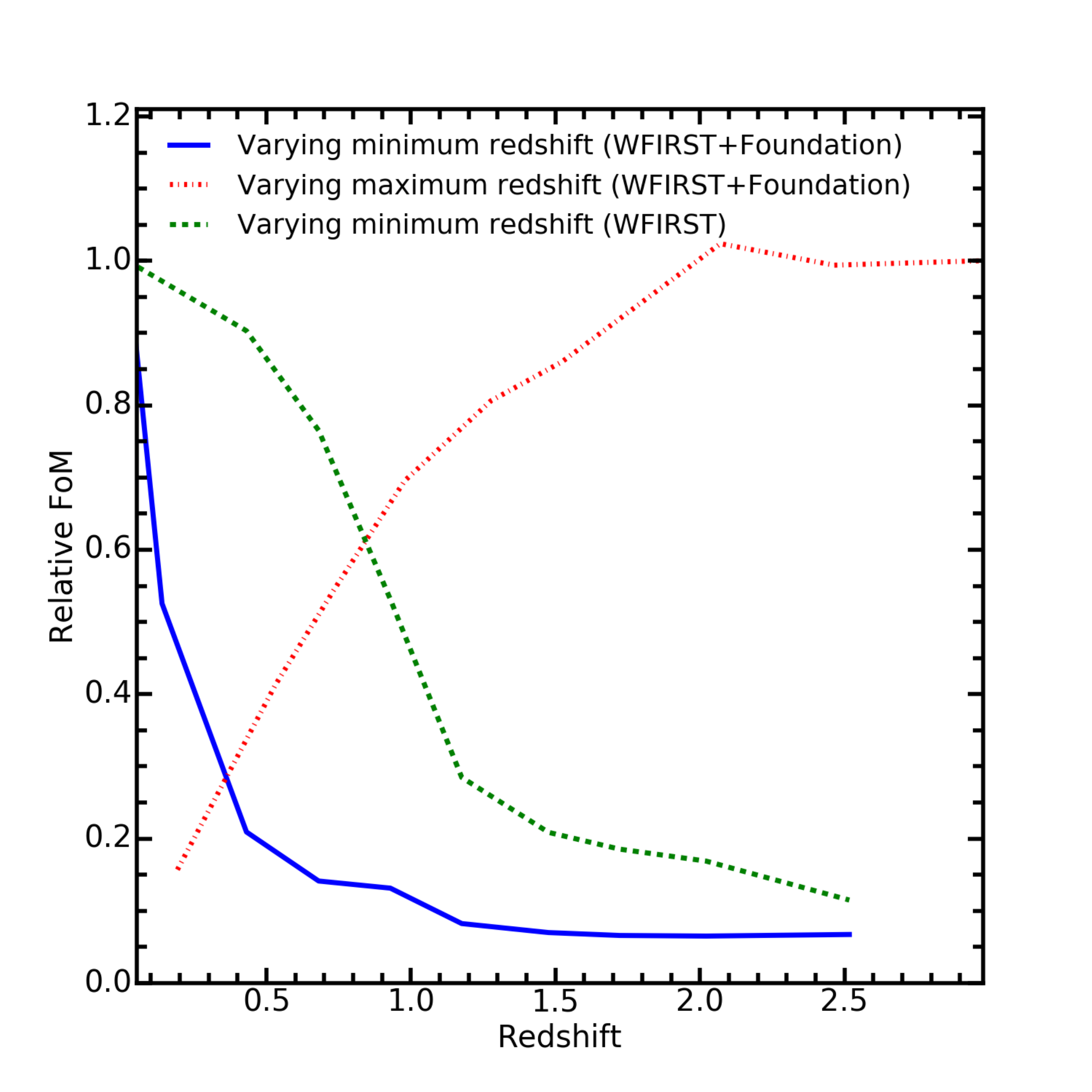}
\caption{This plot illustrates the effect on $\FoMstat$ if the redshift range of the Imaging:Allz survey is modified. 
The red dot-dash line shows the ratio with a high-$z$ cut and minimum redshift $z \simeq 0.02$. The blue line shows the ratio with a low-$z$ cut and maximum redshift $z\simeq 3$. The green dashed line is similar to the blue line except that the low-$z$ cut applies only to the {\it WFIRST} component of the survey such that the ``Foundation-like'' ($0.01 < z < 0.1$) data always remain.}
\label{bin_fom}
\end{figure}

{\red To demonstrate the dependence of $\FoMstat$ on the redshift range of the Imaging:Allz sample, Figure~\ref{bin_fom} shows
how the FoM changes when we vary the minimum and maximum redshift of the sample.
We find that at a maximum redshift of $z=1.5$ (red line), 
$\FoMstat$ is $\sim 86\%$ of the full-sample $\FoMstat$, 
and that $\FoMstat$ no longer increases with additional SNe past $z \approx 2.0$. The dependence on the minimum redshift (blue line) shows the importance of a low-$z$ sample. 
For instance, a cut at $z=0.4$ reduces $\FoMstat$ by $\sim 79\%$.  In addition, we note the importance of the ground-based ``Foundation-like'' data (i.e., $z\leq 0.1$); their removal from the survey reduces  $\FoMstat$ by $47\%$.

When comparing the Imaging:Allz and SDT* Highz $\FoMstat$ values, it is interesting to note that even though the final cosmology sample of the Imaging:Allz survey contains a factor of 6 more SNe~Ia at $z \geq 1.2$, the addition of these SNe increases the relative $\FoMstat$ value by only $\sim$79\%.
The reason for this relatively small improvement in $\FoMstat$ is that the majority of additional SNe in the Imaging:Allz strategy are at $z\geq 1.5$, 
and as shown in Figure~\ref{bin_fom}, these high-$z$ SNe contribute little to $\FoMstat$. This effect is likely due to the fact that at high $z$, dark energy becomes dynamically unimportant in conventional models, and within our chosen $w_{0}$--$w_{a}$ parameterization there is little room to vary $w(z)$ in a way that substantially changes high-redshift distances relative to those at lower redshift. Statistical and systematic uncertainties are also greater at higher $z$ (see Figure~\ref{scat}), and the inclusion of {\it zero-point uncertainties} within each imaging scenario also means that such 
strategies are hitting more of a systematic floor than those which are IFC-focused.}

We have examined four low-$z$ imaging-only strategies with two tiers, in which time from the IFC-S and deep tier have been redistributed amongst the shallow and medium tiers of the discovery survey. This method has allowed for the addition of several filters and an increase in each tier's observational area. Each of these strategies  
failed to meet the required number of SNe~Ia (as outlined in the SDT report) at $z > 1.2$. The Imaging:Lowz survey lacks the desired number at $z < 0.6$ and $z > 1.1$, resulting in the small $\FoMopt = 133$.

For the two high-$z$ imaging-only strategies, time from IFC-S and the shallow-tier observations was re-distributed to the medium and deep tiers and observations were made with additional filters. The Imaging:Highz* survey is one of the most successful imaging-only strategies 
with $\FoMopt = 352$. It also has very small statistical uncertainties with $\FoMstat = 704$, the largest statistical-only FoM for any strategy examined.

Figure~\ref{allsys} shows the FoM estimate for each strategy. The SDT Imaging and Imaging:Lowz strategies are less precise than other strategies since their $\FoMopt$ values are close to or below the $\FoMcurr$ values of most other surveys. These strategies are clearly inferior to other options. 

Excluding the SDT Imaging and Imaging:Lowz strategies, IFC-focused strategies are the least successful. Since the SDT and SDT* strategies are equivalent except in classification, the final systematic uncertainty for either strategy would be essentially equivalent. Therefore, the SDT* strategy is superior to that of the SDT report. Of the IFC-focused strategies
however, the SDT* Highz survey is the most successful. 

For the remaining strategies, it is difficult to assign a clear ranking. The effectiveness of each strategy has different dependencies on specific improvements in systematic uncertainties. The $\FoMcurr$ value for many WFC-focused strategies are comparable, an effect that can likely be attributed to the fact that at this point we are becoming systematics limited.

Our simulations provide important information about where to focus efforts. Considering the $\FoMopt$ values, the top 3 strategies are Imaging:Allz, Imaging:Highz*, and Imaging:Highz+, which all have similar $\FoMopt$ values. There is no obvious optimal strategy among those investigated here and with current knowledge. Importantly, imaging-only strategies have been shown to constrain dark energy as well as, and even better than, our current IFC-S strategies. Figure~\ref{pl12} (right) presents the $w_{0}$--$w_{a}$ 68$\%$ and 95$\%$ confidence contours for the simulated SDT, Imaging:Allz, and Imaging:Highz* surveys. These contours illustrate how competitive imaging-only strategies are with respect to an IFC-focused strategy.

The {\it wavelength dependent calibration uncertainty} for the IFC-S system is currently large enough to significantly hamper the effectiveness of any IFC-focused strategy. We have optimistically assumed
that by launch it will improve by a factor of 17 (see Figure~\ref{diffsys}). 
{\red However, since no clear path has been presented for this improvement, we have also investigated how factors of {\blue 2 and 10} improvement (i.e., 25 and 5~mmag per 7000~\AA) affect the final FoM values of the SDT* strategy. For improvement factors of 1 (no improvement; current value), 5, 10, and 17 (optimistic value), we find $\FoMtot =$166, 171, 209, and 216, respectively}.  
For these calculations, the values of the other systematic uncertainties (i.e., {\it nonlinearity}, {\it host-mass evolution}, {\it population drift}, and {\it intrinsic scatter}) are set to their optimistic values. It is clear that a precision of at least 5~mmag per 7000~\AA\ is required for optimal implementation of an IFC-focused strategy.

In addition, imaging-only strategies like Imaging:Allz and Highz* may have an advantage because their data can be divided into subsamples for further systematic studies, including high and low-$z$ host-mass and high and low-$z$ Galactic extinction studies. If new effects are found such as $\beta(z)$ or a better host-mass function, then imaging-only strategies with superior statistics will prove better for measuring these additional parameters.  

\section{Discussion and Future Work}
\label{diss}

The strategies outlined in this paper illustrate how the {\it WFIRST} SN survey can be modified to increase the number of SNe~Ia 
and to increase the redshift range over which they are found. These strategies are intended as reference options that can be updated and expanded upon to perform more rigorous optimizations. 

Future optimization of the survey may include trading depth or area and adjusting the cadence of the light curves. In addition, the current redshift distribution proposed by the SDT report could be further optimized with relatively small modifications to the survey. Below we discuss in more detail some of the ways in which the survey could be optimized.

Our analysis currently assumes that the redshift of each SN or its host galaxy is perfectly known. In reality, we will know the redshift of the SN with varying levels of accuracy
based on how well it is determined. The accuracy of the redshift affects observation choices such as exposure times, the precision of classification routines, and potential biases that propagate to the Hubble diagram. Meanwhile, the uncertainty in the redshift propagates directly to constraining cosmological parameters.

{\red To obtain redshifts the {\it WFIRST} strategy will likely use a combination of high-resolution spectroscopic host-galaxy redshifts, lower-resolution {\it WFIRST} grism host-galaxy redshifts, SN+galaxy photometric redshifts, and spectroscopic redshifts from the SNe themselves. Further complicating the issue, the redshifts (and their uncertainties) will be updated and improved during the course of the survey.}

{\red If redshifts were to be obtained primarily via photometry of the SN~Ia and its host galaxy (i.e., photo-$z$), then it is important to understand how well these photo-$z$ values can be determined.

The photometric redshift noise, $\Delta z/(1+z)$, scales roughly as SNR$^{-1}$, and at some point a ``degeneracy floor'' is reached because of the degeneracy between intrinsic color and redshift.
Ideally, we would obtain uncertainties no greater than $0.02\times (1+z)$, which may be possible for a subset ($\sim 20\%$) of SNe found within luminous red galaxies where the SNR is high and calibration biases are well controlled. The majority of SN galaxies, however, will possess photo-$z$ uncertainties within the range of (0.03--0.04) $\times (1+z)$ (see the weak-lensing requirements of the SDT report), the extra scatter in the Hubble diagram is high, and there are huge pathologies between redshift and color leading to significant issues. Modern photo-$z$ codes contain many assumptions and are constantly evolving, and thus further work is required within this field to improve the results found.}

A full analysis of the various effects associated with each method of redshift determination (e.g., photo-$z$, high-resolution host-galaxy spectroscopy, host-galaxy grism data, SN spectra) requires an accurate assessment of the redshift catalogs present at the beginning of the SN survey, the ground-based resources available during the survey, the exact {\it WFIRST} survey strategy, and resources available upon completion  of the survey.  With estimates of the available resources, we can assign redshifts with appropriate accuracy to each simulated SN and determine how each survey strategy is affected.

Our simulations have followed the current NASA mandate that all SN discovery and follow-up observations be performed exclusively by {\it WFIRST}. However, we will likely perform follow-up observations of {\it WFIRST}-discovered SNe using ground-based facilities; furthermore there will be {\it WFIRST} follow-up observations of SNe discovered in ground-based surveys. There could be significant efficiency gains, for example, by using
{\red SNe~Ia discovered by the Large Synoptic Survey Telescope \citep[LSST;][]{Ivezic08} and choosing some
{\it WFIRST}-SN and LSST fields to overlap. In IFC-S focused strategies, the inclusion of LSST data would essentially remove the need for a shallow imaging tier (potentially up to $z \approx 0.8$), meaning that at low $z$ {\it WFIRST} would be used for {\blue spectrophotometric follow-up} observations of the LSST SN sample only. 
The discovery search time from the shallow tier could then be redistributed to the higher-$z$ imaging tiers, increasing the likelihood of detection for SNe at these redshift. Conversely, time gained from the shallow imaging tier could be used to obtain additional low-$z$ {\it WFIRST} spectra and thus improve the spectrophotometric calibration between the two surveys.
In addition, if SN classifications can be obtained using ground-based 8--30~m telescopes (e.g., Gemini, TMT), then low-$z$ IFC-S spectra are no longer required (assuming that the relative calibration between the facilities is well known) and additional time can be spent observing SNe with $z \geq 0.8$.}

{\red For imaging-driven strategies, the inclusion of LSST data would remove the shallow imaging tier, allowing more time to be dedicated to SNe at high $z$. Alternatively, a series of short {\it WFIRST} WFC exposures similar to that of LSST might enable the reduction of calibration systematics
(including variable seeing, extended periods of bad weather affecting cadence, etc., for ground-based data) between the two surveys, providing a necessary template between ground-based and space-based datasets. Future simulations will examine the possibility of multiple scenarios for using ground-based observatories to enhance the {\it WFIRST} SN survey.}

The use of the grism has not yet been fully explored or simulated.  We performed preliminary simulations, finding that grism spectroscopy would be effective for classification, but only with longer exposure times than for the IFC-focused strategies \citep[see ][for a detailed examination of the {\it HST} WFC3 grism for this purpose]{Jones13}. More extensive simulations are necessary to determine if the grism is useful for the SN survey.

Usage of both the IFC-S and WFC imaging components with their current 5~day cadence 
would require considerable resources for scheduling.

Within 5~days, data would have to be downloaded, processed, searched for transients, objects fit and selected, IFC-S follow-up observation schedules built and sent, and finally the instrument set to observing again. 

A longer cadence of 7 days, or even a flexible cadence, may simplify some of the scheduling
issues and have little to no scientific impact on the mission. Modified cadence investigations are needed for strategies using the IFC-S.

The idea of using parallel observing for the IFC-S and WFC must also be considered. Parallel observing would allow {\it WFIRST} to operate both the WFC and IFC-S at the same time. Preliminary calculations suggest that given the huge number of IFC-S observations, the WFC imaging fields will be almost completely covered. However, within this basic calculation we have ignored the possibility of selecting particular roll angles, or the likely correlations between angles for a given SN. This means that there will be different cadences for each SN, and thus patchy and incomplete imaging of that SN.

{\red A WFC imaging-focused survey which utilizes parallelization to obtain
IFC-S spectra during deep $H$ or $F$-band observations (required to obtain SNe at high $z$ with ${\rm SNR} \geq 10$) could, however,
be an important hybrid strategy. Obtaining even a small fraction (10--15\%) of SN~Ia spectra may provide data vital to the analysis; such as improving the underlying SALT2 SED model, {\red obtaining detailed information about the host-galaxy environment, allowing characterization of systematic effects like population drift, and helping us to potentially explore the effects of unknown systematics.}
Future work will include the use of parallel fields within our simulations and the creation of such hybrid strategies.}

An important limitation in our analysis
is the training sample used to determine the underlying
SED model. As described by \citet{Astier14}, the SN model uncertainty can be reduced by using the same rest-frame wavelength range at all redshifts. For a rest-frame wavelength range of 2000--25000~\AA, corresponding to the current {\red extended} SALT2 spectral model {\red (the nominal model is 2800--7000~\AA)}, 
the mean effective wavelength for $RZYJHF$ filters will fall in redshift ranges of
$z < 2.10$,  
$z < 3.35$,  
$z < 4.45$,  
$z < 5.5$,  
$z < 7.00$, and  
$z < 8.4$,      
respectively (for the nominal case the redshift ranges are $z < 1.21$, $0.24 < z < 2.11$, $0.56 < z < 2.89$, $0.86 < z < 3.64$, $1.29 < z < 4.71$, and $1.69 < z < 5.71$ for the $RZYJHF$ filters, respectively).

There have been several efforts to obtain NIR SN~Ia data {\red \citep[e.g.,][]{Krisciunas04:sn, Wood-Vasey08, Stritzinger11, Friedman15, Contreras10, Krisciunas17}. In total, there are $\approx$200 SNe~Ia with published NIR light-curves, with some objects observed independently by multiple groups including the Center for Astrophysics \citep{Friedman15} and the Carnegie Supernova Project \citep{Krisciunas17}.} While most of these data are for low-$z$ SNe, the RAISIN (anagram of SN IA in the IR) project \citep{Kirshner12}
has collected rest-frame NIR data on $\sim 45$ moderate-redshift ($0.2 < z < 0.6$) SNe~Ia with {\it HST}/WFC3. In addition to contributing to the NIR model, these data will be useful for investigating systematic uncertainties related to intrinsic scatter, dust, and color.

There are additional calibration issues for the WFC and IFC-S that need to be taken into consideration when examining systematic uncertainties. Required instrumental characterizations include persistence, flat fields, astrometric mapping of detector to the sky, out-of-band stray light, etc. Initial assessments show that these calibration uncertainties are all second-order systematics that are significantly below the ones included in our current analysis, but should 
be reviewed in future work.

When assessing multiple survey strategies, it is best to have a single, pre-defined metric by which one can compare. With multiple metrics, one can generally choose the metric that is optimal for a particular strategy. That said, there can be critical aspects of a problem that do not affect a metric.  For instance, the DETF FoM that we use to compare strategies does not contain any information related to mission cost/risk or enabling ancillary science.

Furthermore, the DETF FoM is not the only metric by which we can optimize our understanding of dark energy. For instance, eigenvectors \citep{Huterer03} have been a popular approach (although \citealt{Linder05} argue that something like the DETF FoM is sufficient for most needs). It will be straightforward to implement modified  
dark-energy characterizations into our simulations, 
but additional cosmology-fitting tools may be needed to implement a more comprehensive analysis.

The SN survey defined within the SDT report limits the number of SNe~Ia at high $z$ and focuses on achieving a larger sample within $0.2 \leq z \leq 0.6$. Our imaging-only strategies, on the other hand, place no limit on the number of SNe~Ia within a given redshift bin, and explore out to $z \leq 3.0$. As our surveys have not been optimized, we have not specifically considered the effects of focusing observations within any given redshift range.  However, our preliminary studies have indicated (see Section~\ref{compss}) that an increase in the fraction of SNe~Ia with higher redshifts (i.e., $z > 1.2$) does not necessarily provide the same fractional
increase in a survey's FoM (see Figure~\ref{bin_fom}). This is likely due to the nature of dark energy and the $w_{0}$--$w_{a}$ parameterization. Variations on the redshift distribution should
be considered in
future optimization studies.

Our work has used constraints on the cosmological parameters from both the BAO \citep{Anderson14} and CMB \citep{Planck15} datasets. However, there is ongoing work to include external constraints from projections of CMB S4 \citep{Abazajian16} and future BAO missions such as \citet{DESI16}. \citet{Weinberg13} showed that the impact of Stage 4 SN constraints on the FoM has a strong dependence on the relative constraints from the Stage 4 BAO and future weak lensing probes. Future work is needed 
to replicate this analysis in the context of various SN strategies.

{\red Although the key objective of the {\it WFIRST} SN survey is to obtain data on SNe~Ia, it will image a significant fraction of the sky with a cadence of 5 days. Within this data a plethora of other transient events may be found which could lead to significant scientific gains \citep[e.g.,][]{Scolnic17K}. When designing the survey such synergies with other areas of astrophysics should be considered.}

\section{Conclusion}
\label{con}

Using open-source tools, including newly created ones\footnote{See \url{https://jet.uchicago.edu/blogs/WFIRST/}}, we have produced the first  realistic simulations
of the {\it WFIRST} SN survey.  We examined 11 
strategies in detail, including the strategy presented in the SDT report. For each simulated SN survey strategy, several statistical and systematic uncertainties have been examined and included in order to calculate the $\FoMopt$ value, which we have used as our final measure of success.

{\red One key aspect of this paper is that the proper incorporation of slew-and-settle overheads (which were omitted in the forecasts in the SDT report) drastically reduces the efficiency of the low-redshift, wide-area tier of the SDT's proposed SN survey. These overheads remain an uncertain aspect of mission performance, and optimal design of the {\it WFIRST} SN survey will have a strong dependence on their values.

For our forecasts of the SDT survey and all additional strategies examined, we have retained the SDT's proposed exposure times minus the 42 s overhead (see Table~\ref{exptim}). For each strategy presented the total observing time is always 6 months. 
Recovered survey time from
not using the IFC, or the removal of a survey tier, goes into increasing the survey area of the remaining tiers and/or the addition of filters.}

Examination of the results produced by our SDT simulation (see Section~\ref{SDTSN}) shows that this strategy results in fewer SNe~Ia than outlined in the SDT report. The selection efficiency of the SDT strategy is low, and the noise is significantly underestimated, resulting in many SNe~Ia being cut or misclassified in the final sample by the strict SDT classification routine outlined in Section~\ref{classification}. With $\FoMopt = 158$, this is 
one of the least successful survey strategies investigated (see Figure~\ref{allsys} and Table~\ref{fomvalss}). 
Modification of the selection criteria (SDT*) to 
account for statistical fluctuations
in the rise in flux between epochs,
and restriction of colors to reduce CC SN contamination (and exotic events),
increased $\FoMopt$ to 216. Even within the SDT* simulation, however, there are still too few SNe~Ia selected at $z < 0.6$, due in part to the short 13~s exposure of the shallow imaging tier. Time devoted to the shallow tier of the SDT* survey is placed into the medium tier to produce our SDT* Highz scenario. For this strategy, $\FoMopt$ {\red improves to} 236.

Our imaging-only strategies include
the shallow+medium tiers (suffix Lowz), all three imaging tiers (SDT Imaging and Imaging:Allz), and the medium+deep tiers (suffix Highz). Tier areas were increased and additional filters added to compensate for removing 
the IFC-S and/or discovery tiers. Additional filters 
resulted in
a broader coverage of the rest-frame optical wavelengths (via addition of $R$ and $Z$ bands), a region where the SALT2 SED model is
well defined, and extension into the rest-frame NIR (via addition of the $F$ band). The Imaging:Allz and Imaging:Highz* surveys have $\FoMopt$ values of 388 and 352 (respectively), making them some of the most successful strategies. The Imaging:Highz* survey also has the highest
$\FoMstat$ value at 704. For many of the imaging-only strategies, the number of SNe~Ia within the final sample is significantly higher than that obtained by the SDT strategy. 

Using $\FoMopt$ to measure the success of each strategy, there is no clear winner. The Imaging:Allz, Imaging:Highz*, Imaging:Highz+ simulated surveys all have similar current and optimistic FoM values. 
Although our work suggests that WFC-focused strategies are more successful than IFC-focused, further constraints on systematic uncertainties are needed, especially for the IFC-S. 

There are several additional concerns related to 
an IFC-focused strategy.
Specifically, an IFC-focused strategy requires 
active target selection (likely with human decisions included), which increases operational resources 
and locks in selection bias at the time of target selection. The ability to produce high-precision spectrophotometry with an IFC has yet to be demonstrated, resulting in a higher risk of reaching systematic uncertainty goals than for an imaging-only strategy. A further limitation of IFC-focused strategies is the limited ability to divide their relatively small sample sizes into subsamples for systematic studies.

{\red A parallel imaging survey conducted during IFC observations could mitigate these concerns, but the power of such a survey relative to an optimized imaging-only survey has not yet been assessed.}

{\red One virtue of an IFC-focused strategy that has not yet been considered is the stacking of multiple low-SNR spectra. Each individual spectrum may not be useful for classification, but stacked spectra may be useful for measuring the evolution of spectral diversity with redshift. 
In general, IFC-focused strategies may afford
advantages in calibration, in that they will have less sensitivity than filter photometry to SN SED evolution, and spectra can provide more information in determining population drift systematics.}

While the strategies we have presented are not fully optimized, they provide a broader understanding of the possibilities for the survey. Moreover, at this stage in the mission, such an investigation is critical for mitigating risk and ensuring the ultimate success of {\it WFIRST}. Our initial investigations have determined that there is no single correct survey scenario for the mission, yet all our top-performing strategies provide a significant improvement in comparison to current surveys which utilize SNe~Ia as cosmological probes, and they progress toward that which is expected by a Stage 4 experiment. 

{\red Our work has focused on establishing a reliable and reproducible set of baseline strategies, using well-defined methods and software tools. In future work we plan to continue optimizing the {\WFIRST} mission to achieve a more successful strategy.}

\section*{Acknowledgements}
{\blue We would like to dedicate this paper to the memory of Neil Gehrels. His hard work and enthusiasm as the WFIRST Project Scientist helped make the mission possible.}

This manuscript is based upon work supported by the National Aeronautics and Space Administration (NASA) under Contract No.\ NNG16PJ34C issued through the {\it WFIRST} Science Investigation Teams Program.
It was also supported in part by the U.S. Department of Energy under contract DE--AC02--76CH03000.  Analysis was done using the Midway-RCC computing cluster at the University of Chicago.

R.H., D.S., and R.J.F.\ were supported in part by NASA grant 14--WPS14--0048. The UCSC group is supported in part by fellowships to R.J.F.\ from the Alfred P.\ Sloan Foundation and the David and Lucile Packard Foundation. 
D.S.\ and R.K.\ acknowledge support from the Kavli Institute for Cosmological Physics at the University of Chicago through National Science Foundation (NSF) grant PHY--1125897 and an endowment from the Kavli Foundation and its founder Fred Kavli. 
Hubble/KICP Fellow D.S.\ is also supported by NASA through Hubble Fellowship grant HST--HF2--51383.001 awarded by the Space Telescope Science Institute (STScI), which is operated by the Association of Universities for Research in Astronomy, Inc., for NASA, under contract NAS 5-26555.
V.M.\ was supported in part by the Charles E.\ Kaufman Foundation, a supporting organization of the Pittsburgh Foundation.
Supernova cosmology at the Harvard College Observatory is supported in part by the NSF through grants AST--1516854 and AST--1211196, and NASA grant NNX15AJ55G. R.P.K.\ and A.A\ were supported in part by the RAISIN1 GO-13046 and RAISIN2 GO-14216 {\it HST} grants, which were administered by STScI.
A.V.F.\ and P.L.K.\ received support from the Christopher R.                              
Redlich Fund, the TABASGO Foundation, the Miller Institute for
Basic Research in Science (U.C. Berkeley), and NASA/{\t HST} grants                                
GO-14041 and GO-14199.

We would like to thank A.\ Friedman for providing his compilation of public NIR photometry, {\red and C.\ Heinrich for their contribution to our cosmological analysis.}

{\red In addition we would also like to thank our referee, {\blue Prof. D. Weinberg. Professor Weinberg's} thoughtful comments and insight served to significantly improve the paper.}



\bibliographystyle{aasjournal_v1.16}
\bibliography{astro_refs}


\appendix
\section{IFC-S Extended Table:}
\label{ap1}
Table~\ref{ifuextend} presented within this Appendix provides a complete listing of each IFC-S bin used within our simulations. A machine-readable version of this table is available online. 

\startlongtable
\begin{deluxetable}{ccccc}
\tabletypesize{\scriptsize}
\tablecaption{Each IFC-S bin between 0.42 and 2.1 $\mu m$. Minimum and maximum wavelength ranges for each bin are {\blue given} along with the FWHM in pixels and associated sources of noise. 
\label{ifuextend}}
\tablehead{
\colhead{Maximum} & \colhead{Minimum} & \colhead{PSF} & \colhead{Zodiacal Noise} & \colhead{Thermal Noise}\\
\colhead{Wavelength} & \colhead{Wavelength} & \colhead{FWHM} & \colhead{(e$^{-}$~s$^{-1}$~pixel$^{-1}$)} & \colhead{(e$^{-}$~s$^{-1}$~pixel$^{-1}$)}\\
\colhead{(\AA)} & \colhead{(\AA)} & \colhead{(pixels)}  & \colhead{} & \colhead{}
}
\startdata
4200.00	&	4209.35	&	1.540	&	0.000	&	0.000	\\
4209.35	&	4218.76	&	1.541	&	0.000	&	0.000	\\
4218.76	&	4228.23	&	1.542	&	0.000	&	0.000	\\
4228.23	&	4237.76	&	1.543	&	0.000	&	0.000	\\
4237.76	&	4247.35	&	1.543	&	0.000	&	0.000	\\
4247.35	&	4257.00	&	1.544	&	0.000	&	0.000	\\
4257.00	&	4266.71	&	1.545	&	0.000	&	0.000	\\
4266.71	&	4276.49	&	1.546	&	0.000	&	0.000	\\
4276.49	&	4286.34	&	1.546	&	0.000	&	0.000	\\
4286.34	&	4296.24	&	1.547	&	0.000	&	0.000	\\
4296.24	&	4306.22	&	1.548	&	0.000	&	0.000	\\
4306.22	&	4316.26	&	1.549	&	0.000	&	0.000	\\
4316.26	&	4326.37	&	1.550	&	0.000	&	0.000	\\
4326.37	&	4336.56	&	1.551	&	0.000	&	0.000	\\
4336.56	&	4346.81	&	1.551	&	0.000	&	0.000	\\
4346.81	&	4357.13	&	1.552	&	0.000	&	0.000	\\
4357.13	&	4367.53	&	1.553	&	0.000	&	0.000	\\
4367.53	&	4378.00	&	1.554	&	0.000	&	0.000	\\
4378.00	&	4388.55	&	1.555	&	0.000	&	0.000	\\
4388.55	&	4399.18	&	1.556	&	0.000	&	0.000	\\
4399.18	&	4409.88	&	1.557	&	0.000	&	0.000	\\
4409.88	&	4420.67	&	1.558	&	0.000	&	0.000	\\
4420.67	&	4431.53	&	1.559	&	0.000	&	0.000	\\
4431.53	&	4442.48	&	1.559	&	0.000	&	0.000	\\
4442.48	&	4453.51	&	1.560	&	0.000	&	0.000	\\
4453.51	&	4464.63	&	1.561	&	0.000	&	0.000	\\
4464.63	&	4475.83	&	1.562	&	0.000	&	0.000	\\
4475.83	&	4487.13	&	1.563	&	0.000	&	0.000	\\
4487.13	&	4498.51	&	1.564	&	0.000	&	0.000	\\
4498.51	&	4509.99	&	1.565	&	0.000	&	0.000	\\
4509.99	&	4521.56	&	1.566	&	0.000	&	0.000	\\
4521.56	&	4533.23	&	1.567	&	0.000	&	0.000	\\
4533.23	&	4544.99	&	1.568	&	0.000	&	0.000	\\
4544.99	&	4556.84	&	1.569	&	0.000	&	0.000	\\
4556.84	&	4568.78	&	1.570	&	0.000	&	0.000	\\
4568.78	&	4580.81	&	1.571	&	0.000	&	0.000	\\
4580.81	&	4592.93	&	1.572	&	0.001	&	0.000	\\
4592.93	&	4605.14	&	1.573	&	0.001	&	0.000	\\
4605.14	&	4617.44	&	1.574	&	0.001	&	0.000	\\
4617.44	&	4629.84	&	1.575	&	0.001	&	0.000	\\
4629.84	&	4642.34	&	1.576	&	0.001	&	0.000	\\
4642.34	&	4654.94	&	1.578	&	0.001	&	0.000	\\
4654.94	&	4667.63	&	1.579	&	0.001	&	0.000	\\
4667.63	&	4680.43	&	1.580	&	0.001	&	0.000	\\
4680.43	&	4693.34	&	1.581	&	0.001	&	0.000	\\
4693.34	&	4706.35	&	1.582	&	0.001	&	0.000	\\
4706.35	&	4719.47	&	1.583	&	0.001	&	0.000	\\
4719.47	&	4732.71	&	1.584	&	0.001	&	0.000	\\
4732.71	&	4746.05	&	1.586	&	0.001	&	0.000	\\
4746.05	&	4759.51	&	1.587	&	0.001	&	0.000	\\
4759.51	&	4773.09	&	1.588	&	0.001	&	0.000	\\
4773.09	&	4786.79	&	1.589	&	0.001	&	0.000	\\
4786.79	&	4800.62	&	1.590	&	0.001	&	0.000	\\
4800.62	&	4814.57	&	1.592	&	0.001	&	0.000	\\
4814.57	&	4828.64	&	1.593	&	0.001	&	0.000	\\
4828.64	&	4842.85	&	1.594	&	0.001	&	0.000	\\
4842.85	&	4857.19	&	1.595	&	0.001	&	0.000	\\
4857.19	&	4871.67	&	1.597	&	0.001	&	0.000	\\
4871.67	&	4886.29	&	1.598	&	0.001	&	0.000	\\
4886.29	&	4901.06	&	1.599	&	0.001	&	0.000	\\
4901.06	&	4915.97	&	1.601	&	0.001	&	0.000	\\
4915.97	&	4931.03	&	1.602	&	0.001	&	0.000	\\
4931.03	&	4946.25	&	1.603	&	0.001	&	0.000	\\
4946.25	&	4961.60	&	1.605	&	0.001	&	0.000	\\
4961.60	&	4977.07	&	1.606	&	0.001	&	0.000	\\
4977.07	&	4992.68	&	1.608	&	0.001	&	0.000	\\
4992.68	&	5008.41	&	1.609	&	0.001	&	0.000	\\
5008.41	&	5024.28	&	1.611	&	0.001	&	0.000	\\
5024.28	&	5040.28	&	1.612	&	0.001	&	0.000	\\
5040.28	&	5056.43	&	1.614	&	0.001	&	0.000	\\
5056.43	&	5072.71	&	1.615	&	0.001	&	0.000	\\
5072.71	&	5089.14	&	1.617	&	0.001	&	0.000	\\
5089.14	&	5105.71	&	1.618	&	0.001	&	0.000	\\
5105.71	&	5122.44	&	1.620	&	0.001	&	0.000	\\
5122.44	&	5139.32	&	1.621	&	0.001	&	0.000	\\
5139.32	&	5156.35	&	1.623	&	0.001	&	0.000	\\
5156.35	&	5173.55	&	1.624	&	0.001	&	0.000	\\
5173.55	&	5190.91	&	1.626	&	0.001	&	0.000	\\
5190.91	&	5208.43	&	1.628	&	0.001	&	0.000	\\
5208.43	&	5226.13	&	1.629	&	0.001	&	0.000	\\
5226.13	&	5244.00	&	1.631	&	0.001	&	0.000	\\
5244.00	&	5262.05	&	1.633	&	0.001	&	0.000	\\
5262.05	&	5280.28	&	1.635	&	0.001	&	0.000	\\
5280.28	&	5298.70	&	1.636	&	0.001	&	0.000	\\
5298.70	&	5317.31	&	1.638	&	0.001	&	0.000	\\
5317.31	&	5336.12	&	1.640	&	0.001	&	0.000	\\
5336.12	&	5355.12	&	1.642	&	0.001	&	0.000	\\
5355.12	&	5374.34	&	1.644	&	0.001	&	0.000	\\
5374.34	&	5393.76	&	1.645	&	0.001	&	0.000	\\
5393.76	&	5413.41	&	1.647	&	0.001	&	0.000	\\
5413.41	&	5433.27	&	1.649	&	0.001	&	0.000	\\
5433.27	&	5453.37	&	1.651	&	0.001	&	0.000	\\
5453.37	&	5473.70	&	1.653	&	0.001	&	0.000	\\
5473.70	&	5494.27	&	1.655	&	0.001	&	0.000	\\
5494.27	&	5515.09	&	1.657	&	0.001	&	0.000	\\
5515.09	&	5536.16	&	1.659	&	0.001	&	0.000	\\
5536.16	&	5557.50	&	1.661	&	0.001	&	0.000	\\
5557.50	&	5579.04	&	1.664	&	0.001	&	0.000	\\
5579.04	&	5600.80	&	1.666	&	0.001	&	0.000	\\
5600.80	&	5622.78	&	1.668	&	0.002	&	0.000	\\
5622.78	&	5644.99	&	1.670	&	0.002	&	0.000	\\
5644.99	&	5667.43	&	1.672	&	0.002	&	0.000	\\
5667.43	&	5690.10	&	1.675	&	0.002	&	0.000	\\
5690.10	&	5713.02	&	1.677	&	0.002	&	0.000	\\
5713.02	&	5736.18	&	1.679	&	0.002	&	0.000	\\
5736.18	&	5759.60	&	1.682	&	0.002	&	0.000	\\
5759.60	&	5783.27	&	1.684	&	0.002	&	0.000	\\
5783.27	&	5807.22	&	1.686	&	0.002	&	0.000	\\
5807.22	&	5831.43	&	1.689	&	0.002	&	0.000	\\
5831.43	&	5855.93	&	1.691	&	0.002	&	0.000	\\
5855.93	&	5880.72	&	1.694	&	0.002	&	0.000	\\
5880.72	&	5905.80	&	1.696	&	0.002	&	0.000	\\
5905.80	&	5931.19	&	1.699	&	0.002	&	0.000	\\
5931.19	&	5956.89	&	1.702	&	0.002	&	0.000	\\
5956.89	&	5982.90	&	1.704	&	0.002	&	0.000	\\
5982.90	&	6009.19	&	1.707	&	0.002	&	0.000	\\
6009.19	&	6035.76	&	1.710	&	0.002	&	0.000	\\
6035.76	&	6062.64	&	1.713	&	0.002	&	0.000	\\
6062.64	&	6089.81	&	1.715	&	0.002	&	0.000	\\
6089.81	&	6117.29	&	1.718	&	0.002	&	0.000	\\
6117.29	&	6145.09	&	1.721	&	0.002	&	0.000	\\
6145.09	&	6173.21	&	1.724	&	0.002	&	0.000	\\
6173.21	&	6201.67	&	1.727	&	0.002	&	0.000	\\
6201.67	&	6230.47	&	1.730	&	0.002	&	0.000	\\
6230.47	&	6259.62	&	1.733	&	0.002	&	0.000	\\
6259.62	&	6289.13	&	1.736	&	0.002	&	0.000	\\
6289.13	&	6319.02	&	1.740	&	0.002	&	0.000	\\
6319.02	&	6349.28	&	1.743	&	0.002	&	0.000	\\
6349.28	&	6379.94	&	1.746	&	0.002	&	0.000	\\
6379.94	&	6411.01	&	1.749	&	0.002	&	0.000	\\
6411.01	&	6442.49	&	1.753	&	0.002	&	0.000	\\
6442.49	&	6474.40	&	1.756	&	0.002	&	0.000	\\
6474.40	&	6506.76	&	1.760	&	0.003	&	0.000	\\
6506.76	&	6539.58	&	1.763	&	0.003	&	0.000	\\
6539.58	&	6572.86	&	1.767	&	0.003	&	0.000	\\
6572.86	&	6606.55	&	1.771	&	0.003	&	0.000	\\
6606.55	&	6640.62	&	1.774	&	0.003	&	0.000	\\
6640.62	&	6675.07	&	1.778	&	0.003	&	0.000	\\
6675.07	&	6709.92	&	1.782	&	0.003	&	0.000	\\
6709.92	&	6745.17	&	1.786	&	0.003	&	0.000	\\
6745.17	&	6780.83	&	1.790	&	0.003	&	0.000	\\
6780.83	&	6816.91	&	1.794	&	0.003	&	0.000	\\
6816.91	&	6853.43	&	1.798	&	0.003	&	0.000	\\
6853.43	&	6890.40	&	1.802	&	0.003	&	0.000	\\
6890.40	&	6927.82	&	1.806	&	0.003	&	0.000	\\
6927.82	&	6965.72	&	1.811	&	0.003	&	0.000	\\
6965.72	&	7004.09	&	1.815	&	0.003	&	0.000	\\
7004.09	&	7042.96	&	1.819	&	0.003	&	0.000	\\
7042.96	&	7082.34	&	1.824	&	0.003	&	0.000	\\
7082.34	&	7122.24	&	1.828	&	0.003	&	0.000	\\
7122.24	&	7162.69	&	1.833	&	0.003	&	0.000	\\
7162.69	&	7203.68	&	1.838	&	0.003	&	0.000	\\
7203.68	&	7245.25	&	1.843	&	0.003	&	0.000	\\
7245.25	&	7287.41	&	1.847	&	0.003	&	0.000	\\
7287.41	&	7330.18	&	1.852	&	0.003	&	0.000	\\
7330.18	&	7373.57	&	1.857	&	0.003	&	0.000	\\
7373.57	&	7417.56	&	1.863	&	0.003	&	0.000	\\
7417.56	&	7462.07	&	1.868	&	0.003	&	0.000	\\
7462.07	&	7507.12	&	1.873	&	0.003	&	0.000	\\
7507.12	&	7552.73	&	1.878	&	0.003	&	0.000	\\
7552.73	&	7598.90	&	1.884	&	0.004	&	0.000	\\
7598.90	&	7645.66	&	1.889	&	0.004	&	0.000	\\
7645.66	&	7693.00	&	1.895	&	0.004	&	0.000	\\
7693.00	&	7740.96	&	1.901	&	0.004	&	0.000	\\
7740.96	&	7789.55	&	1.907	&	0.004	&	0.000	\\
7789.55	&	7838.79	&	1.913	&	0.004	&	0.000	\\
7838.79	&	7888.68	&	1.919	&	0.004	&	0.000	\\
7888.68	&	7939.26	&	1.925	&	0.004	&	0.000	\\
7939.26	&	7990.53	&	1.931	&	0.004	&	0.000	\\
7990.53	&	8042.41	&	1.937	&	0.004	&	0.000	\\
8042.41	&	8094.86	&	1.944	&	0.004	&	0.000	\\
8094.86	&	8147.89	&	1.950	&	0.004	&	0.000	\\
8147.89	&	8201.51	&	1.957	&	0.004	&	0.000	\\
8201.51	&	8255.73	&	1.963	&	0.004	&	0.000	\\
8255.73	&	8310.56	&	1.970	&	0.004	&	0.000	\\
8310.56	&	8366.02	&	1.977	&	0.004	&	0.000	\\
8366.02	&	8422.12	&	1.984	&	0.004	&	0.000	\\
8422.12	&	8478.88	&	1.991	&	0.004	&	0.000	\\
8478.88	&	8536.30	&	1.998	&	0.004	&	0.000	\\
8536.30	&	8594.41	&	2.006	&	0.004	&	0.000	\\
8594.41	&	8653.15	&	2.013	&	0.004	&	0.000	\\
8653.15	&	8712.45	&	2.021	&	0.004	&	0.000	\\
8712.45	&	8772.33	&	2.028	&	0.004	&	0.000	\\
8772.33	&	8832.79	&	2.036	&	0.004	&	0.000	\\
8832.79	&	8893.84	&	2.044	&	0.004	&	0.000	\\
8893.84	&	8955.50	&	2.052	&	0.004	&	0.000	\\
8955.50	&	9017.77	&	2.060	&	0.004	&	0.000	\\
9017.77	&	9080.65	&	2.068	&	0.004	&	0.000	\\
9080.65	&	9144.17	&	2.076	&	0.004	&	0.000	\\
9144.17	&	9208.33	&	2.084	&	0.004	&	0.000	\\
9208.33	&	9273.15	&	2.093	&	0.004	&	0.000	\\
9273.15	&	9338.62	&	2.101	&	0.004	&	0.000	\\
9338.62	&	9404.77	&	2.110	&	0.004	&	0.000	\\
9404.77	&	9471.53	&	2.119	&	0.004	&	0.000	\\
9471.53	&	9538.83	&	2.127	&	0.004	&	0.000	\\
9538.83	&	9606.67	&	2.136	&	0.004	&	0.000	\\
9606.67	&	9675.05	&	2.145	&	0.004	&	0.000	\\
9675.05	&	9743.98	&	2.155	&	0.004	&	0.000	\\
9743.98	&	9813.48	&	2.164	&	0.004	&	0.000	\\
9813.48	&	9883.53	&	2.173	&	0.004	&	0.000	\\
9883.53	&	9954.16	&	2.183	&	0.004	&	0.000	\\
9954.16	&	10025.36	&	2.192	&	0.004	&	0.000	\\
10025.36	&	10097.14	&	2.202	&	0.004	&	0.000	\\
10097.14	&	10169.51	&	2.212	&	0.004	&	0.000	\\
10169.51	&	10242.46	&	2.222	&	0.004	&	0.000	\\
10242.46	&	10315.89	&	2.232	&	0.004	&	0.000	\\
10315.89	&	10389.74	&	2.242	&	0.004	&	0.000	\\
10389.74	&	10463.99	&	2.252	&	0.004	&	0.000	\\
10463.99	&	10538.65	&	2.262	&	0.004	&	0.000	\\
10538.65	&	10613.74	&	2.272	&	0.004	&	0.000	\\
10613.74	&	10689.23	&	2.283	&	0.004	&	0.000	\\
10689.23	&	10765.14	&	2.293	&	0.004	&	0.000	\\
10765.14	&	10841.47	&	2.304	&	0.004	&	0.000	\\
10841.47	&	10918.22	&	2.314	&	0.004	&	0.000	\\
10918.22	&	10995.39	&	2.325	&	0.004	&	0.000	\\
10995.39	&	11072.97	&	2.336	&	0.004	&	0.000	\\
11072.97	&	11150.81	&	2.347	&	0.004	&	0.000	\\
11150.81	&	11228.89	&	2.358	&	0.004	&	0.000	\\
11228.89	&	11307.20	&	2.369	&	0.004	&	0.000	\\
11307.20	&	11385.75	&	2.380	&	0.004	&	0.000	\\
11385.75	&	11464.53	&	2.391	&	0.004	&	0.000	\\
11464.53	&	11543.54	&	2.402	&	0.004	&	0.000	\\
11543.54	&	11622.79	&	2.413	&	0.004	&	0.000	\\
11622.79	&	11702.27	&	2.425	&	0.004	&	0.000	\\
11702.27	&	11781.97	&	2.436	&	0.004	&	0.000	\\
11781.97	&	11861.91	&	2.447	&	0.004	&	0.000	\\
11861.91	&	11942.07	&	2.459	&	0.004	&	0.000	\\
11942.07	&	12022.46	&	2.470	&	0.004	&	0.000	\\
12022.46	&	12103.08	&	2.482	&	0.004	&	0.000	\\
12103.08	&	12183.92	&	2.494	&	0.004	&	0.000	\\
12183.92	&	12264.98	&	2.505	&	0.004	&	0.000	\\
12264.98	&	12346.09	&	2.517	&	0.004	&	0.000	\\
12346.09	&	12427.16	&	2.529	&	0.004	&	0.000	\\
12427.16	&	12508.19	&	2.541	&	0.004	&	0.000	\\
12508.19	&	12589.19	&	2.552	&	0.004	&	0.000	\\
12589.19	&	12670.15	&	2.564	&	0.004	&	0.000	\\
12670.15	&	12751.07	&	2.576	&	0.004	&	0.000	\\
12751.07	&	12831.95	&	2.588	&	0.004	&	0.000	\\
12831.95	&	12912.80	&	2.600	&	0.004	&	0.000	\\
12912.80	&	12993.62	&	2.612	&	0.004	&	0.000	\\
12993.62	&	13074.40	&	2.624	&	0.004	&	0.000	\\
13074.40	&	13155.14	&	2.635	&	0.004	&	0.000	\\
13155.14	&	13235.86	&	2.647	&	0.004	&	0.000	\\
13235.86	&	13316.54	&	2.659	&	0.004	&	0.000	\\
13316.54	&	13397.18	&	2.671	&	0.004	&	0.000	\\
13397.18	&	13477.80	&	2.683	&	0.004	&	0.000	\\
13477.80	&	13558.38	&	2.695	&	0.004	&	0.000	\\
13558.38	&	13638.93	&	2.707	&	0.004	&	0.000	\\
13638.93	&	13719.45	&	2.719	&	0.004	&	0.000	\\
13719.45	&	13799.94	&	2.731	&	0.004	&	0.000	\\
13799.94	&	13880.40	&	2.743	&	0.004	&	0.000	\\
13880.40	&	13960.76	&	2.755	&	0.004	&	0.000	\\
13960.76	&	14040.95	&	2.767	&	0.003	&	0.000	\\
14040.95	&	14120.99	&	2.779	&	0.003	&	0.000	\\
14120.99	&	14200.86	&	2.791	&	0.003	&	0.000	\\
14200.86	&	14280.58	&	2.803	&	0.003	&	0.000	\\
14280.58	&	14360.15	&	2.815	&	0.003	&	0.000	\\
14360.15	&	14439.57	&	2.827	&	0.003	&	0.000	\\
14439.57	&	14518.84	&	2.839	&	0.003	&	0.000	\\
14518.84	&	14597.96	&	2.851	&	0.003	&	0.000	\\
14597.96	&	14676.95	&	2.863	&	0.003	&	0.000	\\
14676.95	&	14755.79	&	2.875	&	0.003	&	0.000	\\
14755.79	&	14834.49	&	2.887	&	0.003	&	0.000	\\
14834.49	&	14913.06	&	2.899	&	0.003	&	0.000	\\
14913.06	&	14991.49	&	2.911	&	0.003	&	0.000	\\
14991.49	&	15069.80	&	2.923	&	0.003	&	0.000	\\
15069.80	&	15147.97	&	2.935	&	0.003	&	0.000	\\
15147.97	&	15226.01	&	2.947	&	0.003	&	0.000	\\
15226.01	&	15303.93	&	2.959	&	0.003	&	0.000	\\
15303.93	&	15381.66	&	2.971	&	0.003	&	0.000	\\
15381.66	&	15459.16	&	2.983	&	0.003	&	0.000	\\
15459.16	&	15536.43	&	2.994	&	0.003	&	0.000	\\
15536.43	&	15613.49	&	3.006	&	0.003	&	0.000	\\
15613.49	&	15690.32	&	3.018	&	0.003	&	0.000	\\
15690.32	&	15766.94	&	3.030	&	0.003	&	0.000	\\
15766.94	&	15843.35	&	3.042	&	0.003	&	0.000	\\
15843.35	&	15919.56	&	3.053	&	0.003	&	0.000	\\
15919.56	&	15995.57	&	3.065	&	0.003	&	0.000	\\
15995.57	&	16071.37	&	3.077	&	0.003	&	0.000	\\
16071.37	&	16146.98	&	3.089	&	0.003	&	0.000	\\
16146.98	&	16222.40	&	3.100	&	0.003	&	0.000	\\
16222.40	&	16297.63	&	3.112	&	0.003	&	0.000	\\
16297.63	&	16372.68	&	3.124	&	0.003	&	0.000	\\
16372.68	&	16447.54	&	3.135	&	0.003	&	0.000	\\
16447.54	&	16522.23	&	3.147	&	0.003	&	0.000	\\
16522.23	&	16596.71	&	3.158	&	0.003	&	0.000	\\
16596.71	&	16670.98	&	3.170	&	0.003	&	0.000	\\
16670.98	&	16745.06	&	3.181	&	0.003	&	0.000	\\
16745.06	&	16818.93	&	3.193	&	0.003	&	0.000	\\
16818.93	&	16892.61	&	3.204	&	0.003	&	0.000	\\
16892.61	&	16966.10	&	3.216	&	0.003	&	0.000	\\
16966.10	&	17039.40	&	3.227	&	0.002	&	0.000	\\
17039.40	&	17112.51	&	3.239	&	0.002	&	0.000	\\
17112.51	&	17185.43	&	3.250	&	0.002	&	0.000	\\
17185.43	&	17258.18	&	3.262	&	0.002	&	0.000	\\
17258.18	&	17330.75	&	3.273	&	0.002	&	0.000	\\
17330.75	&	17403.06	&	3.284	&	0.002	&	0.000	\\
17403.06	&	17475.09	&	3.296	&	0.002	&	0.000	\\
17475.09	&	17546.85	&	3.307	&	0.002	&	0.000	\\
17546.85	&	17618.34	&	3.318	&	0.002	&	0.000	\\
17618.34	&	17689.58	&	3.329	&	0.002	&	0.000	\\
17689.58	&	17760.55	&	3.341	&	0.002	&	0.000	\\
17760.55	&	17831.27	&	3.352	&	0.002	&	0.000	\\
17831.27	&	17901.74	&	3.363	&	0.002	&	0.000	\\
17901.74	&	17971.97	&	3.374	&	0.002	&	0.000	\\
17971.97	&	18041.97	&	3.385	&	0.002	&	0.000	\\
18041.97	&	18111.72	&	3.396	&	0.002	&	0.000	\\
18111.72	&	18181.25	&	3.407	&	0.002	&	0.000	\\
18181.25	&	18250.55	&	3.418	&	0.002	&	0.000	\\
18250.55	&	18319.63	&	3.429	&	0.002	&	0.000	\\
18319.63	&	18388.49	&	3.440	&	0.002	&	0.001	\\
18388.49	&	18457.14	&	3.451	&	0.002	&	0.001	\\
18457.14	&	18525.57	&	3.461	&	0.002	&	0.001	\\
18525.57	&	18593.80	&	3.472	&	0.002	&	0.001	\\
18593.80	&	18661.82	&	3.483	&	0.002	&	0.001	\\
18661.82	&	18729.64	&	3.494	&	0.002	&	0.001	\\
18729.64	&	18797.26	&	3.505	&	0.002	&	0.001	\\
18797.26	&	18864.69	&	3.515	&	0.002	&	0.001	\\
18864.69	&	18931.93	&	3.526	&	0.002	&	0.001	\\
18931.93	&	18998.95	&	3.537	&	0.002	&	0.001	\\
18998.95	&	19065.75	&	3.547	&	0.002	&	0.001	\\
19065.75	&	19132.34	&	3.558	&	0.002	&	0.001	\\
19132.34	&	19198.71	&	3.568	&	0.002	&	0.002	\\
19198.71	&	19264.86	&	3.579	&	0.002	&	0.002	\\
19264.86	&	19330.81	&	3.590	&	0.002	&	0.002	\\
19330.81	&	19396.56	&	3.600	&	0.002	&	0.002	\\
19396.56	&	19462.10	&	3.611	&	0.002	&	0.002	\\
19462.10	&	19527.45	&	3.621	&	0.002	&	0.002	\\
19527.45	&	19592.60	&	3.631	&	0.002	&	0.002	\\
19592.60	&	19657.57	&	3.642	&	0.002	&	0.003	\\
19657.57	&	19722.34	&	3.652	&	0.002	&	0.003	\\
19722.34	&	19786.92	&	3.662	&	0.002	&	0.003	\\
19786.92	&	19851.33	&	3.673	&	0.002	&	0.003	\\
19851.33	&	19915.55	&	3.683	&	0.002	&	0.004	\\
19915.55	&	19979.60	&	3.693	&	0.002	&	0.004	\\
19979.60	&	20043.47	&	3.704	&	0.002	&	0.004	\\
20043.47	&	20107.16	&	3.714	&	0.002	&	0.004	\\
20107.16	&	20170.69	&	3.724	&	0.002	&	0.005	\\
20170.69	&	20234.01	&	3.734	&	0.002	&	0.005	\\
20234.01	&	20297.13	&	3.744	&	0.002	&	0.006	\\
20297.13	&	20360.05	&	3.754	&	0.002	&	0.006	\\
20360.05	&	20422.78	&	3.764	&	0.001	&	0.006	\\
20422.78	&	20485.32	&	3.775	&	0.001	&	0.007	\\
20485.32	&	20547.67	&	3.785	&	0.001	&	0.007	\\
20547.67	&	20609.83	&	3.795	&	0.001	&	0.008	\\
20609.83	&	20671.81	&	3.805	&	0.001	&	0.008	\\
20671.81	&	20733.62	&	3.815	&	0.001	&	0.009	\\
20733.62	&	20795.24	&	3.824	&	0.001	&	0.009	\\
20795.24	&	20856.69	&	3.834	&	0.001	&	0.010	\\
20856.69	&	20917.97	&	3.844	&	0.001	&	0.011	\\
20917.97	&	20979.08	&	3.854	&	0.001	&	0.011	\\
20979.08	&	21000.00	&	3.861	&	0.001	&	0.012	\\		
\enddata
\end{deluxetable}     

%

\section{Host-galaxy Surface Brightness Library}
\label{ap1a}

\begin{figure*}
\centering
\includegraphics[keepaspectratio=true, width=\textwidth]{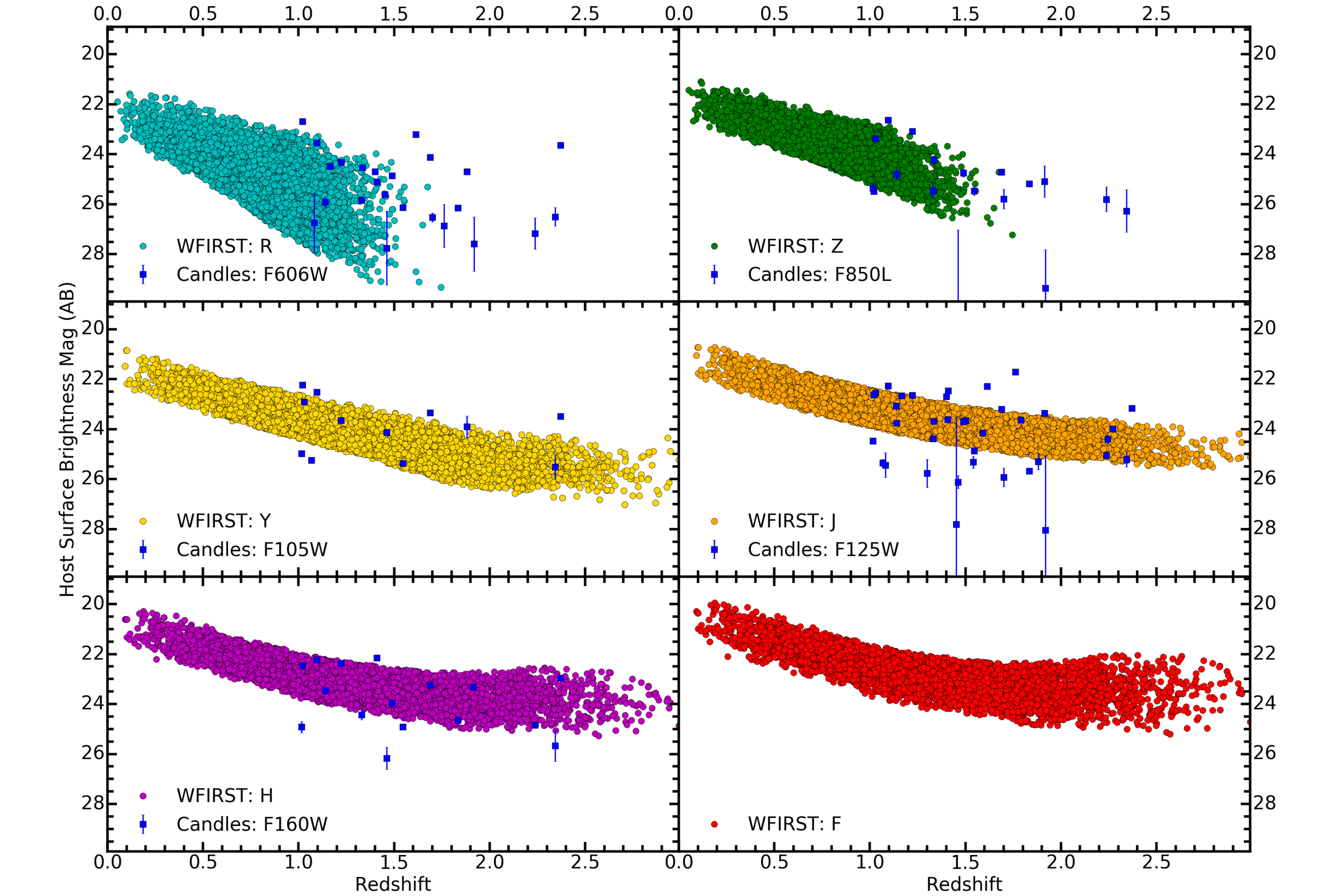}
\caption{Simulated host-galaxy surface brightness 
 vs. redshift
for each {\it WFIRST} filter, 
and comparison with data from CANDLES.}
\label{hostsb}
\end{figure*}

{\red Using CANDLES data we obtained host-galaxy surface brightnesses at the site of $\sim$40 SN events. Local surface brightness fluxes were obtained 
on images without SN light, using a 0.2\arcsec\ radius aperture with the following {\it HST} filters: $F606W$, $F775W$, $F850L$, $F105W$, $F125W$, $F140W$, and $F160W$. Rest-frame SED galaxy models were redshifted and warped so that their synthetic photometry agreed with the observed {\it HST} photometry. The resulting best-fit SEDs were used to compute synthetic photometry in the {\WFIRST} filters as a function of redshift.

SNANA can generate galaxies with arbitrary profiles and magnitudes; however, since our {\it HST} data were measured only at the site of the SN, we forced our simulated SN location to be at the center of its host galaxy with a Sersic profile index of 0.5 and a slight offset applied to the magnitude. To test how well this approximation works, 
Figure~\ref{hostsb} compares the 
simulated {\it WFIRST} host-galaxy surface brightnesses to that of corresponding data in a CANDLES filter. 
Although not perfect, this simple approximation was implemented throughout our study.}

\section{Photometric Cuts in the IFC-S data}
\label{ap1b}

\begin{figure*}
\centering
\includegraphics[keepaspectratio=true, width=\textwidth]{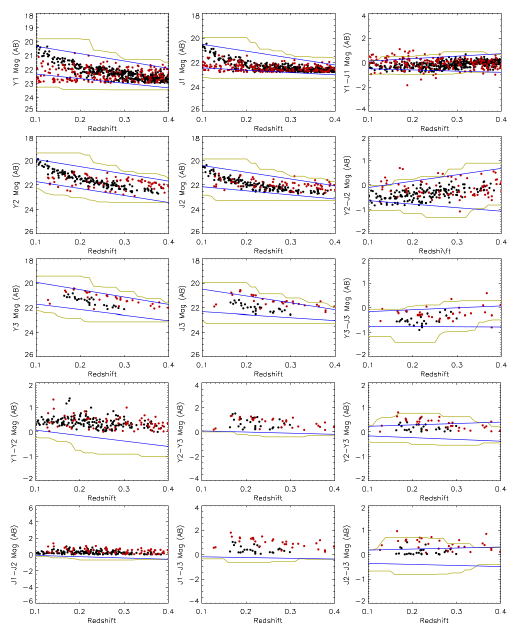}
\caption{Photometric cuts applied to SNe Ia (black circles) and SNe CC (red circles) discovered within the shallow imaging tier of the SDT survey strategy. The top row shows the $Y$ and $J$ magnitudes within the first {\red imaging} epoch ($Y$1/$J$1), and their corresponding colors vs. redshift. The second and third rows represent the same information but for the second and third epochs, respectively ($Y$2/$J$2 and $Y$3/$J$3). The fourth row shows rise data for the $Y$ band, with the fifth showing rise values for the $J$. The blue and gold lines represent rise or color-cut regions, defined by the scatter within the SN sample.}
\label{photc}
\end{figure*}

{\red The SDT report describes a number of photometric cuts to be implemented upon imaging data used in the discovery of SNe. These cuts are listed within Section~\ref{classification}, and require that both the color and flux of a SN be examined from one epoch to the next.

Within our analysis of IFC-S focused strategies, the photometric cuts were applied to imaging data by defining acceptable color and rise value (i.e., increase in flux) ranges from one epoch to the next via an iterative process. 
The ranges were defined using the simulated SN sample as a whole, and encompassed the impact of both the photometric and intrinsic scatter of the SNe. 
Figure~\ref{photc} depicts the
photometric cuts applied to SNe within the shallow tier of the SDT survey. A SN that occupies the resultant ranges has a color which is consistent (with respect to the simulated sample) with a SN~Ia at its host-galaxy redshift, and has an acceptable flux increase from one epoch to the next.

The photometric uncertainties associated with each individual SN, however, were not included when defining these ranges. We note that the exclusion of individual photometric uncertainties leads to rather restrictive color and rise values, but at the time this was our best interpretation of the SDT report. The method of selection adopted within our SDT* strategy is likely more representative of what was originally {\blue planned}, although not stated.}

\section{Measuring the Population Drift} 
\label{ap2}
For a fixed light-curve shape, the evolutionary change of {\it intrinsic color} with redshift 
has been shown to occur with marginal significance 
\citep{Foley12:sdss, Maguire12, Milne15}. 
The color variation has been demonstrated to be
empirically correlated with SN ejecta velocity \citep{Foley11:vel, Foley11:vgrad, Foley12:vel, Mandel14} and therefore
measuring the ejecta velocity can
remove this potential redshift-dependent bias and improve the distance precision.

Since the color change correlation with ejecta velocity is restricted to $\lambda < 4500$~\AA\  in the rest frame, SNe across the {\it WFIRST} redshift range will be affected by different amounts.

Although we could exclude all data blueward of 4500~\AA, this would result in using only $35\%$ of the pixels for a $z = 1.5$ SN~Ia, greatly diminishing the distance precision of these SNe. Alternatively, we should be able to use all data if we can measure a precise velocity. 
We have already shown that this measurement is possible for spectral resolution $R > 75$, if the data are of high quality \citep{Foley13:ca}. 
In fact, an $R = 130$ spectrum has already been used to measure
a \ion{Si}{2} velocity for a $z = 1.55$ SN~Ia \citep{Rodney12}.

The most important feature for measuring the ejecta velocity is Si~II $\lambda$6355, the hallmark feature of SNe~Ia. This feature is blueshifted to $\sim$6100~\AA\ in the rest frame, making it accessible for all IFC-S spectra and all grism spectra at $z > 1.2$. To determine if we can measure the ejecta velocity with realistic IFC-S data, we measured the Si~II velocity for all long-exposure spectra in the final SDT sample. 

We found that the typical velocity uncertainty is
1000~km~s$^{-1}$, with a $\sim 5\%$ failure rate. The ejecta velocities are biased low by $\sim 500$~km~s$^{-1}$, although this bias
can be corrected with measurements from higher-resolution spectra (perhaps from the ground) or from simulations.

This large velocity uncertainty propagates into a 0.10 mag distance modulus uncertainty \citep{Foley11:vgrad}. This relatively large uncertainty (comparable to the total distance uncertainty), is caused by a combination of low resolution and low SNR of the IFC-S spectra. For instance, at infinite SNR, we find a scatter of 340~km~s$^{-1}$ (close to the limit from galactic rotation) and a bias of 180~km~s$^{-1}$. For the grism, the uncertainty decreases to 800~km~s$^{-1}$ for the same (binned) SNR as the long-exposure IFC-S spectrum, indicating that most of the uncertainty is caused by the low SNR.

This spectroscopic study shows that {\it WFIRST} has the potential to measure the SN ejecta velocity, but for this velocity to be helpful for improving distance estimates, we have found that we require 
${\rm SNR} > 20$, beyond the current SDT design. However, a slight modification to the survey design and/or strategy (higher resolution and/or a higher SNR spectrum) would alleviate this problem while simultaneously improving spectral classification. Although this modification 
requires additional exposure time per SN, it would reduce the statistical uncertainty of each SN. 

Other studies have indicated that additional spectral features, including flux ratios from narrow wavelength bands, can improve distance measurements slightly \citep{Bailey09, Blondin11}. Using the simulated long-exposure spectra, we measured the \citet{Bailey09} flux ratios and find
that the uncertainties are generally 20\%, which propagates into a $\sim 0.4$~mag distance modulus uncertainty. This large uncertainty rules out using such flux ratios from the IFC-S at the current SNR.

\end{document}